\documentclass{article}[12pt]

\begin{document}

\begin{center}
\large{\textbf{LECTURE NOTES ON CHERN-SIMONS (SUPER-)GRAVITIES \\ Second Edition - February 2008}}

\vspace{0.5cm}

Jorge Zanelli$^{*}$

\vspace{0.5cm}

Centro de Estudios Cient\'{\i}ficos (CECS), Casilla 1469, Valdivia, Chile
\end{center}

\begin{abstract}
This is intended to be a broad introduction to Chern-Simons gravity and supergravity. The motivation for these theories lies in the desire to have a gauge invariant system --with a fiber bundle formulation-- in more than three dimensions, which could provide a firm ground for constructing a quantum theory of the gravitational field. The starting point is a gravitational action which generalizes the Einstein theory for dimensions $D>4$ --Lovelock gravity. It is then shown that in odd dimensions there is a particular choice of the arbitrary parameters of the action that makes the theory gauge invariant under the (anti-)de Sitter or the Poincar\'{e} groups. The resulting lagrangian is a Chern-Simons form for a connection of the corresponding gauge groups and the vielbein and the spin connection are parts of this connection field. These theories also admit a natural supersymmetric extension for all odd $D$ where the local supersymmetry algebra closes off-shell and without a need for auxiliary fields. No analogous construction is available in even dimensions. A cursory discussion of the unexpected dynamical features of these theories and a number of open problems are also presented.

These notes were prepared for the Fifth CBPF Graduate School, held in Rio de Janeiro in July 2004, published in Portuguese \cite{CBPF}. These notes were in turn based on a lecture series presented at the \emph{Villa de Leyva Summer School 2001} \cite{Villa de Leyva}, at the \emph{La Hechicera School 1999} in M\'{e}rida (Venezuela), and at the 20th Particles and Fields Meeting, Sao Louren\c{c}o, Brazil, Oct. 1999 \cite{hepth0010049}. This second edition contains a number of important corrections of errors and obscure points in the previous versions. A number of new references have been added to improve the text.

\end{abstract}

\vspace{0.5cm}

$^{*}$e-mail: z[at]cecs[dot]cl

\newpage
\begin{flushright}
\begin{tabular}{|c|}
\hline
\textbf{MEN WANTED} \\
\textbf{FOR HAZARDOUS JOURNEY.} \\
LOW WAGES, BITTER COLD, \\
LONG MONTHS OF COMPLETE \\
DARKNESS. CONSTANT DANGER, \\
SAFE RETURN DOUBTFUL. \\
HONOUR AND RECOGNITION \\
IN CASE OF SUCCESS. \\
\verb"             ERNEST SHACKLETON"\cite{shackleton}\\
\hline
\end{tabular}
\end{flushright}

\vskip 0.5cm

\section{The Quantum Gravity Puzzle}
The construction of the action principle for general relativity is reviewed from a modern perspective. The analysis avoids as much as possible to discuss coordinates and invariance under coordinate transformations. Instead, it uses the assumption that the spacetime is a smooth manifold and hence it is endowed with a tangent space which is isomorphic to the Minkowski spacetime. In this description of the spacetime geometry, the metric and affine properties are represented by independent fields, an idea that goes back to the works of Cartan and Palatini. This leads naturally to a formulation of gravity in terms of two independent 1-form fields: the vielbein, $e^{a}$, and the spin connection $\omega_{\;b}^{a}$. The construction is valid for a theory of gravity in any number of dimensions and makes the similarities and contrasts between gravity and a gauge theory explicit.

\subsection{Renormalizability and the triumph of gauge theory}

\subsubsection{Quantum Field Theory} 

The Standard Model of high energy physics is a remarkably successful, enormously precise and predictive theory for particle interactions. With this model, three of the four forces of nature (electromagnetism, weak, and strong interactions) are explained and accurately described. The dynamical structure of the model is a Yang-Mills action, built on the assumption that nature should be invariant under a group of transformations acting independently at each point of spacetime: a local, or ``gauge", symmetry. This symmetry fixes almost uniquely the types of couplings among all the fields that describe both the basic constituents of matter, and the carriers of their interactions.

A most important feature of the standard model is the way in which it avoids inconsistencies: renormalizability, absence of anomalies, etc. In fact, it is this capability which makes it a believable tool and at the same time, it is the prime criterion for its construction.

The remarkable thing is that renormalizability and lack of anomalies are tremendously restrictive conditions, so that a very limited number of possible actions pass the test, which is reassuring. The opposite would be embarrassing: having a large a number of physically sensible theories would prompt the question, how come we don't see the others?

Although not all gauge invariant theories are guaranteed to be renormalizable, the only renormalizable theories that describe our universe are gauge theories. That this is so is an unexpected bonus of the gauge principle since gauge invariance was not introduced to cure the renormalizability problem but rather as a systematic way to bring about interactions that would respect a given symmetry.

Thus, gauge invariance seems to be a crucial ingredient in the construction of physically testable (renormalizable) theories. Symmetry principles, then, are not only useful in constructing the right (classical) action functionals, but they are often sufficient to ensure the viability of a quantum theory built from a given classical action. An intuitive way to understand the usefulness of gauge invariance in a quantum setting is that the gauge symmetry does not depend on the field configurations. Now, if this symmetry relates the divergences appearing in the scattering amplitudes in such a way that they can be absorbed by a redefinition of the parameters in the action at a certain order in the semiclassical (loop) expansion, it should do the same at all orders, as the symmetry is not spoiled by quantum corrections. In contrast with this, some symmetries are only realized if the fields obey the classical equations of motion, something that is often referred to as an \emph{on shell symmetry}. In general, on shell symmetries are not respected by quantum mechanics.

The underlying structure of the gauge principle is mathematically captured through the concept of {\bf fiber bundle}, which is a systematic way to implement a group acting on a set of fields that carry a particular representation of the group. For a discussion of the physical applications, see \cite{Nakahara,EGH}.

\subsubsection{Enter Gravity} 

The fourth interaction of nature, the gravitational attraction, has stubbornly resisted quantization. Attempts to set up a semiclassical or perturbative expansion of the theory have systematically failed and the progress in this direction has remained rather formal, and after more than 70 years of efforts, no consistent quantum theory of gravity in four dimensions is known\footnote{In three spacetime dimensions, the analogous of Einstein gravity is quantizable, although it does not possess local, propagating degrees of freedom \cite{Witten}.}.

The straightforward perturbative approach to quantum gravity was explored by many authors, starting with Feynman \cite{Feynman} and it was soon realized that the perturbative expansion is nonrenormalizable (see, e.g., \cite{'t Hooft}). One way to see this is that if one splits the gravitational lagrangian as a kinetic term plus interaction, the coupling constant for the interaction among gravitons is Newton's constant $G$, which has dimensions of $mass^{-2}$. This means that the diagrams with larger number of vertices require higher powers of momentum in the numerators to compensate the dimensions, and this in turn means that one can expect ultraviolet divergences of all powers to be present in the perturbative expansion. The lesson one can learn from this frustrating exercise is that General Relativity in its most naive interpretation as an ordinary field theory for the metric is, at best, an effective theory. For an interesting discussion of how one can live with an effective theory of gravity, see the recent review by Burgess \cite{Burgess}.

The situation with quantum gravity is particularly irritating because we have been led to think that the gravitational attraction is a fundamental interaction. The dynamical equations of complex systems, like fluids and dispersive media, are not amenable to a variational description, they are not truly fundamental and therefore one should not expect to have a quantum theory for them. The gravitational field on the other hand, is governed by the Einstein equations, which in turn can be derived from an action principle. This is why we could expect, in principle, to define a path integral for the gravitational field, even if calculating with it could be a very nontrivial issue \cite{Teitelboim}.

General Relativity seems to be the only consistent framework that describes gravitational phenomena, compatible with the principle that physics should be insensitive to the state of motion of the observer. This principle is formally translated as invariance under general coordinate transformations, or general covariance. This invariance is a local symmetry, analogous to the gauge invariance of the other three forces and one could be tempted to view it as a gauge symmetry, with interesting consequences. Unfortunately, general relativity doesn't qualify as a gauge theory, except for a remarkable accident in three spacetime dimensions.

One of the differences between a coordinate transformation and a proper gauge transformation is most manifest by the way in which they act on the fields: coordinate transformations change the arguments as well as the field components, whereas a gauge transformation leaves the arguments unchanged. But this is not a very serious obstruction and one can find the right combination of fields so that a change of frame does not change the coordinates. This can be done adopting the tangent space representation and that is what we will do here. A more serious problem is to prove that the action for gravity is invariant under the group of local translations, which is the analog in the tangent space of a local shift in spacetime coordinates. It turns out that, under such transformations, the gravitational action changes by a term which vanishes if the field equations hold and not otherwise. This means that this is at best an \emph{on-shell symmetry}.

In these lectures we attempt to shed some light on this issue, and will show how  the three dimensional accident can be generalized to higher dimensions.

\subsection{Minimal Couplings and Connections} 

Gauge symmetry fixes the form in which matter couples to gauge fields. In electrodynamics, for example, the ordinary derivative in the kinetic term for the matter fields, $\partial_{\mu }$, is replaced by the covariant derivative, \begin{equation}
\nabla _{\mu }=\partial _{\mu } -iA_{\mu }, \label{minimal}
\end{equation}
which is the quantum version of the minimal substitution of classical electrodynamics, $p_{\mu}\rightarrow p_{\mu}-eA_{\mu}$. It is remarkable that this little trick works equally well for either relativistic or non relativistic, classical and quantum theories, accounting for all electromagnetic interactions of matter.

Thus, the gauge coupling provides a unique way to describe interactions with charged fields. At the same time, it doesn't require the introduction of dimensionful constants in the action, which is a welcome feature in perturbation theory since the expansion is likely to be well behaved. Also, gauge symmetry severely restricts the type of counterterms that can be added to the action, as there are very few gauge invariant expressions in a given number of spacetime dimensions. Thus, if the lagrangian contains all possible terms allowed by the symmetry, perturbative corrections could only lead to rescalings of the coefficients in front of each term in the lagrangian. These rescalings, in turn can always be absorbed in a redefinition of the parameters of the action, which is why the renormalization procedure works in gauge theories and is the key to their internal consistency.

The ``vector potential'' $A_{\mu }$ is a connection 1-form, which means that, under a gauge transformation,
\begin{equation}
{\bf A(x)} \rightarrow {\bf A(x)}^{\prime}= {\bf U}(x){\bf A(x)U}(x)^{-1} +
{\bf U}(x)d{\bf U}^{-1}(x), \label{gauge}
\end{equation}
where ${\bf U}(x)$ represents a position-dependent group element. The value of ${\bf A}$ depends on the choice of gauge ${\bf U}(x)$ and it can even be made to vanish at a given point by an appropriate choice of ${\bf U}(x)$. The combination $\nabla_{\mu }$ is the covariant derivative, a differential operator that, unlike the ordinary derivative and ${\bf A}$ itself, transforms homogeneously under the action of the gauge group,
\begin{equation}
\nabla_{\mu} \rightarrow \nabla_{\mu }^{\prime}={\bf U}(x)\nabla_{\mu}{\bf U}(x)^{-1}.
\end{equation}

The connection is in general a matrix-valued object. For instance, in the case of nonabelian gauge theories, ${\bf \nabla }_{\mu}$ is an operator 1-form,
\begin{eqnarray}
{\bf \nabla} &=&d+{\bf A}  \label{covariantDer-1} \\
&=&dx^{\mu }(\partial _{\mu }+{\bf A}_{\mu }).  \nonumber
\end{eqnarray}
Acting on a function $\phi(x)$, which is in a vector representation of the gauge group ($\phi(x)\rightarrow \phi^{\prime }(x)={\bf U}(x)\cdot \phi(x)$), the covariant derivative reads
\begin{equation}
{\bf \nabla}\phi = d\phi + {\bf A}\wedge\phi.\label{delphi}
\end{equation}
The covariant derivative operator ${\bf \nabla}$ has a remarkable property: its square is not a differential operator but a multiplicative one, as can be seen
from (\ref{delphi})
\begin{eqnarray}
{\bf \nabla}{\bf \nabla}\phi&=&d({\bf A}\phi)+ {\bf A}d\phi+{\bf A\wedge A}\phi
\\ \nonumber &=& (d {\bf A}+{\bf A}\wedge{\bf A})\phi \\ \nonumber &=& {\bf
F}\phi
\end{eqnarray}
The combination ${\bf F}= d{\bf A} + {\bf A}\wedge{\bf A}$ is the field strength of the nonabelian interaction. This generalizes the electric and magnetic fields of electromagnetism and it indicates the presence of energy.

One can see now why the gauge principle is such a powerful idea in physics: the covariant derivative of a field, ${\bf \nabla}\phi$, defines the coupling between $\phi$ and the gauge potential ${\bf A}$ in a unique way. Furthermore, ${\bf A}$ has a uniquely defined field strength ${\bf F}$, which in turn defines the dynamical properties of the gauge field. In 1954, Robert Mills and Chen-Nin Yang grasped the beauty and the power of this idea and constructed what has been since known as the nonabelian Yang-Mills theory \cite{Yang-Mills}.  On a curved manifold, there is another operator analogous to ${\bf \nabla}$, also called the covariant derivative in differential geometry,
\begin{eqnarray}
{\bf D} &=&d+{\bf \Gamma }  \label{covariantDer-2}
\\ &=&dx^{\mu }(\partial _{\mu }+{\bf \Gamma}_{\mu }),  \nonumber
\end{eqnarray}
where the components of the connection ${\bf \Gamma}$ are functions of the metric and its derivatives, known as the Christoffel connection or Christoffel symbol. The transformations properties of ${\bf \Gamma}$ under diffeomorphisms are such that the differential operator ${\bf D}$ transforms homogeneously (it is covariant) under the diffeomorphism group. To that extent ${\bf \Gamma}$ acts as a connection, however this is not enough to turn gravity into a gauge
theory. The problem is that this group acts on the coordinates of the manifold as $x^{\mu} \rightarrow x'^{\mu}(x) = x^{\mu}+\xi^{\mu}(x)$, which is a shift in the arguments of the fields (tensors) on which it acts. On the other hand, a gauge transformation in the sense of fiber bundles, acts on the functions and not on their arguments, i.e., it generates a motion along the fiber at a fixed point on the base manifold. For this reason, ${\bf \Gamma }$ is not a connection in the sense of fiber bundles.

\subsection{Gauge Symmetry and General Coordinate Transformations}

The difference between gravity and a standard gauge theory for a Yang-Mills system is aggravated by the fact that the action for gravity in four dimensions cannot be written as that of a gauge invariant system for the diffeomorphism group. In YM theories the connection ${\bf A}_{\mu }$ is an element of a Lie algebra $\mathcal{L}$, but the algebraic properties of $\mathcal{L}$ (structure constants and similar invariants, etc.) are independent of the dynamical fields or their equations. In electroweak and strong interactions, the connection $A$ is dynamical, while both the base manifold and the symmetry groups are fixed, regardless of the values of $A$ or the position in spacetime. This implies that the structure constants are neither functions of the field $A$, or the position $x$. If $G^{a}(x)$ are the gauge generators in a YM theory, they obey an algebra of the form
\begin{equation}
\lbrack G^{a}(x),G^{b}(y)]=C_{c}^{ab}\delta (x,y)G^{c}(x),  \label{LieG}
\end{equation}
where $C_{c}^{ab}$ are the structure constants.

In contrast with this, the algebra of diffeomorphisms takes the form
\begin{equation}
\begin{array}{lll}
\lbrack {\cal H}_{\perp }(x),{\cal H}_{\perp }(y)] & = & g^{ij}(x)
\delta(x,y),_{i}{\cal H}_{j}(y)-g^{ij}(y)\delta (y,x),_{i}{\cal H}_{j}(x) \\
\lbrack {\cal H}_{i}(x),{\cal H}_{j}(y)] & = & \delta
(x,y),_{i}{\cal H}_{j}(y)-\delta (x,y),_{j}{\cal H}_{i}(y) \\
\lbrack {\cal H}_{\perp }(x),{\cal H}_{i}(y)] & = & \delta(x,y),_{i}{\cal
H}_{\perp }(y)
\end{array},  \label{LieH}
\end{equation}
where the ${\cal H}_{\perp }(x)$, ${\cal H}_{i}(x)$ are the generators of time and space translations, respectively, also known as the hamiltonian constraints of gravity, and $\delta (y,x),_{i}=\frac{\partial \delta (y,x)}{\partial x^{i}}$ \cite{Teitelboim-PhD}. Clearly, these generators do not form a Lie algebra but an {\bf open algebra}, which has {\em structure functions} instead of {\em structure constants} \cite{Henneaux}. Here one now finds functions of the dynamical fields, $g^{ij}(x)$ playing the role of the structure constants $C_{c}^{ab}$, which identify the symmetry group in a gauge theory. In this case, the structure ``constants'' may change from one point to another, which means that the symmetry group is not uniformly defined throughout spacetime. This invalidates an interpretation of gravity in terms of fiber bundles, in which the base is spacetime and the symmetry group is the fiber.

\subsubsection{Invariance under general coordinate transformations: an overrated trivial symmetry}

The use of coordinates introduced by Descartes was a great step in mathematics since it allowed to translate geometrical notions into analysis and algebra. The whole Euclidean geometry was reduced to analytic equations. The price was dear, though. Ever since, mathematicians and especially physicists had to live with labels for spacetime points, so much so that space itself was viewed as a number field (${\bf R^3}$). When the geometry of curved surfaces was developed by Gauss, Lobachevsky and Riemann, the use of coordinates became more a nuisance than an advantage as they obscured the intrinsic content of geometric relations.

Tensor calculus was developed as an aid to weed out the meaningful aspects of the geometry from those resulting from the use of particular coordinates: genuine geometrical properties should be insensitive to the changes of coordinate labels. The mere existence of this covariant language goes to show that all reference to coordinates can be consistently hidden from the analysis, in the same way that one can avoid employing units in the derivation of physical laws. Such a langauge exists and it is called exterior calculus. While it may be necessary at some point to refer to a particular coordinate system --as one may also need to refer to \textbf{MKS} units sometimes--, most of the discussion can be made in a coordinate-free manner. An additional advantage of using exterior calculus is that since there are no coordinates, there are no coordinate indices in the expressions. This not only makes life easier for TeX fans, but also simplifies formulas and makes their content more transparent.

So, what is the role of diffeomorphism invariance in this language? None. It is built in, in the same way as the invariance under changes of units is implicit in all physical laws. There is no mathematical content in the former as there is no physical content in the latter. In fact, all of physics and mathematics should be invariant under diffeomorphisms, to the extent that coordinates are human constructs. Lagrange was a mong the first who realized this and declared that a lagrangian for a mechanical system can be written in any coordinates one may choose and the resulting orbits would not depend on that choice.

\newpage
\section{Gravity as Geometry}

Let us now briefly review the standard formulation of General Relativity. On November 25 1915, Albert Einstein presented to the Prussian Academy of Natural Sciences the equations for the gravitational field in the form we now know as Einstein equations \cite{Einstein}. Curiously, five days before, David Hilbert had proposed the correct action principle for gravity, based on a communication in which Einstein had outlined the general idea of what should be the form of the equations \cite{Hilbert}. As we shall see, this is not so surprising in retrospect, because there is a unique action which is compatible with the postulates of general relativity in four dimensions that admits flat space as a solution. If one allows constant curvature geometries, there is essentially a one-parameter family of actions that can be constructed: the Einstein-Hilbert form plus a cosmological term,
\begin{equation} \label{EinsteinAction}
I[g]=\int \sqrt{-g}(\alpha_{1}R+\alpha_{2})d^{4}x,
\end{equation}
where $R$ is the scalar curvature, which is a function of the metric $g_{\mu \nu }$, its inverse $g^{\mu \nu }$, and its derivatives (for the definitions and conventions used here, see Ref.\cite{MTW}). The action $I[g]$ is the only functional of the metric which is invariant under general coordinate transformations and gives second order field equations in four dimensions. The coefficients $\alpha_{1}$ and $\alpha_{2}$ are related to the gravitational constant and the cosmological constant through
\begin{equation}
G=\frac{1}{16\pi \alpha_{1}}\;,\;\Lambda=-\frac{\alpha_{2}}{2\alpha_{1}}.
\label{G-Constants}
\end{equation}

The Einstein equations are obtained by extremizing this action, (\ref{EinsteinAction})
\begin{equation}
R^{\mu}_{\nu} - \frac{1}{2} \delta^{\mu}_{\nu}(R-2\Lambda) =0,
\end{equation} and
and they are unique in that: \\
({\bf i}) they are tensorial equations \\
({\bf ii}) they involve only up to second derivatives of the metric \\
({\bf iii}) they reproduce Newtonian gravity in the nonrelativistic weak field approximation.

The first condition implies that the equations have the same meaning in all coordinate systems. This follows from the need to have a coordinate independent (covariant) formulation of gravity in which the gravitational force is replaced by the nonflat geometry of spacetime. The gravitational field being a geometrical entity implies that it cannot depend on the coordinate choice or, in physical terms, a preferred choice of observers.

The second condition means that Cauchy data are necessary (and sufficient in most cases) to integrate the equations. This condition is a concession to the classical physics tradition: the possibility of determining the gravitational field at any moment from the knowledge of the configuration of space, $g_{ij}(x)$, and its time derivative $\dot{g}_{ij}(x)$, at a given time. This requirement is also the hallmark of Hamiltonian dynamics, which is the starting point for canonical quantum mechanics and therefore suggests that a quantum version of the theory could exist.

The third requirement is the correspondence principle, which accounts for our daily experience that an apple and the moon fall the way they do.

If one further assumes that Minkowski space be among the solutions of the matter-free theory, then one must set $\Lambda =0$, as most sensible particle physicists would do. If, on the other hand, one believes in static homogeneous and isotropic cosmologies, as Einstein did, then $\Lambda$ must have a finely tuned nonzero value. Experimentally, $\Lambda$ has a value of the order of $10^{-120}$ in ``natural'' units ($\hbar=c=G=1$) \cite{Weinberg}. Furthermore, astrophysical measurements seem to indicate that $\Lambda $ must be positive \cite{LambdaExp}. This presents a problem because there seems to be no theoretical way to predict this ``unnaturally small'' nonzero value.

As we will see in what follows, for other dimensions the Einstein-Hilbert action is not the only possibility in order to satisfy conditions ({\bf i-iii}).

\subsection{Metric and Affine Structures} 

We would like  to discuss now what we mean by spacetime geometry. Geometry is sometimes understood as the set of assertions that can be made about points, lines and higher dimensional submanifolds embedded on a given manifold. This broad (and vague) idea, is often viewed as encoded in the metric tensor, $g_{\mu \nu }(x)$, which provides the notion of distance between nearby points with coordinates $x^{\mu}$ and $x^{\mu}+ dx^{\mu}$,
\begin{equation}
ds^{2}=g_{\mu \nu }\;dx^{\mu }dx^{\nu }.  \label{Metric}
\end{equation}

This is the case in Riemannian geometry, where all relevant objects defined on the manifold (distance, area, angles, parallel transport operations, curvature, etc.) can be constructed from the metric. However, a distinction should be made between metric and affine features of spaceime. \textbf{Metricity} refers to measurements of lengths, angles, areas, volumes, of objects which are locally defined in spacetime. \textbf{Affinity} refers to properties which remain invariant under translations --or more generally, affine transformations--, such as parallelism. This distinction is useful because the two notions are logically independent and reducing one to the other is an unnecessary form of violence.

Euclidean geometry was constructed using two elementary instruments: the compass and the unmarked straightedge. The first is a metric instrument because it allows comparing lengths and, in particular, drawing circles. The second is used to draw straight lines which, as will be seen below, is a basic affine operation.

Pythagoras' famous theorem is a metric statement; it relates the lengths of the sides of a triangle with two sides forming a particular angle. Affine properties on the other hand, do not change if the scale is changed, as for example the shape of a triangle or the angle between two straight lines, or the length of a segment. A typical affine statement is, for instance, the fact that when two parallel lines intersect a third, the corresponding angles are equal.

Of course parallelism can be reduced to metricity. As we all learned in school, one can draw a parallel to a line using a compass and an unmarked straightedge: One draws two circles of equal radii with centers at two points of one line and then draws the tangent to the two circles with the ruler. Thus, given a way to measure distances and straight lines in space, one can define parallel transport.

As any child knows from the experience of stretching a string or a piece of rubber band, a straight line is the shape that minimizes the distance between two points. This is clearly a metric feature because it requires {\em measuring} lengths. Orthogonality is also a metric notion that can be defined using the scalar product obtained from the metric. A right angle is a metric feature because in order to build one, one should be able to {\em measure} angles, or measure the sides of triangles\footnote{The Egyptians knew how to {\em use} Pythagoras' theorem to make a right angle, although they didn't know how to prove the theorem (and probably never worried about it). Their recipe was probably known long before, and all good construction workers today still know it: make a loop of rope with 12 segments of equal length. Then, the triangle formed with the loop so that its sides are 3, 4 and 5 segments long is such that the shorter segments are perpendicular to each other \cite{Dunham}.}. We will now argue that {\em parallelism does not require metricity}.

There is something excessive about using a compass for the construction of a parallel to a straight line through a given point. In fact, a rigid wedge of any fixed angle would suffice: align one of the sides of the wedge with the straight line and rest the straightedge on other side; then slide the wedge along the straightedge to reach the desired point. The only requirement is for the wedge {\em not to change} in the process. Thus, the essence of parallel transport is a wedge and a straightedge to connect two points. The only requirement being that the wedge rigid, that is, that the angle between its sides should be preserved, but no measurements of lengths or angles are needed.

There is still some cheating in this argument because we took the construction of a straightedge for granted. How do we construct a straight line if we had no notion of distance? After all, as we discussed above, a straight line is often defined as the shortest path between two points. Fortunately, there is a purely affine way to construct a straight line: Take two short enough segments (two short sticks, matches or pencils would do), and slide them one along the other, as a cross country skier would do. In this way a straight line is generated by {\em parallel transport of a vector along itself}, and we have not used distance anywhere. It is essentially this {\em affine} definition of a straight line that can be found in Book I of Euclid's \emph{Elements}. This definition could be regarded as the {\em straightest line}, which need not coincide with the {\em line of shortest distance}. In fact, if the two sticks one uses are two identical arcs, one could construct families of ``straight" lines by parallel transport of a segment along itself, but they would not correspond to shortest lines. They are conceptually (i.e., logically) independent objects. The fact that this purely affine construction is logically acceptable means that parallel transport is not necessarily a metric concept.

In a space devoid of a metric structure the ``straightest" line could be a rather weird looking curve, but it could still be used to define parallelism. If a ruler has been constructed by transporting a vector along itself, then one can use it to define parallel transport, completely oblivious of the fact that the straight lines are not necessarily the shortest. There would be nothing wrong with such construction except that it will probably not coincide with the standard metric construction.

\subsubsection{Parallelism in an open neighborhood and curvature} 

A general feature of any local definition of parallelism is that it cannot be consistently defined on an open set of an arbitrarily curved manifold. The problem is that in general, parallelism is not a transitive relation if the parallel transport is performed along different curves. Take, for example, an ordinary 2-dimensional sphere. The construction of a straight line by parallel transport of a small segment along itself yields a great circle defined by the intersection of the sphere with a plane that passes through the center of the sphere. If we take three of segments of these great circles and form a triangle, it is possible to transport a vector around the triangle in such a way that the angle between the vector and the arc is kept constant. This notion of parallel transport is the one outlined above, and it is easy to see that the vector obtained by parallel transport around the triangle on a curved surface does not coincide with the original vector. This lack of transitivity on an open neighborhood makes the notion of parallelism fail as an equivalence relation in a curved space.

Here the problem arises when parallel transport is performed along different curves, so one could expect to see no conflict for parallelism along a single ``straight line". However, this is also false as can be checked by parallel transporting a vector along a straight line that encloses the apex of a cone. Now the problem stems from trying to define parallel transport on a region that encloses a singularity of infinite curvature. Note that this effect of curvature only requires a definition of parallel along a segment, which is, again, a purely affine notion. This ia a completely general feature: the curvature of a manifold is completely determined by the notion of parallel transport, independently of the metric.

\subsection{Metric connection} 

In differential geometry, parallelism is encoded in the affine connection $\Gamma_{\beta \gamma}^{\alpha}(x)$: a vector $u_{||}$ at the point of coordinates $x$ is said to be parallel to the vector $u$ at a point with coordinates $x+dx$, if their components are related by ``parallel transport'',
\begin{eqnarray}
u_{||}^{\alpha}(x) &=& u^{\alpha}(x+dx) + dx^{\mu} \Gamma^{\alpha}_{\mu\beta} u^{\beta}(x)  \\ \nonumber &=& u^{\alpha}(x)+ dx^{\mu}[\partial_{\mu} u^{\alpha} + \Gamma^{\alpha}_{\mu\beta} u^{\beta}(x)] . \label{Affine}
\end{eqnarray}

AS we have mentioned, the affine connection $\Gamma _{\mu \beta}^{\alpha }(x)$ need not be functionally related to the metric tensor $g_{\mu \nu }(x)$. However, Einstein formulated General Relativity adopting the point of view that the spacetime metric should be the only dynamically independent field, while the affine connection should be a function of the metric given by the Christoffel symbol,
\begin{equation}
\Gamma_{\mu\beta}^{\alpha}=\frac{1}{2}g^{\alpha \lambda}(\partial_{\mu} g_{\lambda\beta}+
\partial_{\beta}g_{\lambda\mu}-\partial_{\lambda} g_{\mu \beta}).  \label{Christoffel}
\end{equation}

This was the starting point of a controversy between Einstein and Cartan, which is vividly recorded in the correspondence they exchanged between May 1929 and May 1932 \cite{Cartan-Einstein}. In his letters, Cartan politely but forcefully insisted that metricity and parallelism could be considered as independent, while Einstein pragmatically replied that since the space we live in seems to have a metric, it would be more economical to assume the affine connection to be a function of the metric. Cartan advocated economy of assumptions. Einstein argued in favor of economy of independent fields.

Here we adopt Cartan's point of view. It is less economical in dynamical variables but is more economical in assumptions and therefore more general. This alone would not be sufficient ground to adopt Cartan's philosophy, but it turns out to be more transparent in many ways, and lends itself better to discuss the differences and similarities between gauge theories and gravity. Moreover, Cartan's point of view emphasizes the distinction between the spacetime manifold $M$ as the base of a fiber bundle, and the tangent space at every point $T_x$, where Lorentz vectors, tensors and spinors live.

In physics, the metric and the connection play different roles as well. The metric enters in the definition of the kinetic energy of particles and of the stress-energy tensor density in a field theory, $T^{\mu \nu}$. The spin connection instead, appears in the coupling of fermionic fields with the geometry of spacetime. All forms of matter are made out of fermions (leptons and quarks), so if we hope someday to attain a unified description of all interactions including gravity and matter, it is advisable to grant the affine structure of spacetime an important role, on equal footing with the metric structure.

\newpage
\section{First Order Formulation for Gravity} 

\subsection{The equivalence Principle}

As early as 1907, Einstein observed that the effect of gravity can be neutralized by free fall. In a freely falling laboratory, the effect of gravity can be eliminated. This trick is a local one: the lab has to be \emph{small enough} and the time span of the experiments must be \emph{short enough}. Under these conditions, the experiments will be indistinguishable from those performed in absence of gravity, and the laws of physics that will be reflected by the experiments will be those valid in Minkowski space, \emph{i. e.}, Lorentz invariance. This means that, in a local neighborhood, spacetime possesses Lorentz invariance. In order to make manifest this invariance, it is necessary to perform an appropriate coordinate transformation to a particular reference system, viz., a freely falling one.

Conversely, Einstein argued that in the absence of gravity, the gravitational field could be mocked by applying an acceleration the laboratory. This idea is known as the principle of equivalence, meaning that gravitation and acceleration are equivalent effects in a small spacetime region.

A freely falling observer defines a locally inertial system. For a small enough region around him or her the trajectories of projectiles (freely falling as well) are straight lines and the discrepancies with Euclidean geometry are negligible. Particle collisions mediated by short range forces, such as those between billiard balls, molecules or subnuclear particles, satisfy the conservation laws of energy and momentum that are valid in special relativity.

How large is the neighborhood where this approximation is good enough? Well, it depends on the accuracy one is expecting to achieve, and that depends on how curved spacetime really is in that neighborhood. In the region between the earth and the moon, for instance, the deviation from flat geometry is about one part in $10^9$. Whether this is good enough or not, it depends on the experimental accuracy of the tests involved (see \cite{CMWill}).

Thus, we view spacetime as a smooth $D$-dimensional manifold $M$. At every point $x \in M$ there exists a $D$-dimensional flat tangent space $T_x$ of lorentzian signature $(-,+,...,+)$. This tangent space $T_x$ is a good approximation of the manifold $M$ on an open set in the neighborhood of $x$. This means that there is a way to represent tensors on $M$ by tensors on $T_x$, and vice versa. The precise relation between the tensor spaces on $M$ and on $T_{x}$ is an isomorphism represented by means of a linear map $e$.

\subsection{The Vielbein} 

The isomorphism between $M$ and the collection $\{T_{x} \}$ can be realized as a transformation between some orthonormal coordinates in the Minkowski space $T_x$ and the a system of local coordinates on an open neighborhood of $x$, the Jacobian matrix
\begin{equation}
\frac{\partial z^a}{\partial x^{\mu }}=e_{\mu }^{a}(x).  \label{vielbein-0}
\end{equation}
It is sufficient to define its action on a complete set of vectors such as the coordinate separation $dx^{\mu }$ between two infinitesimally close points on $M$. The corresponding separation $dz^{a}$ in $T_{x}$ is
\begin{equation}
dz^{a}=e_{\mu }^{a}(x)dx^{\mu}.  \label{vielbein}
\end{equation}

The family $\{ e_{\mu }^{a}(x), a=1,...,D=dim M \}$ defines a local orthonormal frame on $M$, also called {\it``soldering form''}, {\it``moving frame"}, or simply, \textbf{vielbein}. The definition (\ref{vielbein-0}) implies that $e_{\mu }^a(x)$ transforms as a covariant vector under diffeomorphisms on $M $ and as a contravariant vector under local Lorentz rotations of $T_{x}$, $SO(D-1,1)$ (the signature of the manifold $M$ to be Lorentzian).
This one-to-one correspondence between vectors can be extended to tensors on $M$ and on $T_{x}$: if $\Pi $ is a tensor with components $\Pi^{\mu _{1}...\mu _{n}}(x)$ on $M$, then the corresponding tensor on the tangent space $T_{x}$ is\footnote{The inverse vielbein, $e_{a}^{\mu }(x)$, where $e_{a}^{\nu }(x)e_{\nu }^{b}(x)=\delta _{a}^{b}$, and $e_{a}^{\nu }(x)e_{\mu }^{a}(x)=\delta _{\mu }^{\nu }$, relates lower index tensors, $P_{a_1...a_n}(x)=e_{a_1}^{\mu_1}(x)\cdot \cdot \cdot e_{a_n}^{\mu_n}(x)\Pi_{\mu_1...\mu_n}(x)$.}
\begin{equation}
P^{a_1...a_n}(x)=e_{\mu_1}^{a_1}(x)\cdot \cdot \cdot e_{\mu_n}^{a_n}(x)\Pi ^{\mu_1...\mu_n}(x).
\label{T-map}
\end{equation}

The Lorentzian metric defined in Minkowski space can be used to induce a metric on $M$ through the isomorphism $e_{\mu }^{a}$. The arc length in $T_x$, $ds^2=\eta_{ab}dz^a dz^b$, can also be expressed as $\eta_{ab}e_{\mu}^a(x)e_{\nu}^b(x)dx^{\mu} dx^{\nu}$, where the metric in $M$ is identified,
\begin{equation}
g_{\mu \nu }(x)=e_{\mu }^{a}(x)e_{\nu }^{b}(x)\eta _{ab}.  \label{metric}
\end{equation}
This relation is an example of the map between tensors on $T_x$ and on $M$ that can be read as to mean that the vielbein is the ``square root" of the metric. Since $e_{\mu}^a(x)$ determines the metric, all the metric properties of spacetime are contained in the vielbein. The converse, however, is not true: given a metric $g_{\mu \nu }(x)$, there exist infinitely many choices of vielbein that reproduce the same metric. This infinitely many choices of vielbein correspond to the different choice of local orthonormal frames that can be used as bases for the tangent space vectors at $T_x$. It is possible to rotate the vielbein by a Lorentz transformation and this should be undetectable from the point of view of the manifold $M$. Under a Lorentz transformation, the vielbein are transformed as
\begin{equation}
e_{\mu }^{a}(x)\longrightarrow e_{\mu}^{\prime a}(x)=\Lambda_{\;b}^{a}(x)e_{\mu }^{b}(x),
\label{Transf-e}
\end{equation}
where the matrix ${\bf \Lambda}(x) \in SO(D-1,1)$. By definition of $SO(D-1,1)$,  ${\bf \Lambda}(x)$ leaves the metric in the tangent space unchanged,
\begin{equation}
\Lambda_{\;c}^{a}(x)\Lambda_{\;d}^{b}(x)\eta_{ab}=\eta_{cd}. \label{Lorentz}
\end{equation}
The metric $g_{\mu \nu }(x)$ is clearly unchanged by this transformation. This means, in particular, that there are many more components in $e_{\mu }^{a}$ than in $g_{\mu \nu }$. In fact, the vielbein has $D^{2}$ independent components, whereas the metric has only $D(D+1)/2$. The mismatch is exactly $D(D-1)/2$, the number of independent rotations in $D$ dimensions.

Another way to read (\ref{metric}) is that in this framework the metric is not a fundamental field, but it emerges as a composite object, more or less as a correlator of the vielbein at two points $x$ and $x'$ in the limit $x\rightarrow x'$. In this sense, the metric in this representation in analogous to the coincidence limit of the correlator of the vielbein field which, in a quantum version of the theory would look like,
\begin{equation}\label{correlator}
  g_{\mu \nu}(x)= \lim_{x'\rightarrow x}<e^a_{\mu}(x)e^b_{\nu}(x')>\eta_{a b}.
\end{equation}

\subsection{The Lorentz connection}

The collection of tangent spaces at each point of the manifold defines the action of the symmetry group of the tangent space (the Lorentz group) at each point of $M$, thereby endowing the manifold with a fiber bundle structure, the tangent bundle. In order to define a derivative operator on the manifold, a connection field is required so that the differential structure remains invariant under local Lorentz transformations that act independently at each spacetime point. This field is the Lorentz connection, also called ``spin connection'' in the physics literature, but Lorentz connection could be a more appropriate name. The word ``spin" is due to the fact that $\omega _{\;b\mu }^{a}$ arises naturally in the discussion of spinors, which carry a special representation of the group of rotations in the tangent space, but that is irrelevant at the moment\footnote{Here, only the essential ingredients are given. For a more extended discussion, there are several texts such as those of Refs.\cite{Nakahara}, \cite{EGH}, \cite{Schutz}, \cite{Goeckeler-Schuecker} and \cite{Darling}.}.

The $SO(D-1,1)$ group acts independently on Lorentz tensors at each $T_{x}$. This why the matrices ${\bf \Lambda}$ are functions of $x$. In order to define a derivative of a tensor $\phi$ in $T_{x}$, one would like to have a definition of $D\phi$ to be a tensor of the same rank and nature as $\phi$. This is what happens in all gauge theories: one needs to introduce a connection in order to compensate for the fact that for the gauge group acts independently at neighboring points. In this case, the gauge field is the Lorentz connection, $\omega_{\;b\mu }^{a}(x)$. Suppose $\phi^{a}(x)$ is a field that transforms as a vector under the Lorentz group, $SO(D-1,1)$, its covariant derivative,
\begin{equation}
D_{\mu}\phi^{a}(x)=\partial _{\mu }\phi^{a}(x)+\omega _{\;b\mu}^{a}(x) \phi^{b}(x),  \label{D-mu}
\end{equation}
is defined so that it also transforms like a Lorentz vector at $x$. This requires that under a $SO(D-1,1)$ rotation, $\Lambda_{\;c}^a(x)$, the connection transforms as [see (\ref{gauge})]
\begin{equation}
\omega _{\;b\mu }^{a}(x)\rightarrow \omega_{\;\;b\mu}'^a(x)= \Lambda_{\;c}^a(x) \Lambda_b^{\;d}(x)
\omega_{\;d\mu}^c(x) +\Lambda_{\;c}^{a}(x)\partial_{\mu}\Lambda_b^{\;c}(x), \label{LorConn}
\end{equation}
where $\Lambda_b^{\;d}=\eta_{ab}\eta^{cd} \Lambda^a_{\;c}$ is the inverse (transpose) of $\Lambda_{\;d}^b$.

The connection $\omega_{\;b\mu }^a(x)$ defines the {\em parallel transport} of Lorentz tensors in the tangent space, between nearby points $T_{x}$ and $T_{x+dx}$. The parallel transport of the vector field $\phi^{a}(x)$ from $x+dx$ to $x$, is a vector $\phi_{||}^{a}(x)$, defined as
\begin{eqnarray}
\phi_{||}^{a}(x) &\equiv& \phi^{a}(x+dx)+ dx^{\mu}\omega_{\;b\mu }^{a}(x)\phi^{b}(x) \\ \nonumber
&=& \phi^{a}(x) + dx^{\mu}[\partial_{\mu} \phi^{a}(x)+ \omega_{\;b\mu}^{a}(x)\phi^{b}(x)] \\
\nonumber &=:& \phi^{a}(x)+dx^{\mu} D_{\mu}\phi^{a}(x). \label{ParallelTransp}
\end{eqnarray}
Thus, the covariant derivative $D_{\mu}$ measures the change in a tensor produced by parallel transport between neighboring points,
\begin{eqnarray}
dx^{\mu }D_{\mu }\phi^{a}(x) &=& \phi_{||}^{a}(x)-\phi^{a}(x) \\ \nonumber &=&dx^{\mu}
[\partial_{\mu} \phi^{a}+\omega_{\;b\mu }^{a}(x)\phi^{b}(x)] \label{Lor-cov-der}.
\end{eqnarray}
In this way, the affine properties of space are encoded in the components $\omega_{\;b\mu}^a(x)$, which are, until further notice, totally arbitrary and independent from the metric. Note the similarity between the notion of parallelism in (\ref{ParallelTransp}) and that for vectors whose components are referred to a coordinate basis (\ref{Affine}). These two definitions are independent as they refer to objects on different spaces, but they could be related using the local isomorphism between the base manifold and the tangent space, provided by $e^a_{\mu}$.

\subsection{Invariant tensors}

The group $SO(D-1,1)$ has two invariant tensors, the Minkowski metric, $\eta_{ab}$, and the totally antisymmetric Levi-Civitta tensor, $\epsilon_{a_1 a_2 \cdots a_D}$. These tensors are defined by the algebraic structure of the Lorentz group and therefore they are the same in every tangent space and, consequently, they must be constant throughout the manifold $M$: $d\eta_{ab}=0=d\epsilon_{a_1 a_2 \cdots a_D}$. Moreover, since they are invariant, they must also be covariantly constant,
\begin{eqnarray}
d\eta_{ab}=D\eta_{ab}&=&0, \\
d\epsilon_{a_1 a_2 \cdots s_D}=D\epsilon_{a_1 a_2 \cdots a_D} &=&0.
\end{eqnarray}
This implies that the Lorentz connection satisfies two identities,
\begin{eqnarray}
\eta_{ac} \omega^c_{\;\;b} &=& -\eta_{bc} \omega^c_{\;\;a}, \label{metricity} \\
\epsilon_{b_1 a_2 \cdots a_D}\omega^{b_1}_{\;\;a_1}+\epsilon_{a_1 b_2 \cdots  a_D}\omega^{b_2}_{\;\;a_2}+ &\cdots& + \epsilon_{a_1 a_2 \cdots
b_D}\omega^{b_D}_{\;\;a_D} =0 \label{Chebichev}.
\end{eqnarray}
The requirement that the Lorentz connection be compatible with the metric structure of the tangent space (\ref{metricity}) restricts $\omega^{ab}$ to be antisymmetric. The second relation, (\ref{Chebichev}), does not impose further restrictions on the components of the Lorentz connection. Then, the number of independent components of $\omega _{\;b\mu }^{a}$ is $D^{2}(D-1)/2$, which is{\em \ less} than the number of independent components of the Christoffel symbol ($D^{2}(D+1)/2$). This can be understood as due to the fact that the Christoffel symbols are determined by both the vielbein and the Lorentz connection, $\Gamma^{\lambda}_{\mu \nu}= \Gamma^{\lambda}_{\mu \nu}(e, \omega)$.

\subsection{Building structures}

It can be observed that both the vielbein and the spin connection arise through the combinations
\begin{equation}
e^{a}(x)\equiv e_{\mu }^{a}(x)dx^{\mu },  \label{1Form-e}
\end{equation}
\begin{equation}
\omega _{\;b}^{a}(x)\equiv \omega _{\;b\mu }^{a}(x)dx^{\mu }, \label{1Form-w}
\end{equation}
that is, they are local 1-forms. This is no accident. It turns out that all the geometric properties of $M$ can be expressed with these two 1-forms, their exterior products and their exterior derivatives only. Since both $e^{a}$ and $\omega^{a}_{\;b}$ carry no coordinate indices ($ \mu$,  $\nu$, etc.), they are scalars under coordinate diffeomorphisms of $M$. Like all exterior forms, they are invariant under general coordinate transformations. This is why a description of the geometry that only uses these forms, their exterior products and exterior derivatives is naturally coordinate-free.

\subsubsection{Curvature}

The 1-form exterior derivative operator, $dx^{\mu}\partial_{\mu}\wedge$ is such that acting on a p-form, $\alpha_p$, it yields a (p+1)-form, $d\alpha_p$. One of the fundamental properties of exterior calculus is that the second exterior derivative of a differential form vanishes identically,
\begin{equation}\label{d2}
d(d\alpha_p) =: d^2 \alpha_p=0.
\end{equation}
This is trivially so because, when acting on continuously differentiable forms, the partial derivatives commute $\partial_{\mu} \partial_{\nu} \alpha = \partial_{\nu} \partial_{\mu} \alpha$, while $dx^{\mu}\wedge dx^{\nu} = -dx^{\nu}\wedge dx^{\mu}$. A consequence of this is that the square of the covariant derivative operator is not a differential operator, but an algebraic operator known as the curvature two-form. For instance, the second covariant derivative of a vector yields
\begin{eqnarray}\label{D2}
D^2\phi^a &=& D[d\phi^a+\omega^a_{\;b}\phi^b] \\ \nonumber &=& d[d\phi^a+\omega^a_{\;b}\phi^b] +
\omega^a_{\;b}[d\phi^b+\omega^b_{\;c}\phi^c] \\ \nonumber &=& [d\omega^a_{\;b}
+\omega^a_{\;c}\wedge \omega^c_{\;b}]\phi^b.
\end{eqnarray}
The two-form within brackets in this last expression is a second rank Lorentz tensor (the curvature two-form)
\begin{eqnarray}\label{Curvature}
R_{\;b}^a &=& d\omega_{\;b}^a + \omega_{\;c}^a \wedge \omega_{\;b}^c. \\
\nonumber &=& \frac{1}{2}R_{\;b\mu \nu }^{a}dx^{\mu }\wedge dx^{\nu }
\end{eqnarray}

For a formal definition of this operator see, e.g., \cite{Goeckeler-Schuecker,EGH}. The curvature two form defined by (\ref{Curvature}) is a Lorentz tensor on the tangent space, and is related to
the Riemann tensor, $R_{\;\beta \mu \nu }^{\alpha}$, through
\begin{equation}
R^a_{\;b} = \frac{1}{2}e_{\;\alpha}^a e_{\; b}^{\beta}R_{\;\beta \mu \nu}^{\alpha}dx^{\mu} \wedge
dx^{\nu}.
\end{equation}

The fact that $\omega_{\;b}^{a}(x)$ and the gauge potential in Yang-Mills theory, $A_{\;b}^{a}=A_{\;b\mu }^{a}dx^{\mu }$, are both 1-forms and have similar properties is not an accident since they are both connections of a gauge group\footnote{In the precise language of mathematicians, $\omega$ is ``a locally defined Lie-algebra valued 1-form on $M$, which is also a connection in the principal $SO(D-1,1)$-bundle over $M$", while $A$ is ``a Lie-algebra valued 1-form on $M$, which is also a connection in the vector bundle of some gauge group $G$".}. Their transformation laws have the same form, and the curvature $R_{\;b}^{a}$ is completely analogous to the field strength in Yang-Mills,
\begin{equation}
\textbf{F}=d\textbf{A}+\textbf{A}\wedge \textbf{A}. \label{FStrength}
\end{equation}

\subsubsection{Torsion}

The fact that the two independent geometrical ingredients, $\omega$ and $e$, play different roles is underscored by their different transformation rules under the Lorentz group: the vielbein transforms as a vector and not as a connection. In gauge theories this is reflected by the fact that vector fields play the role of matter, while the connection irepresents the carrier of interactions. 

Another important consequence of this asymmetry is the impossibility to construct a tensor two-form solely out of $e^a$ end its exterior derivatives, incontrast with the curvature which is uniquely defined by the connection. The only tensor obtained by differentiation of $e^{a}$ is its covariant derivative, also known the Torsion 2-form,
\begin{equation}
T^{a}=de^{a}+\omega _{\;b}^{a}\wedge e^{b},  \label{Torsion}
\end{equation}
which involves both the vielbein and the connection. In contrast with $T^a$, the curvature $R_{\;b}^{a}$ is not a covariant derivative of anything and depends only on $\omega$.

\subsubsection{Bianch identity}

As we saw in \textbf{3.5.1}, taking the second covariant derivative of a vector amounts to multiplying by the curvature 2-form. As a consequence of a this relation between covariant differentiation and curvature, there exists an important property known as Bianchi identity,
\begin{equation}\label{Bianchi}
DR_{\;b}^{a} = dR_{\;b}^{a} + \omega_{\;c}^{a}\wedge R_{\;b}^{c} + \omega_{\;b}^{c}\wedge R_{\;c}^{a}\equiv 0\;.
\end{equation}
This is an identity and not a set of equations because it is satisfied for \emph{any} well defined connection 1-form whatsoever, and it does not restrict in any way the form of the field $\omega_{\;b \mu}^{a}$, which can be checked explicitly by substituting (\ref{Curvature}) in the second term of (\ref{Bianchi}). If conditions $DR_{\;b}^{a}=0$ were a set of equations instead, they would define a subset of connections that have a particular form, corresponding to some class of geometries.

As a consequence of this identity, taking successive covariant derivatives does not produce new independent tensors. 

\subsubsection{Building blocks}

The basic building blocks of first order gravity are $e^{a}$, $\omega_{\;b}^{a}$, $R_{\;b}^{a}$, $T^{a}$. With them one must put together an action principle. But, are there other building blocks? The answer is no and the proof is by exhaustion. As a cowboy would put it, "if there were any more of them 'round here, we would have heard..." And we haven't. However, there is a more subtle
argument to rule out the existence of other building blocks. We are interested in objects that transform in a controlled way under Lorentz rotations (vectors, tensors, spinors, etc.). The existence of certain identities implies the taking the covariant derivatives of $e^{a}$, $R_{\;b}^{a}$, and $T^{a}$, one finds always combinations of the same objects, or zero:
\begin{eqnarray}
De^{a} &=&de^{a}+\omega _{\;b}^{a}\wedge e^{b}=T^{a}  \label{Torsion2}
\\
DR_{\;b}^{a} &=&dR_{\;b}^{a} + \omega_{\;c}^{a}\wedge R_{\;b}^{c} +
\omega_{\;b}^{c}\wedge R_{\;c}^{a}=0
\label{Bianchi1} \\
DT^{a} &=&dT^{a}+\omega_{\;b}^{a}\wedge T^{b}=R_{\;b}^{a}\wedge e^{b}.
\label{Bianchi2}
\end{eqnarray}
The first relation is just the definition of torsion and we leave it to the reader to prove these identities, which are direct consequences of the fact that the exterior derivative is nilpotent, $d^{2}=\partial _{\mu }\partial _{\nu }dx^{\mu }\wedge dx^{\nu }=0$.

In the next sections we discuss the construction of the possible actions for gravity using these ingredients. In particular, in 4 dimensions, the Einstein
action can be written as
\begin{equation}
I[g]=\int \epsilon_{abcd}(\alpha R^{ab}e^{c}e^{d} + \beta e^{a}e^{b}e^{c}e^{d}). \label{Einstein1st}
\end{equation}
This is basically the only action for gravity in dimension four, but many more options exist in higher dimensions.

\subsection{Gravity as a gauge theory}

Symmetry principles help in constructing the right classical action and, more importantly, they are often sufficient to ensure the viability of the quantum theory obtained from a classical action. In particular, gauge symmetry is the key to prove consistency (renormalizability) of the quantum field theories that correctly describe three of the four basic interactions of nature. The gravitational interaction has stubbornly escaped this rule in spite of the fact that, as we saw, it is described by a theory based on general covariance, which is a local invariance quite analogous to gauge symmetry. Here we try to shed some light on this puzzle.

Roughly one year after C. N. Yang and R. Mills proposed their model for nonabelian gauge invariant interactions \cite{Yang-Mills}, R. Utiyama showed that the Einstein theory can also be written as a gauge theory for the Lorentz group \cite{Utiyama}. This can be checked directly from the lagrangian in (\ref{Einstein1st}), which is a Lorentz scalar and hence, trivially invariant
under (local) Lorentz transformations. This generated the expectation to construe gravity as a gauge theory for the Poincar{\'e} group, $G=ISO(3,1)$, which is the standard symmetry group in particle physics, that includes both Lorentz transformations and translations. The inclusion of translations seems natural in view of the fact that a general coordinate transformation 
\begin{equation}
x^{\mu} \rightarrow x^{\mu}  + \xi^{\mu} ,  \label{Diffeo}
\end{equation}
looks like a local translation. This is undoubtedly a local symmetry in the sense that it is a transformation that leaves the action invariant and whose parameters are functions of the position in spcetime.  This suggests that diffeomorphism invariance could be identified with a symmetry under local translations that could extend the Lorentz group into its Poincar{\'e} embedding. Although this looks plausible, writing down a local action invariant under local translations has been impossible so far.  The problem is how to implement this symmetry as a gauge transformation of the dynamical fields ($e$, $\omega$ or $g_{\mu \nu}$). Under a coordinate diffeomorphism (\ref{Diffeo}), the metric changes by a Lie derivative,
\begin{eqnarray}
 \delta g_{\mu\nu}(x)&&= g_{\mu\nu}(x+ \xi(x))-g_{\mu\nu}(x) \nonumber \\
&&=-\left[ \partial_{\mu} \xi^{\alpha} g_{\alpha \nu} + \partial_{\nu} \xi^{\alpha} g_{\alpha \mu} + \xi^{\alpha}( \partial_{\alpha} g_{\mu \nu}+ \partial_{\mu}g_{\alpha \nu} +\partial_{\nu}g_{\alpha \mu}) \right]
\end{eqnarray}
This is the transformation law for a second rank covariant tensor. The connection for general coordinate diffeomorphisms is the Christoffel symbol, but this is not a fundamental field in the second order formalism. In the first order formalism, on the other hand, the fields are $e$ and $\omega$ are differential forms, and therefore are invariant under (\ref{Diffeo}). 

Attempts to identify the coordinate transformations with local translations, however, have systematically failed. The problem is that there is no known action for general relativity in four dimensions which is invariant under local $ISO(3,1)$ transformations \cite{Kibble,Yang,Mansouri,MacDowell-Mansouri}. In other words, although the fields $\omega^{ab}$ and $e^a$ have the right tensor properties to match the generators of the Poincar{\'e} group, there is no Poincar{\'e}-invariant 4-form available constructed with the connection for the Lie algebra of $iso(3,1)$.  We see that although superficially correct, the assertion that gravity is a gauge theory for the translation group is crippled by the profound differences between a gauge theory with fiber bundle structure and another with an open algebra, such as gravity.

The best way to convince ourselves of this obstruction is by assuming the existence of a connection for the group of local translations, analogous to the spin connection, which is the gauge field corresponding to the freedom to change Lorentz frames in a local neighborhood. Since the new gauge field corresponds to the invariance under translations in the tangent space, it should have an index structure to match that of the generator of translations, $P_a$. Thus, the connection for local translations must be a field with the same indeces as the vielbein. Therefore, a gauge theory for the Poincar{\'e} group should be based on the connection
\begin{equation}
 A=e^aP_a + \omega ^{ab} J_{ab}\; . \label{Poincare connection}
\end{equation}
It is easy to convince oneself that there is no Poincar{\'e}-invariant 4-form that can be constructed with this field, and therefore no  Poincar{\'e}-invariant gravity action can be constructed in four dimensions\footnote{Some attempts to improve this ansatz introduce an auxiliary nondynamical field, similar to the Stueckelberg compensating field in gauge theory, that renders the vielbein translation-invariant \cite{Stelle-West}. At the end of the day, however, the resulting gauge theory is one where the translation symmetry is ``spontaneously broken'', which points to the fact that this purported symmetry was never there.}.

A more sophisticated approach could be to replace the Poincar{\'e} group by another group $G$ which contains the Lorentz transformations as a subgroup. The idea is as follows: Our conviction that the space we live in is four dimensional and approximately flat stems from our experience that we can act with the group of four-dimensional translations to connect any two points in spacetime. But we know that this to be only approximately true. Like the surface of the Earth, our spacetime could be curved but with a radius of curvature so large we wouldn't notice the deviation from flatness except in very delicate observations. So, instead of the symmetries of a four-dimensional \emph{flat} spacetime, we might be experiencing the symmetries of a four dimensional spacetime of \emph{nonzero constant curvature}, also known as a pseudosphere.

It is therefore natural to expect the local symmetries of spacetime to be compatible with a larger group which contains both $SO(3,1)$ and some symmetries analogous to translations,
\begin{equation}
SO(3,1) \hookrightarrow G.  \label{Embedding}
\end{equation}
The smallest nontrivial choices for $G$ --which are not a direct product of the form $SO(3,1) \times G_0$--, are:
\begin{equation}
G=\left\{
\begin{array}{cc}
SO(4,1) & \textrm{de Sitter (dS)} \\
SO(3,2) & \textrm{anti-de Sitter (AdS)} \\
ISO(3,1) & \textrm{Poincar\'{e}}
\end{array}
\right.  \label{Lorentzembeddings}
\end{equation}
Both de Sitter and anti-de Sitter groups are semisimple, while the Poincar\'{e} group is not --it can be obtained as a contraction of either dS or AdS. This technical detail could mean that $SO(4,1)$ and $SO(3,2)$ have better chances than the Poincar\'{e} group, to become physically relevant for gravity.  Semisimple groups are preferred as gauge groups because they have an invariant
in the group, known as the \emph{Killing metric}, which can be used to define kinetic terms for the gauge fields\footnote{Non semisimple groups contain \emph{abelian} invariant subgroups. The generators of the abelian subgroups commute among themselves, and the fact that they are invariant subgroups implies that too many structure constants in the Lie algebra vanish. This in turn makes the Killing metric to acquire zero eigenvalues preventing its invertibility.}.

In spite of this improved scenario, it is still impossible to express gravity in four dimensions as a gauge theory for the dS or AdS groups.  However, as we shall see next, in odd dimensions ($D=2n-1$), and only in that case, gravity can be cast as a gauge theory of the groups $SO(D,1)$, $SO(D-1,2)$, or $ISO(D-1,1)$, in contrast with what one finds in dimension four, or in any other even dimension.

\newpage
\section{Gravity in higher dimensions}

We now turn to the construction of an action for gravity as a local functional of the one-forms $e^{a}$, $\omega _{\;b}^{a}$ and their exterior derivatives. The fact that $d^2 \equiv0$ implies that the lagrangian must involve at most first derivatives of these fields through the two-forms $R^{a}_{\; b}$ and $T^{a}$. We need not worry about invariance under general
coordinate transformations as exterior forms are coordinate scalars by construction. On the other hand, the action principle cannot depend on the choice of basis in the tangent space and hence Lorentz invariance should be ensured. A sufficient condition to respect Lorentz invariance is to demand the lagrangian to be a Lorentz scalar (although, as we will see, this is not
strictly necessary) and for this construction, the two invariant tensors of the Lorentz group, $\eta _{ab}$, and $\epsilon _{a_{1}\cdot \cdot \cdot \cdot a_{D}}$ can be used to raise, lower and contract indices.

Finally, since the action must be an integral over the $D$-dimensional spacetime manifold, the problem is to construct a Lorentz invariant $D$-form with the following ingredients:
\begin{equation}
e^{a},\;\;\; \omega _{\;b}^{a}, \;\;\; R_{\; b}^{a}, \;\;\; T^{a},\;\;\; \eta
_{ab}, \;\;\; \epsilon _{a_{1}\cdot \cdot \cdot \cdot a_{D}}. \nonumber
\end{equation}

Thus, we tentatively postulate the lagrangian for gravity to be a $D$-form made of linear combinations of products of the above ingredients in a Lorentz invariant way. We exclude from the ingredients functions such as the metric, since the metric is not an exterior product and it is certainly not an elementary object to deserve its inclusion in the action as an independent field. This rules rules out the inclusion of the inverse metric and of the Hodge $\star $-dual. This rule is additionally justified since it reproduces the known cases, and it explicitly excludes inverse fields, like $e_{a}^{\mu }(x)$, which would be like $A_{\mu }^{-1}$ in Yang-Mills theory (see \cite{Zumino} and \cite{Regge} for more on this). This postulate rules out the possibility of including tensors like the Ricci tensor $R_{\mu \nu}=e_{a}^{\lambda}\eta_{bc}e_{\mu}^{c}R_{\; \lambda \nu}^{ab}$, or $R_{\alpha \beta}R_{\mu \nu}R^{\alpha \mu \beta \nu}$, etc. This is sufficient and necessary to account for all sensible theories of gravity in $D$ dimensions. 

\subsection{Lovelock theorem}

The natural extension of the Einstein0Hilbert action for $D\neq 4$ is provided by the following 

\smallskip{\bf Theorem }[Lovelock, 1970 \cite{Lovelock}-Zumino, 1986 \cite{Zumino}]: The most general action for gravity that does not involve torsion and gives at most second order field equations for the metric \footnote{These conditions can be translated to mean that the Lovelock theories possess the same degrees of freedom as the Einstein Hilbert lagrangian in each dimension, that is, $D(D-3)/2$ (see, e.g., \cite{HTZ})}, is of the form
\begin{equation}
I_{D}=\kappa \int_{M} \sum_{p=0}^{[D/2]}a _{p}L^{(D,p)}, \label{LL-action}
\end{equation}
where the $a_{p}$s are arbitrary constants, and $L^{(D,p)}$ is given by
\begin{equation}
L^{(D,\;p)}=\epsilon _{a_{1}\cdots a_{d}}R^{a_{1}a_{2}}\!\cdot \!\cdot \!\cdot \!R^{a_{2p-1}a_{2p}}e^{a_{2p+1}}\!\cdot \!\cdot \!\cdot \!e^{a_{D}}.
\label{Lovlag}
\end{equation}
Here and in what follows we omit the wedge symbol in the exterior products.

For $D=2$ this action reduces to a linear combination of the $2$-dimensional Euler character, $\chi _{2}$, and the spacetime volume (area),
\begin{eqnarray}
I_{2} &=&\kappa \int_{M} \epsilon_{ab} [a_{1} R^{ab} + a_{0} e^a e^b] \nonumber \\
          &=&\kappa \int_M \sqrt{|g|}\left( a_{1}R + 2 a_{0}\right) d^{2}x  \label{2DGrav} \\
          &=&\kappa a_{1}\cdot \chi_{2}+ 2\kappa a _{0}\cdot V_{2}. \nonumber
\end{eqnarray}
This action has as a local extremum for $V_2=0$, which reflects the fact that, unless other matter sources are included, $I_{2}$ does not make a very interesting dynamical theory for the geometry. If the manifold $M$ has Euclidean metric and a prescribed boundary, the first term picks up a boundary term and the action is extremized by a surface in the shape of a soap bubble. This is the famous Plateau problem, which consists of establishing the shape of the surface of minimal area bounded by a certain fixed closed curve (see, e., g., http://www-gap.dcs.st-and.ac.uk/history/Mathematicians/Plateau.html).

For $D=3$, (\ref{LL-action}) reduces to the Hilbert action with a volume term, whose coefficient is the cosmological constant. For $D=4$ the action has, in addition, the four dimensional Euler invariant $\chi_{4}$,
\begin{eqnarray}
I_{4} &=&\kappa \int_M \epsilon_{abcd} \left[ a_2 R^{ab}R^{cd}+ a_1 R^{ab}e^c e^d + a_{0}e^a e^b e^ c e^d \right] \nonumber \\
 &=&-\kappa \int_M \sqrt{|g|}\left[a_2\left(R^{\alpha \beta \gamma  \delta }R_{\alpha \beta \gamma \delta}-4R^{\alpha \beta}R_{\alpha \beta} +R^{2}\right) + 2a_1R + 24 a_0 \right] d^4 x \nonumber \\
&=&-\kappa a_{2}\cdot \chi_{4}- 2a _1\int_M \sqrt{|g|}R d^4 x  - 24\kappa a_0 \cdot V_{4}. \label{4DGrav}
\end{eqnarray}

For all dimensions, the lagrangian is a polynomial of degree $d\leq D/2$ in the curvature 2-form. In general, each term $L^{(D,\;p)}$ in the lagrangian (\ref{LL-action}) is the continuation to $D$ dimensions of the Euler density from dimension $p\leq D$ \cite{Zumino}. In particular, for even $D$ the highest power in the curvature is the Euler character $\chi_{D}$. In four dimensions, the term $L^{(4,\;2)}$ in (\ref{4DGrav}) can be identified as the Gauss-Bonnet density, whose integral over a closed compact four dimensional manifold $M_4$ equals the Euler characteristic $\chi(M_4)$. This term also provides the first nontrivial generalization of Einstein gravity occurring in five dimensions, where the quadratic term that can be added to the lagrangian is the Gauss-Bonnet 5-form,
\begin{equation}
\epsilon _{abcde}R^{ab}\!R^{cd}e^{e}\!= \sqrt{|g|}\left[R^{\alpha \beta \gamma \delta }R_{\alpha \beta \gamma \delta }-4R^{\alpha \beta }R_{\alpha \beta} +R^{2}\right] d^5 x  \label{G-B}.
\end{equation}

The fact that Gauss-Bonnet term could be added to the Einstein-Hilbert action in five dimensions seems to have been known for many years. This is commonly attributed to Lanczos \cite{Lanczos}, but the original source is unclear. The generalization to arbitrary $D$ in the form (\ref{LL-action}) was obtained more than 30 years ago in \cite{Lovelock} and is known as the Lovelock lagrangian. This lagrangian was also identified as describing the only ghost-free\footnote{Physical states in quantum field theory have positive probability, which means that they are described by positive norm vectors in a Hilbert space. Ghosts instead, are unphysical states of negative norm. A lagrangian containing arbitrarily high derivatives of fields generally leads to ghosts. The fact that a gravitational action such as (\ref{LL-action}) leads to a ghost-free theory was unexpected and is highly nontrivial.} effective theory for a spin two field, generated from string theory at low energy \cite{Zwiebach,Zumino}. From our perspective, the absence of ghosts is only a reflection of the fact that the Lovelock action yields at most second
order field equations for the metric, so that the propagators behave as $\alpha k^{-2}$, and not as $\alpha k^{-2}+\beta k^{-4}$, as would be the case in a general higher derivative theory.

\subsubsection{Dynamical Content}

Extremizing the action (\ref{LL-action}) with respect to $e^{a}$ and $\omega^{ab}$, yields
\begin{equation}
\delta I_{D}=\int [\delta e^{a}{\cal E}_{a}+\delta \omega^{ab}{\cal E}_{ab}]=0,
\label{Var-action}
\end{equation}
modulo surface terms. The condition for $I_{D}$ to have an extreme under arbitrary first order variations is that ${\cal E}_{a}$ and ${\cal E}_{ab}$ vanish. This implies that the geometry satisfies
\begin{equation}
{\cal E}_{a}=\sum_{p=0}^{[\frac{D-1}{2}]} a_p(D-2p){\cal E}_{a}^{(p)}=0, \label{D-curvature}
\end{equation}
and
\begin{equation}
{\cal E}_{ab}=\sum_{p=1}^{[\frac{D-1}{2}]} a_p p(D-2p){\cal E}_{ab}^{(p)}=0, \label{D-torsion}
\end{equation}
where we have defined
\begin{eqnarray}
{\cal E}_{a}^{(p)}&:=&\epsilon_{ab_2 \cdots b_D} R^{b_2 b_3} \cdots R^{b_{2p}b_{2p+1}}e^{b_{2p+2}}\cdots e^{b_D},\hspace{-0.06in}  \label{ELL1}
\\
{\cal E}_{ab}^{(p)}&:=&\epsilon_{aba_3 \cdots a_D} R^{a_3 a_4 } \cdots R^{a_{2p-1}a_{2p}} T^{a_{2p+1}} e^{a_{2p+2}} \cdots e^{a_D}.
\label{ELL2}
\end{eqnarray}
These equations involve only first derivatives of $e^{a}$ and $\omega_{\;b}^{a}$, simply because $d^{2}=0$. If one furthermore assumes --as is usually done-- that the torsion vanishes identically,
\begin{equation}
T^{a}=de^{a}+\omega _{\;b}^{a}e^{b}=0,  \label{0-Torsion}
\end{equation}
then Eq. (\ref{D-torsion}) is automatically satisfied and the torsion-free condition (\ref{0-Torsion}) can be solved for $\omega$ as a function of the inverse vielbein ($e^{\mu}_a$) and its
derivative as
\begin{equation}
\omega^a_{\; b \mu} =-e^{\nu}_b(\partial_{\mu}e^a_{\nu} - \Gamma^{\lambda}_{\mu \nu} e^a)_{\lambda}),
\end{equation}
where $\Gamma^{\lambda}_{\mu \nu}$ is symmetric in $\mu \nu$ and can be identified as the Christoffel symbol (torsion-free affine connection). Substituting this expression for the spin connection back into (\ref{ELL1}) yields second order field equations for the metric. These equations are identical to the ones obtained from varying the Lovelock action written in terms of the Riemann tensor and the metric,
\begin{equation}
I_{D}[g]=\int_M d^{D}x\sqrt{g}\left[ a' _0 + a'_1 R + a'_2 (R^{\alpha \beta \gamma \delta} R_{\alpha \beta \gamma\delta}- 4R^{\alpha \beta }R_{\alpha \beta} + R^2) + \cdots
\right] . \label{LL-metric}
\end{equation}
This purely metric form of the action is also called second order formalism, because it contains up to second derivatives of the metric. The fact that the lagrangian contains second derivatives of $g_{\mu \nu}$ has induced some authors to refer to these actions as \emph{higher derivative theories of gravity}. This, however, is incorrect. The Einstein-Hilbert action, as well as
its Lovelock generalization, both yield second order field equations and for the same reason: the second derivatives of the metric enter through a total derivative in the lagrangian and therefore the field equations remain second order. Lagrangians that do give rise to higher order field equations for the metric are those that contain arbitrary powers of the curvature tensor and their contractions, like $R^{s}$, with $s\neq 1$, or $\left(a R^{a \beta \gamma \delta }R_{\alpha \beta \gamma \delta }+ b R^{\alpha \beta }R_{\alpha \beta }+c R^{2}\right)$, with $a:b:c \neq 1:-4:1$, etc.

Higher derivatives equations for the metric would mean that the initial conditions required to uniquely determine the time evolution are not those of General Relativity and hence the theory would have different degrees of freedom from standard gravity. It would also make the propagators in the quantum theory to develop poles at imaginary energies: \emph{ghosts}. Ghost states spoil the unitarity of the theory, making it hard to make sense of it and to interpret its predictions.

One important feature of the Lovelock theories, that makes their behavior very different for $D\leq4$ and for $D> 4$ is that in the former case the field equations (\ref{D-curvature}, \ref{D-torsion}) are linear in $R^{ab}$, while in the latter case the equations are nonlinear in the curvature tensor. In particular, while for $D\leq 4$ the equations (\ref{ELL2}) imply the vanishing of torsion, this is no longer true for $D>4$. In fact, the field equations evaluated in some configurations may leave some components of the curvature and torsion tensors completely undetermined. For example, Eq.(\ref{D-torsion}) has the form of a polynomial in $R^{ab}$ times $T^{a}$, and it is possible that the polynomial vanishes identically, imposing no conditions on the torsion tensor. However, the configurations for which the equations do not determine $R^{ab}$ and $T^{a}$ form sets of measure zero in the space of geometries. In a generic case, outside of these degenerate configurations, the Lovelock theory has the same $D(D-3)/2$ degrees of freedom as in ordinary gravity \cite{Te-Z}.

\subsection{Torsional series}

Lovelock's theorem assumes torsion to be identically zero. If equation (\ref{0-Torsion}) is assumed as an identity, it means that $e^a$ and $\omega_{\;b}^a$ are not independent fields, contradicting the assumption that these fields correspond to two independent features of the geometry on equal footing. Moreover, for $D\leq 4$, equation (\ref{0-Torsion}) coincides with
(\ref{ELL2}), so that imposing the torsion-free constraint is at best unnecessary.

In general, if the field equation for some field $\phi$ can be solved algebraically as $\phi =f(\psi)$ in terms of the remaining fields, then by the implicit function theorem, the original action principle $I[\phi,\psi]$ is identical to the reduced one obtained by substituting $f(\psi)$ in the action, $I[f(\psi),\psi]$. This occurs in 3 and 4 dimensions, where the spin connection
can be algebraically obtained from its own field equation and $I[\omega ,e] =I[\omega (e,\partial e),e]$ . In higher dimensions, however, the torsion-free condition is not necessarily a consequence of the field equations and although (\ref{ELL2}) is algebraic in $\omega $, it is practically impossible to solve for $\omega $ as a function of $e$. Therefore, it is not clear in general whether the action $I[\omega ,e]$ is equivalent to the second order form of the action, $I[\omega (e,\partial e),e]$.

As can be seen from  (\ref{D-torsion}), the torsion-free condition does not automatically follow from the field equations. It could be that the curvature is such that the torsion is completely indeterminate, as it happens for instance if the geometry has constant curvature for the choice of $a_p$'s to be discussed below. Therefore, it seems natural to consider the generalization of the Lovelock action in which torsion \emph{is not} assumed to vanish identically. This generalization consists of adding of all possible Lorentz invariants involving $T^a$ explicitly. This includes combinations like $R^{ab}e_b$, which do not involve torsion explicitly but which vanish for $T=0$ ($DT^a=R^{ab}e_b$). The general construction was worked out in \cite{M-Z}. The main difference with the torsion-free case is that now, apart from the dimensional continuation of the Euler densities, one encounters the Pontryagin (or Chern) classes as well.

For $D=3$, the only new torsion term not included in the Lovelock family is
\begin{equation}
e^{a}T_{a},  \label{eT}
\end{equation}
while for $D=4$, there are three terms not included in the Lovelock series,
\begin{equation}
e^{a}e^{b}R_{ab}, \;\; T^{a}T_{a}, \;\; R^{ab}R_{ab}.\;\;  \label{e2R+T2}
\end{equation}
The last term in (\ref{e2R+T2}) is the Pontryagin density, whose integral also yields a topological invariant. A linear combination of the other two terms is a topological invariant known as the Nieh-Yan density, given by \cite{Nieh-Yan}
\begin{equation}
N_{4}=T^{a}T_{a}-e^{a}e^{b}R_{ab}.  \label{NY}
\end{equation}
The properly normalized integral of (\ref{NY}) over a 4-manifold is an integer related to the $SO(5)$ and $SO(4)$ Pontryagin classes \cite{ChaZ}.

In general, the terms related to torsion that can be added to the action are combinations of the form
\begin{eqnarray}
A_{2n} &=&e_{a_1}R_{\;a_2}^{a_1}R_{\;a_3}^{a_2}\cdots R_{\;a_n}^{a_{n-1}}e^{a_{n}},\mbox{ even} \: n\geq 2, \label{eRe} \\ 
B_{2n+1} &=&T_{a_1}R_{\;a_2}^{a_1}R_{\;a_3}^{a_2}\cdots R_{\;a_{n}}^{a_{n-1}}e^{a_n},\mbox{ any } \: n\geq 1,  \label{eRT}\\ 
C_{2n+2} &=&T_{a_{1}}R_{\;a_2}^{a_1}R_{\;a_3}^{a_2}\cdots R_{\;a_{n}}^{a_{n-1}}T^{a_n},\rm{ odd } \: n\geq 1, \label{TRT}
\end{eqnarray}
which are $2n$, $2n+1$ and $2n+2$ forms, respectively. These Lorentz invariants belong to the same family with the Pontryagin densities or Chern classes,
\begin{equation}
P_{2n}=R_{\;a_2}^{a_1} R_{\;a_3}^{a_2}\cdots R_{\;a_1}^{a_n},\rm{ even} \: n.  \label{Pontryagin}
\end{equation}

The lagrangians that can be constructed now are much more varied and there is no uniform expression that can be provided for all dimensions. For example, in 8 dimensions, in addition to the Lovelock terms, one has all possible 8-forms made by taking products among the elements of the set \{$A_{4}$, $A_{8}$, $B_{3}$, $B_{5}$, $B_{7}$, $C_{4}$, $C_{8}$, $P_{4}$, $P_{8}$\}. They are
\begin{equation}
(A_4)^2,  A_8,  (B_3 B_5),  (A_4 C_4),  (C_4)^2, C_8, (A_4 P_4), (C_4 P_4), (P_4)^2, P_8. \label{Torsional 8D}
\end{equation}

To make life even harder, there are some linear combinations of these products which are topological densities, as in (\ref{e2R+T2}). In 8 dimensions there are two Pontryagin forms
\begin{eqnarray*}
P_8 &=&R_{\;a_2}^{a_1} R_{\;a_3}^{a_2} \cdots R_{\;a_1}^{a_4}, \\
(P_4)^2 &=&(R_{\;b}^a R_{\;a}^b)^2,
\end{eqnarray*}
which also occur in the absence of torsion, and there are two generalizations of the Nieh-Yan form,
\begin{eqnarray*}
(N_4)^2 &=& (T^a T_a -e^a e^b R_{ab})^2, \\ 
N_4 P_4  &=& (T^a T_a -e^a e^b R_{ab}) (R_{\;d}^c R_{\; c}^d),
\end{eqnarray*}
etc. (for details and extensive discussions, see Ref.\cite{M-Z}).

\subsection{Lorentz and torsional Chern-Simons series}
The Pontryagin classes $P_{2n}$ defined in (\ref{Pontryagin}), as well as those  that involve torsion, such as the Nieh-Yan invariant (\ref{NY}) and its generalizations, are all closed forms. Therfore, one can look for a locally defined CS form whose exterior derivative yields the corresponding closed form. These CS forms can also be included as lagrangian densities in the appropriate dimension.  

The idea is best illustrated with examples. Consider the Pontryagin and the Nieh-Yan forms in four dimensions, $P_4$ and $N_4$, respectively. The corresponding CS three-forms are
\begin{eqnarray}
C^{Lor}_3 &=& \omega^a_{\;b} d\omega^b_{\;a} + \frac{2}{3} \omega^a_{\;b} \omega^b_{\;c} \omega^c_{\;a} \\
C^{Tor}_3&=& e^aT_a
\end{eqnarray}
Both these terms are invariant under $SO(2,1)$ (Lorentz invariant in three dimensions), and are related to the four-dimensional Pontryagin and Nieh-Yan forms,
\begin{eqnarray}
dC^{Lor}_3 &=& R^{ab}R_{ab}, \\
dC^{Tor}_3&=& T^{a}T_{a}-e^{a}e^{b}R_{ab}.
\end{eqnarray}

\subsection{Overview}

Looking at these expressions one can easily feel depressed. The lagrangians look awkward and the number of terms in them grows wildly with the dimension\footnote{As it is shown in \cite{M-Z}, the number of torsion-dependent terms grows as the partitions of $D/4$, which is given by the Hardy-Ramanujan formula, $p(D/4)\sim \frac{1}{\sqrt{3}D}\exp [\pi \sqrt{D/6}]$.}. This problem is not purely aesthetic. The coefficients in front of each term in the lagrangian are arbitrary and dimensionful. This problem already occurs in 4 dimensions, where Newton's constant and the cosmological constant have dimensions of [length]$^{2}$ and [length]$^{-4}$ respectively. Moreover, as evidenced by the outstanding cosmological constant problem, there is no theoretical argument to predict their values by natural arguments in a way that can be compared with observations.

\newpage

\section{Selecting Sensible Theories}

The presence of dimensionful parameters leaves little room for optimism in a quantum version of the theory. Dimensionful parameters in the action are potentially dangerous because they are likely to acquire uncontrolled quantum corrections. This is what makes ordinary gravity nonrenormalizable in
perturbation theory: In 4 dimensions, Newton's constant has dimensions of [mass]$^{-2}$ in natural units. This means that as the order in perturbation series increases, more powers of momentum will occur in the Feynman graphs, making the ultraviolet divergences increasingly worse. Concurrently, the radiative corrections to these bare parameters require the introduction of infinitely many counterterms into the action to render them finite\cite{'t Hooft}. But an illness that requires infinite amount of medication is synonym of incurable.

--------------------------------- \\
The only safeguard against the threat of uncontrolled divergences in quantum theory is to have some symmetry principle that fixes the values of the
parameters in the action, limiting the number of possible counterterms that could be added to the lagrangian. Obviously, a symmetry endowed with such a high responsibility should be a bona fide quantum symmetry, and not just an approximate feature of its effective classical descendent. A symmetry that is only present in the classical theory but is not a feature of the quantum theory is said to be anomalous. This means that if one conceives the quantum theory as the result of successuve quantum corrections to the classical theory, these corrections ``break" the symmetry. Of  course, we know that the classical theory is a limit of the quantum world, some sort of shadow of an underlying reality that is blurred in the limit. An anomalous symmetry is an artifact of the classical limit, that does not correspond to a true symmetry of the microscopic world.  

Thus, if a ``non anomalous" symmetry fixes the values of the parameters in the action, this symmetry will protect those values under renormalization. A good indication that this might happen would be if all the coupling constants are dimensionless and could be absorbed in the fields, as in Yang-Mills theory. As shown below, in odd dimensions there is a unique combination of terms in the action that can give the theory an enlarged gauge symmetry. The resulting
action can be seen to depend on a unique multiplicative coefficient ($\kappa$), analogous to Newton's constant. Moreover, this coefficient can be shown to be
quantized by an argument similar to the one that yields Dirac's quantization of the product of magnetic and electric charge \cite{QuantumG}.

\subsection{Extending the Lorentz Group}

The coefficients $\alpha _{p}$ in the Lovelock action (\ref{LL-action}) have dimensions $l^{D-2p}$. This is because the canonical dimension of the vielbein is $[e^{a}]=l$, while the Lorentz connection has dimensions that correspond to a true gauge field, $[\omega^{ab}]=$ $l^{0}$. This reflects the fact that gravity is naturally only a gauge theory for the Lorentz group, where $e^a$ plays the role of a matter field, while the vielbein \emph{is not} a connection field but transforms as a vector under Lorentz rotations.

\subsubsection{Poincar\'{e} Group}

Three-dimensional gravity, where $e^{a}$ does play the role of a connection, is an important exception to this statement. This is in part thanks to the coincidence in three dimensions that allows to regard a vector as a connection for the Lorentz group,  
\begin{equation}
\hat{ \omega}^a=\frac{1}{2}\epsilon ^{abc} \omega_{bc}.
\end{equation}
Consider the Einstein-Hilber lagrangian in three dimensions
\begin{equation}
L_{3}=\epsilon_{abc}R^{ab}e^{c}.  \label{D=3}
\end{equation}
Under an infinitesimal Lorentz transformation with parameter $\lambda_{\,\;b}^{a}$, the Lorentz connection transforms as
\begin{eqnarray}
\delta \omega_{\,\;b}^{a} &=&D\lambda_{\,\;b}^{a} \label{delta w}\\
&=&d\lambda_{\,\;b}^{a} + \omega _{\,\;c}^{a}\lambda_{\,\; b}^{c} -
\omega_{\,\;b}^{c}\lambda _{\,\;c}^{a}, \nonumber
\end{eqnarray}
while $e^{c}$, $R^{ab}$ and $\epsilon_{abc}$ transform as tensors,
\begin{eqnarray*}
\delta e^{a} &=&-\lambda _{\,\;c}^{a}e^{c} \\ \delta R_{\,}^{ab} &=&-(\lambda_{\,\;c}^{a}R^{cb}+\lambda_{\,\; c}^{b}R^{ac}), \\
\delta \epsilon_{abc} &=&-(\lambda_{\,\;a}^{d}\epsilon_{dbc}+\lambda_{\,\;b}^{d}\epsilon_{adc}+\lambda_{\,\;c}^{d}\epsilon_{abd})\equiv0.
\end{eqnarray*}
Combining these relations, the Lorentz invariance of $L_{3}$ can be directly checked. What is unexpected is that the action defined by (\ref{D=3}) is also invariant under the group of  local translations in three dimensions. For this additional symmetry  $e^{a}$ transforms as a gauge connection for the translation group. In fact, if the vielbein transforms under ``local translations'' in tangent space, parametrized by $\lambda^{a}$ as 
\begin{eqnarray}
\delta e^{a} &=&D\lambda ^{a}  \nonumber \\
&=&d\lambda ^{a}+\omega _{\,\;b}^{a}\lambda^{b},  \label{LocalTrans-e}
\end{eqnarray}
while the spin connection remains unchanged,
\begin{equation}
\delta \omega^{ab} = 0, \label{LocalTrans-w}
\end{equation}
then, the lagrangian $L_{3}$ changes by a total derivative,
\begin{equation}
\delta L_{3}=d[\epsilon_{abc}R^{ab}\lambda^{c}],  \label{dL3}
\end{equation}
which can be dropped from the action, under the assumpton of standard boundary conditions. This means that in three dimensions ordinary gravity is gauge invariant under the whole Poincar\'{e} group. We leave it as an exercise to the reader to prove this\footnote{Hint: use the infinitesimal transformations $\delta e$ and $\delta \omega $ to compute the commutators of the second variations to obtain the Lie algebra of the Poincar\'{e} group.}

\subsubsection{(Anti-)de Sitter Group}

In the presence of a cosmological constant $\Lambda =\mp \frac{1}{6l^{2}}$ it is also possible to extend the local Lorentz symmetry. In this case, however, the invariance of the appropriate tangent space is not the local Poincar\'e  symmetry, but the local (anti)-de Sitter group. The point is that different spaces $T^{*}M$ can be chosen as tangents to a given manifold $M$, provided they are diffeomorphic to the open neighborhoods of $M$. However, a useful choice of tangent space corresponds to the covering space of a vacuum solution of the Einstein equations. In the previous case, flat space was singled out because it is the maximally symmetric solution of the Einstein equations. If $\Lambda \neq 0$,  flat spacetime is no longer a solution of the Einstein equations, but the de Sitter or anti-de Sitter  space,  for $\Lambda > 0$ or $\Lambda < 0$, respectively.

The three-dimensional lagrangian in (\ref{LL-action}) reads
\begin{equation}
L_{3}^{AdS}=\epsilon _{abc}(R^{ab}e^{c}\pm \frac{1}{3l^{2}}e^{a}e^{b}e^{c}),
\label{L3AdS}
\end{equation}
and the action is invariant --modulo surface terms-- under the infinitesimal transformations,
\begin{eqnarray}
\delta \omega _{\,\;}^{ab} &=&d\lambda _{\;}^{ab}+ \omega_{\,\;c}^{a} \lambda^{cb}+\omega_{\,\;c}^{b} \lambda_{\;}^{ac} 
\pm [e^{a}\lambda ^{b}-\lambda ^{a}e^{b}]l^{-2} \label{dw-AdS} \\ 
\delta e^{a} &= &d\lambda^{a}+\omega _{\,\;b}^{a}\lambda ^{b} - \lambda_{\;b}^ae^b. \label{de-AdS}
\end{eqnarray}
These transformations can be cast in a more suggestive way as
\begin{eqnarray*}
\delta \left[
\begin{array}{cc}
\omega _{\,\;}^{ab} & e^a l^{-l} \\
-e^b l^{-l} & 0
\end{array}
\right] &=& d\left[
\begin{array}{cc}
\lambda _{\;}^{ab} & \lambda^a l^{-l} \\
-\lambda^b l^{-l} & 0
\end{array}
\right] \\
&& +\left[
\begin{array}{cc}
\omega _{\;c}^{a} & e^a l^{-l} \\
-e_{c}l^{-l} & 0
\end{array}
\right] \left[
\begin{array}{cc}
\lambda ^{cb} & \lambda^{c}l^{-1} \\
\pm\lambda^{b}l^{-1} & 0
\end{array}
\right] \\
&&- \left[
\begin{array}{cc}
\lambda ^{ac} & \lambda^{a} l^{-1} \\
-\lambda^{c} l^{-1}& 0
\end{array}
\right] \left[
\begin{array}{cc}
\omega _{c}^{\;b} &  e_c l^{-1} \\
\pm e^bl^{-1} & 0
\end{array}
\right] .
\end{eqnarray*}
This can also be written as
\[
\delta W^{AB}=d\Lambda_{\;}^{AB}+W_{\,\;C}^{A}\Lambda ^{CB}- \Lambda^{AC}W_{C}^{\;B} ,
\]
where the 1-form $W^{AB}$ and the 0-form $\Lambda^{AB}$ stand for the combinations
\begin{eqnarray}
W_{\;}^{AB} &=&\left[
\begin{array}{cc}
\omega _{\,\;}^{ab} & e^{a}l^{-1} \\
-e^{b}l^{-1} & 0
\end{array}
\right]  \label{baticonn} \\
\Lambda ^{AB} &=&\left[
\begin{array}{cc}
\lambda_{\;}^{ab} & \lambda^{a}l^{-1} \\
-\lambda^{b}l^{-1} & 0
\end{array}
\right] ,  \label{batiparam}
\end{eqnarray}
(here $a,b,..=1,2,..D,$ while $A,B,...=1,2,..,D+1$). Clearly, $W_{\; }^{AB}$ transforms as a connection and $\Lambda^{AB}$ can be identified as the
infinitesimal transformation parameters, but for which group? Since $\Lambda^{AB}=-\Lambda^{BA}$, this indicates that the group is one that leaves invariant a symmetric, real bilinear form, so it must be a group in the $SO(r,s)$ family. The signs ($\pm $) in the transformation above can be traced back to the sign of the cosmological constant. It is easy to check that this structure fits well if indices are raised and lowered with the metric
\begin{equation}
\Pi^{AB}=\left[
\begin{array}{cc}
\eta _{\,\;}^{ab} & 0 \\
0 & \mp 1
\end{array}
\right] ,  \label{tangentAdS}
\end{equation}
so that, for example, $W^{AB}=\Pi^{BC}W^A_{\,\;C}$. Then, the covariant derivative in the connection $W$ of this metric vanishes identically,
\begin{equation}
D_{W}\Pi^{AB}=d\Pi^{AB}+W^{A}_{\,\;C}\Pi^{CB}+W^{B}_{\,\; C}\Pi ^{AC}=0.
\end{equation}
Since $\Pi^{AB}$ is constant, this last expression implies $W^{AB}+W^{BA}=0$, in exact analogy with what happens with the Lorentz connection,
$\omega^{ab}+\omega ^{ba}=0$, where $\omega ^{ab}\equiv\eta ^{bc}\omega_{\;c}^{a}$. Indeed, this is a very awkward way to discover that the 1-form $W_{\;}^{AB}$ is actually a connection for the group which leaves invariant the metric $\Pi^{AB}$. Here the two signs in $\Pi^{AB}$ correspond to the de Sitter ($+$) and anti-de Sitter\ ($-$) groups, respectively.

What we have found here is an explicit way to immerse the three-dimensional Lorentz group into a larger symmetry group, in which oth, the vielbein and the Lorentz connection have been incorporated on equal footing as components of a larger (A)dS connection. The Poincar\'{e} symmetry is obtained in the limit $l\rightarrow \infty $ ($\lambda\rightarrow 0 $). In that case, instead of (\ref{dw-AdS}, \ref{de-AdS}) one has
\begin{eqnarray}
\delta \omega _{\,\;}^{ab} &=&d\lambda _{\;}^{ab}+ \omega_{\,\;c}^{a} \lambda^{cb}+\omega_{\,\;c}^{b}\lambda_{\; }^{ac}  \label{dw-Poinc} \\ \delta e^{a} &=&d\lambda ^{a}+\omega _{\,\;b}^{a}\lambda ^{b} - \lambda _{\;b}^{a}e^{b}. \label{de-Poinc}
\end{eqnarray}
The vanishing cosmological constant limit is actually a deformation of the (A)dS algebra analogous to the deformation that yields the Galileo group from the Poincar\'e symmetry in the limit of infinite speed of light ($c\rightarrow \infty$). These deformations are examples of what is known as a In\"on\"u-Wigner contraction \cite{Gilmore,Inonu} (see also Sect. 8.2). The procedure starts from a semisimple Lie algebra and some generators are rescaled by a parameter ($l$ or $\lambda$ in the above example). Then, in the limit where the parameter is taken to zero (or infinity),  a new nonsemisimple algebra is obtained.  For the Poincar\'e group which is the familiar symmetry of Minkowski space, the representation in terms of $W$ becomes inadequate because the metric $\Pi^{AB}$
should be replaced by the degenerate (noninvertible) metric of the Poincar\'{e} group,
\begin{equation}
\Pi_0^{AB}=\left[
\begin{array}{cc}
\:\eta _{\,\;}^{ab} & 0 \\
0 & 0
\end{array}
\right] ,  \label{tangentPoinc}
\end{equation}
and is no longer clear how to raise and lower indices. Nevertheless, the lagrangian (\ref{L3AdS}) in the limit $l\rightarrow \infty $ takes the usual
Einstein Hilbert form with vanishing cosmological constant,
\begin{equation}
L_{3}^{EH}=\epsilon_{abc}R^{ab}e^{c}, \label{L3EH}
\end{equation}
which can be directly checked to be invariant under (\ref{de-Poinc}). We leave this as an exercise for the reader.

As Witten showed, General Relativity in three spacetime dimensions is a renormalizable quantum system \cite{Witten}. It is strongly suggestive that precisely in 2+1 dimensions GR is also a gauge theory on a fiber bundle. It could be thought that the exact solvability miracle is due to the absence of propagating degrees of freedom in three-dimensional gravity, but the final power-counting argument of renormalizability rests on the fiber bundle structure of the Chern-Simons system and doesn't seem to depend on the absence of propagating degrees of freedom. In what follows we will generalize the gauge invariance of three-dimensional gravity to higher dimensions.

\subsection{More Dimensions}

Everything that has been said about embedding the Lorentz group into the (A)dS group for $D=3$, starting at equation (\ref{dw-AdS}), can be generalized for any $D$. In fact, it is always possible to embed the $D$-dimensional Lorentz group into the de-Sitter, or anti-de Sitter groups,
\begin{equation}
SO(D-1,1)\hookrightarrow \left\{
\begin{array}{cc}
SO(D,1), & \Pi ^{AB}=\rm{diag }(\eta _{\,\;}^{ab},+1) \\
SO(D-1,2), & \Pi ^{AB}=\rm{diag }(\eta _{\,\;}^{ab},-1)
\end{array}
.\right.  \label{embeddingAdS}
\end{equation}
as well as into their Poincar\'{e} limit,
\begin{equation}
SO(D-1,1)\hookrightarrow ISO(D-1,1).  \label{embeddingPoinc}
\end{equation}

The question naturally arises,  are there gravity actions in dimensions $\geq 3$ which are also invariant, not just under the Lorentz group, but under some of its extensions, $SO(D,1)$, $SO(D-1,2)$, $ISO(D-1,1)$? As we will see now, the answer to this question is affirmative in odd dimensions: There exist gravity actions for every $D=2n-1$, invariant under local $SO(2n-2,2)$, $SO(2n-1,1)$ or $ISO(2n-2,1)$ transformations, where the vielbein and the spin connection combine to form the connection of the larger group. In even dimensions, in contrast, this cannot be done.

Why is it possible in three dimensions to enlarge the symmetry from local $SO(2,1)$ to local $SO(3,1)$, $SO(2,2)$ and $ISO(2,1)$? What happens if one
tries to do this in four or more dimensions? Let us start with the Poincar\'{e} group and the Hilbert action for $D=4$,
\begin{equation}
L_{4}=\epsilon _{abcd}R^{ab}e^{c}e^{d}.  \label{D=4}
\end{equation}
Why is this not invariant under local translations $\delta e^{a}= d\lambda^{a}+\omega _{\,\;b}^{a}\lambda ^{b}$? A simple calculation yields
\begin{eqnarray}
\delta L_{4} &=&2\epsilon _{abcd}R^{ab}e^{c}\delta e^{d}  \nonumber \\
&=&d(2\epsilon _{abcd}R^{ab}e^{c}\lambda ^{d})-2\epsilon_{abcd}R^{ab}T^{c}\lambda ^{d}.  \label{delta4}
\end{eqnarray}
The first term in the r.h.s. of (\ref{delta4}) is a total derivative and therefore gives a surface contribution to the action. The last term, however, need not vanish, unless one imposes the field equation $T^{a}=0$. But this means that the invariance of the action only occurs ``on shell''. Now, ``on shell
symmetries" are not real symmetries and they probably don't survive quantization because quantum mechanics doesn't respect equations of motion.

On the other hand, the miracle in 3 dimensions occurred because the lagrangian (\ref{L3EH}) is linear in $e$. In fact, lagrangians of the form
\begin{equation}
L_{2n+1}=\epsilon_{a_{1}\cdots a_{2n+1}}R^{a_{1}a_{2}}\cdots R^{a_{2n-1}a_{2n}}e^{a_{2n+1}}, \label{odd-D-Poinc}
\end{equation}
--which are only defined in odd dimensions--, are also invariant under local Poincar\'{e} transformations (\ref{dw-Poinc}, \ref{de-Poinc}), as can be easily
checked. Since the Poincar\'{e} group is a limit of (A)dS, it seems likely that there should exist a lagrangian in odd dimensions, invariant under local (A)dS
transformations, whose limit for vanishing cosmological constant ($l\rightarrow \infty$) is (\ref{odd-D-Poinc}). One way to find out what that lagrangian
might be, one could take the most general Lovelock lagrangian and select the coefficients by requiring invariance under (\ref{dw-AdS}, \ref{de-AdS}). This
is a long and tedious but sure route. An alternative approach is to try to understand why it is that in three dimensions the gravitational lagrangian with
cosmological constant (\ref{L3AdS}) is invariant under the (A)dS group.

Let us consider the three-dimensional case first. If we take seriously the notion that $W^{AB}$ is a connection, then the associated curvature is
\[
F^{AB}=dW^{AB}+W_{\;C}^{A}W^{CB},
\]
where $W^{AB}$ is defined in (\ref{baticonn}). Then, it is easy to prove that
\begin{equation}
F_{\;}^{AB}=\left[
\begin{array}{cc} R_{\,\;}^{ab}\pm l^{-2}e^{a}e^{b} & l^{-1}T^{a} \\
-l^{-1}T^{b} & 0
\end{array}
\right] .  \label{baticurvature}
\end{equation}
where $a,b$ run from 1 to 3 and $A,B$ from 1 to 4. Since the (A)dS group has an invariant tensor $\epsilon_{ABCD}$, one can construct the 4-form invariant
\begin{equation}
E_{4}=\epsilon _{ABCD}F_{\;}^{AB}F_{\;}^{CD}.  \label{E=F2}
\end{equation}
This is invariant under the (A)dS group and is readily recognized as the Euler density for a four-dimensional manifold\footnote{This identification is formal,
since the differential forms that appear here are defined in three dimensions, but they can be naturally extended to four dimensional forms by simply
extending the range of coordinate indices. This implies that one is considering the three-dimensional manifold as embedded in --or, better, as the boundary
of-- a four dimensional manifold.} whose tangent space is not Minkowski, but has the metric $\Pi ^{AB}= $diag $(\eta _{\,\;}^{ab},\mp 1)$. The Euler density $E_{4}$ can also be written explicitly in terms of $R^{ab}$, $T^{a}$, and $e^{a}$,
\begin{eqnarray}
E_{4} &=&4\epsilon _{abc}(R_{\,\;}^{ab}\pm l^{-2}e^{a}e^{b})l^{-1}T^{a} \label{E4} \\
&=&\frac{4}{l}d\left[ \epsilon _{abc}\left( R_{\,\;}^{ab}\pm \frac{1}{
3l^{2}}e^{a}e^{b}\right) e^{c}\right] ,  \nonumber
\end{eqnarray}
which is, up to constant factors, the exterior derivative of the three-dimensional lagrangian (\ref{L3AdS}),
\begin{equation}
E_{4}=\# dL_{3}^{AdS}.  \label{E4=dL3}
\end{equation}

This is the key point: the l.h.s. of (\ref{E4=dL3}) is invariant under local (A)dS$_3$ by construction. Therefore, the same must be true of the r.h.s.,
\[
\delta \left( dL_{3}^{AdS}\right) =0.
\]
Since the variation ($\delta $) is a linear operation, this can also be written as
\[
d\left( \delta L_{3}^{AdS}\right) =0,
\]
which in turn means, by Poincar\'{e}'s Lemma \cite{Spivak} that, locally, $\delta L_{3}^{AdS}= d(something)$. This explains why the action is (A)dS
invariant up to surface terms, which is exactly what we found for the variation, [see, (\ref{dL3})]. The fact that three dimensional gravity can be
written in this way was observed many years ago in Refs. \cite{Achucarro-Townsend,Witten}.

The key to generalize the (A)dS lagrangian from $3$ to $2n-1$ dimensions is now clear\footnote{The construction we outline here was discussed by Chamseddine \cite{Chamseddine}, M\"{u}ller-Hoissen \cite{Muller-Hoissen}, and by Ba\~{n}ados, Teitelboim and this author in \cite{JJG,BTZ94}.}:

$\bullet$ First, generalize the Euler density (\ref{E=F2}) to a $2n$-form,
\begin{equation}
E_{2n}=\epsilon _{A_{1}\cdot \cdot \cdot A_{2n}}F^{A_{1}A_{2}}\cdot \cdot \cdot
F^{A_{2n-1}A_{2n}}.  \label{E2n=Fn}
\end{equation}

$\bullet$ Second, express $E_{2n}$ explicitly in terms of $R^{ab}$, $T^{a}$ and $e^{a}$ using (\ref{baticurvature}).

$\bullet$ Write this as the exterior derivative of a $(2n-1)$-form $L_{2n-1}$.

$\bullet$ $L_{2n-1}$ can be used as a lagrangian in $(2n-1)$ dimensions and is (A)dS invariant by construction.

Proceeding in this way, directly yields the $(2n-1)$-dimensional (A)dS invariant lagrangian as
\begin{equation}
L_{2n-1}^{(A)dS}=\sum_{p=0}^{n-1}\bar{\alpha}_{p}L^{(2n-1,p)},
\label{(A)dS2n+1}
\end{equation}
where $L^{(D,p)}$ is given by (\ref{Lovlag}). This is a particular case of a Lovelock lagrangian in which all the coefficients $\bar{\alpha}_{p}$ have been
fixed to take the values
\begin{equation}
\smallskip \bar{\alpha}_{p}=\kappa \cdot \frac{(\pm 1)^{p+1}l^{2p-D}}{(D-2p)}
\left(
\begin{array}{c}
n-1 \\
p
\end{array}
\right) ,\;p=1,2,...,n-1=\frac{D-1}{2},  \label{alphasCS}
\end{equation}
where $1\leq p \leq n-1=(D-1)/2$, and $\kappa$ is an arbitrary dimensionless constant. It is left as an exercise to the reader to check that $dL_{2n-1}^{(A)dS} = E_{2n}$ and to show the invariance of $L_{2n-1}^{(A)dS}$ under the (A)dS group.

Another interesting exercise is to show that, for AdS, the action (\ref{(A)dS2n+1}) can also be written as \cite{MOTZ1}
\begin{equation}
I_{2n-1}=\frac{\kappa}{l} \int\limits_{M}\int\limits_{0}^{1}dt\;\epsilon_{a_1\cdots a_{2n-1}} R_{t}^{a_1 a_2}\cdots R_{t}^{a_{2n-3} a_{2n-2}} e^{a_{2n-1}} \;, \label{t-AdSAction}
\end{equation}
where we have defined $R_{t}^{ab}:= R^{ab} +(t^2/l^2) e^{a}e^{b}$.

\textbf{Example:} In five dimensions, the (A)dS lagrangian reads 
\begin{equation}
L_5^{(A)dS}=\frac{\kappa}{l}\epsilon _{abcde}\left[ e^a R^{bc} R^{de} \pm \frac{2}{3l^2} e^a e^b e^c R^{de}
+\frac{1}{5l^4}e^a e^b e^c e^d e^e \right] . \label{(A)dS5}
\end{equation}

The parameter $l$ is a length scale --the Planck length-- and cannot be fixed by theoretical considerations. Actually, $l$ only appears in the combination
\[
\tilde{e}^{a}=l^{-1} e^a,
\]
that could be considered as the ``true'' dynamical field, as is the natural thing to do if one uses $W^{AB}$ instead of $\omega ^{ab}$ and $e^{a} $
separately. In fact, the lagrangian (\ref{(A)dS2n+1}) can also be written in terms of $W^{AB}$ and its exterior derivative, as
\begin{equation}
L_{2n-1}^{(A)dS}=\kappa \cdot \epsilon _{A_1 \cdots A_{2n}}\left[ W(dW)^{n-1}+a_3 W^3 (dW)^{n-2}+\cdots a_{2n-1}W^{2n-1}\right] ,  \label{(A)dS2n+1'}
\end{equation}
where all indices are contracted appropriately and all the coefficients $a_3$, $\cdots a_{2n-1}$, are dimensionless rational numbers fixed by the condition $dL_{2n-1}^{(A)dS}= E_{2n}$.

\subsection{Generic Chern-Simons forms}

There is a more general way to look at these lagrangians in odd dimensions, which also sheds some light on their remarkable enlarged symmetry. This is
summarized in the following

{\bf Lemma:} Let ${\cal P}(F)$ be an invariant $2n$-form constructed with the field strength $F=dA+A^{2}$, where $A$ is the connection for some gauge group $G$. If there exists a $2n-1$ form, ${\cal C}$, depending on $A$ and $dA$, such that $d{\cal C}={\cal P}$, then under a gauge transformation, ${\cal C}$ changes by a total derivative (exact form), $\delta {\cal C}=d($something$)$.

The $(2n-1)-$form ${\cal C}$ is known as the Chern-Simons ({\bf CS}) form, and as it  changes by an exact for), it can be used as a lagrangian for a gauge theory for the connection $A$. It must be emphasized that $C$ defines a nontrivial lagrangian which {\em is not invariant} under gauge transformations, but that changes by a function that only depends on the fields at the boundary: it is quasi-invariant. This is sufficient to define a physical lagrangian since the action principle considers variations of the physical fields subject to some appropriate boundary conditions. So, it is always possible to select the boundary condition on the fields in such a way that $\delta {\cal C}=0$.

This construction is not restricted to the Euler invariant discussed above, but applies to any invariant of similar nature, generally known as characteristic
classes. Other well known characteristic classes are the Pontryagin or Chern classes and their corresponding CS forms were studied first in the context of
abelian and nonabelian gauge theories (see, e. g., \cite{Jackiw,Nakahara}).

The following table gives examples of CS forms which define lagrangians in three dimensions, and their corresponding topological invariants, \\
\newpage
\begin{center}
\textbf{Table 1}
\begin{tabular}{|l|l|l|}
\hline $D=3$ Chern-Simons Lagrangian & Top. Invariant(${\cal P}$) & Group \\
\hline \hline $L_{3}^{(A)dS} =\epsilon_{abc}(R^{ab}\pm\frac{e^a e^b}{3l^2})e^c$
& $E_4=\epsilon_{abc}(R^{ab}\pm\frac{e^a e^b}{l^2})T^c $ &$SO(4)$\dag
\\ \hline  $L_{3}^{Lorentz} =\omega_{\;b}^{a}d\omega
_{\;a}^{b}+\frac{2}{3} \omega _{\;b}^{a}\omega _{\;c}^{b}\omega _{\;a}^{c}$ &
$P_4^{Lorentz} =R_{\;b}^{a}R_{\;a}^{b}$ & $SO(2,1)$ \\ \hline
$L_{3}^{Torsion} = e^{a}T_{a}$ & $N_4 =T^{a}T_{a}-e^{a}e^{b}R_{ab}$
&$SO(2,1)$ \\ \hline $ L_{3}^{U(1)}= AdA$ & $P_4^{U(N)} =FF$ & $ U(1)$ \\
\hline $ L_{3}^{SU(N)} =tr[{\bf
A}d{\bf A+}\frac{2}{3}{\bf AAA]}$ & $P$ $_4^{SU(4)} =tr[{\bf FF}]$ & $ SU(N)$ \\
\hline
\end{tabular}
\end{center}
\dag {\tiny Either this or any of its cousins, $SO(3,1)$, $SO(2,2)$}.\\
Here $R$, $F$, and ${\bf F}$ are the curvatures of the Lorentz, the electromagnetic, and the Yang-Mills ($SU(N)$) connections $\omega _{\;b}^{a}$, $A$ and {\bf A}, respectively; $T$ is the torsion; $E_4$ and $P_4$ are the Euler and the Pontryagin densities for the Lorentz group \cite{EGH}, and $N_4$ is the Nieh-Yan invariant \cite{Nieh-Yan}. The lagrangians are locally invariant (up to total derivatives) under the corresponding gauge groups.

\subsection{Torsional Chern-Simons forms}

So far we have not included torsion in the CS lagrangians, but as we see in the table above, it is also possible to construct CS forms that include torsion.
All the CS forms above are Lorentz invariant (up to an exact form), but there is a linear combination of the second and third which is invariant under the
(A)dS group. This is the so-called exotic gravity \cite{Witten},
\begin{equation}
L_{3}^{Exotic}=L_{3}^{Lor} \pm \frac{2}{l^{2}}L_{3}^{Tor}.  \label{3d-exotic}
\end{equation}
As can be shown directly by taking its exterior derivative, this is invariant under (A)dS:
\begin{eqnarray*}
dL_{3}^{Exotic} &=&R_{\;b}^{a}R_{\;a}^{b}\pm \frac{2}{l^{2}} \left(
T^{a}T_{a}-e^{a}e^{b}R_{ab}\right) \\ &=&F_{\;B}^{A}F_{\;A}^{B}.
\end{eqnarray*}
This exotic lagrangian has the curious property of giving exactly the same field equations as the standard $dL_{3}^{AdS}$, but interchanged: varying with
respect to $e^{a}$ one gives the equation for $\omega ^{ab}$ of the other, and vice-versa.

In five dimensions, there are no Lorentz invariants that can be formed using $T^a$ and hence no new torsional lagrangians. In seven dimensions there are
three torsional CS terms,
\begin{center}
\textbf{Table 2}
\hspace{-1cm}
\begin{tabular}{|l|l|}
\hline $D=7$ Torsional Chern-Simons Lagrangian & ${\cal P}$  \\ \hline \hline
$L_7^{Lorentz}=\omega(d\omega)^3 +\cdot\cdot\cdot+\frac{4}{7}\omega^7$ & $R_{\;b}^{a}R_{\;c}^b R_{\;d}^c R_{\;a}^d $ \\ \hline 
$L_7^A=(L_3^{Lorentz})R_{\;b}^a R_{\;a}^b = \left(\omega_{\; b}^a d\omega_{\;a}^b +\frac{2}{3}\omega_{\;b}^a \omega_{\;c}^b \omega_{\;a}^c
\right) R_{\;b}^a R_{\;a}^b$ & $\left(R_{\;b}^a R_{\;a}^b \right)^2$ \\ \hline 
$L_7^B =(L_3^{Torsion})R_{\;b}^a R_{\;a}^b = (e^a T_a) (R_{\;b}^a R_{\;a}^b)$ & $(T^a T_a -e^a e^b R_{ab})R_{\;c}^d R_{\;d}^c$
\\ \hline
\end{tabular}
\end{center}
\vskip 0.3cm

\subsection{Characteristic Classes and Even $D$}

The CS construction fails in $2n$ dimensions for the simple reason that there are no topological invariants constructed with the ingredients we are using in
$2n+1$ dimensions. The topological invariants we have found so far, also called characteristic classes, are the Euler and the Pontryagin or Chern-Weil classes. The idea of characteristic class is one of the unifying concepts in mathematics that connects algebraic topology, differential geometry and algebraic
geometry. The theory of characteristic classes explains mathematically why it is not always possible to perform a gauge transformation that makes the
connection vanish everywhere even if it is locally pure gauge. The non vanishing value of a topological invariant signals an obstruction to the existence of a gauge transformation that trivializes the connection globally.

There are basically two types of invariants relevant for a Lorentz invariant theory in an even-dimensional manifold

$\bullet$ The Euler class, associated with the $O(D-n,n)$ groups. In two dimensions, the Euler number is related to the genus ($g$) of the surface,
$\chi = 2-2g$.

$\bullet$ The Pontryagin class, associated with any classical semisimple group $G$. It counts the difference between self dual and anti-self dual gauge
connections that are admitted in a given manifold.

The Nieh-Yan invariants correspond to the difference between Pontriagin classes for $SO(D+1)$ and $SO(D)$ in $D$ dimensions \cite{ChaZ}.

As there are no similar invariants in odd dimensions, there are no CS actions for gravity for even $D$, invariant under the (anti-) de Sitter or Poincar\'{e}
groups. In this light, it is fairly obvious that although ordinary Einstein-Hilbert gravity can be given a fiber bundle structure for the Lorentz group, this structure cannot be extended to include local translational invariance.

\subsubsection{Quantization of the gravitation constant}

The only free parameter in a Chern-Simons action is $\kappa$. Suppose a CS lagrangian is used to describe a simply connected, compact $2n-1$ dimensional manifold $M$, which is the boundary of a $2n$-dimensional compact orientable manifold $\Omega$. Then the action for the geometry of $M$ can be expressed as the integral of the Euler density $E_{2n}$ over $\Omega$, multiplied by $\kappa$. But since there can be many different manifolds with the same boundary $M$, the integral over $\Omega $ should give the same physical predictions as that over a different manifold, $\Omega^{\prime}$. In order for this change to leave the path integral unchanged, a minimal requirement would be
\begin{equation}
\kappa \left[ \int_{\Omega }E_{2n}-\int_{\Omega^{\prime }}E_{2n}\right] =2n\pi \hbar .  \label{quantum}
\end{equation}
The quantity in brackets --with the appropriate normalization-- is the Euler number of the manifold obtained by gluing $\Omega $ and $\Omega^{\prime}$ along $M$ in the right way to produce an orientable manifold, $\chi [\Omega \cup \Omega^{\prime }\dot{]}$. This integral can take an arbitrary integer value and from this one concludes that $\kappa $ must be quantized\cite{QuantumG},
\begin{equation}
\kappa =nh,
\end{equation}
where $h$ is\ Planck's constant.

\subsubsection{Born-Infeld gravity}

The closest one can get to a CS theory in even dimensions is with the so-called Born-Infeld ({\bf BI}) theories \cite{JJG,BTZ94,Tr-Z}. The BI lagrangian is
obtained by a particular choice of the $\alpha _{p}$'s in the Lovelock series, so that the lagrangian takes the form
\begin{equation}
L_{2n}^{BI}=\epsilon _{a_{1}\cdot \cdot \cdot a_{2n}}\bar{R}^{a_{1}a_{2}}\cdots \bar{R}^{a_{2n-1}a_{2n}},  \label{2nBI}
\end{equation}
where $\bar{R}^{ab}$ stands for the combination
\begin{equation}
\bar{R}^{ab}=R^{ab}\pm \frac{1}{l^{2}}e^{a}e^{b}. \label{concircular}
\end{equation}
With this definition it is clear that the lagrangian (\ref{2nBI}) contains only one free parameter, $l$, which, as explained above, can always be absorbed in
a redefinition of the vielbein. This lagrangian has a number of interesting classical features like simple equations, black hole solutions, cosmological models, etc. \cite{JJG,BTZ94,BHscan}. The simplification comes about because the equations admit a unique maximally symmetric configuration given by $\bar{R}^{ab}=0$, in contrast with the situation when all $\alpha_p$'s are arbitrary. As already mentioned, for arbitrary $\alpha_p$'s, the field
equations do not determine completely the components of $R^{ab}$ and $T^a$ in general. This is because the high nonlinearity of the equations can give rise to degeneracies. The BI choice is in this respect the best behaved since the degeneracies are restricted to only one value of the radius of curvature
($R^{ab}\pm \frac{1}{l^2} e^a e^b = 0$). At the same time, the BI action has the least number of algebraic constrains required by consistency among the field equations, and it is therefore the one with the simplest dynamical behavior\cite{Tr-Z}.

\subsection{Finite Action and the Beauty of Gauge Invariance}

Classical invariances of the action are defined modulo surface terms because they are usually assumed to vanish in the variations. This is true for boundary conditions that keep the values of the fields fixed at the boundary: Dirichlet conditions. In a gauge theory, however, it may be more relevant to fix gauge invariant properties at the boundary --like the curvature--, which are not precisely Dirichlet  boundary conditions, but rather conditions of the Neumann type.

On the other hand, it is also desirable to have an action which has a finite value, when evaluated on a physically observable configuration --e.g., on a classical solution. This is not just for the sake of elegance, it is a necessity if one needs to study the semiclassical thermodynamic properties of the theory. This is particularly true for a theory possessing black holes with interesting thermodynamic features. Moreover, quasi-gauge invariant actions defined on an infinitely extended spacetime are potentially ill defined. This is because, under gauge tranformations, the boundary terms that they pick could give infinite contributions to the action integral. This would not only cast doubt on the meaning of the action itself, but it would violently contradict the wish to have a gauge invariant action principle.

Changing the lagrangian by a boundary term may seem innocuous but it is a delicate business. The empirical fact is that adding a total derivative to a lagrangian in general changes the expression for the conserved Noether charges, and again, possibly by an infinite amount. The conclusion from this discussion is that some regularization principle must be in place in order that the action be finite on physically interesting configurations, and that assures it remains finite under gauge transformations, and yields well defined conserved charges. 

In \cite{MOTZ1} it is shown that the action has an extremum when the field equations hold, and is finite on classically interesting configuration if the
AdS action (\ref{(A)dS2n+1}) is supplemented with a boundary term of the form
\begin{equation}
B_{2n}=-\kappa n\int\limits_0^1dt\int\limits_0^t ds\;\epsilon \theta e\left( \widetilde{R}+t^2 \theta^2+s^2 e^2\right)^{n-1},  \label{B2n}
\end{equation}
where $\widetilde{R}$ and $\theta$ are the intrinsic and extrinsic curvatures of the boundary. The resulting action attains an extremum for boundary
conditions that fix the extrinsic curvature of the boundary. In that reference it is also shown that this action principle yields finite charges (mass, angular momentum) without resorting to ad-hoc regularizations or background subtractions. It can be asserted that in this case --as in many others--, the demand of gauge invariance is sufficient to cure other seemingly unrelated problems. 

\subsubsection{Transgressions}

The boundary term (\ref{B2n}) that that ensures convergence of the action and charges turns out to have other remarkable properties. It makes the action gauge invariant --and not just quasi-invariant-- under gauge transformations that keep the curvature constant (AdS geometry at the boundary), and the form of the extrinsic curvature fixed at the boundary. The condition of having a fixed AdS geometry asymptotic is natural for localized matter distributions such as black holes. Fixing the extrinsic curvature, on the other hand, implies that the connection aproaches a fixed reference connection at infinity in a prescribed manner. 

On closer examination, this boundary term can be seen to convert the action into the integral of a transgression. A transgression form is a gauge invariant object whose exterior derivative yields the difference of two Chern classes \cite{Nakahara},
\begin{equation}
 d{\cal T}_{2n-1} (A, \bar{A})={\cal P}_{2n}(A)-{\cal P}_{2n}(\bar{A}),  \label{transgression}
\end{equation}
where $A$ and $\bar{A}$ are two connections in the same Lie algebra. There is an explicit expression for the transgression form in terms of the Chern-Simons forms for $A$ and $\bar{A}$, 
\begin{equation}
{\cal T}_{2n+1} (A, \bar{A})={\cal C}_{2n+1}(A)-{\cal C}_{2n+1}(\bar{A}) + d{\cal B}_{2n}(A,\bar{A}). \label{transg}
\end{equation}

The last term in the R.H.S. is uniquely determined by the condition that the transgression form be invariant under simultaneous gauge transformations of both connections throughout the entire manifold $M$
\begin{eqnarray}
A \rightarrow A'=\Lambda^{-1} A \Lambda + \Lambda^{-1}d\Lambda \\ \label{delta A}
\bar{A} \rightarrow \bar{A}'=\bar{\Lambda}^{-1} \bar{A} \bar{\Lambda} + \bar{\Lambda}^{-1}d\bar{\Lambda} \label{delta barA}
\end{eqnarray}
with the matching condition at the boundary,
\begin{equation}
\bar{\Lambda}(x)-\Lambda(x) = 0, \;\; \mbox{ for }\;\;  x \in \partial M. \label{matching}
\end{equation}
It can be seen that the boundary term in (\ref{B2n}) is precisely the boundary term ${\cal B}_{2n}$ in the transgression form. The interpretation now presents some subtleties. Clearly one is not inclined to duplicate the fields by introducing a second dynamically independent set of fields ($\bar{A}$), having exactly the same couplings, spin, gauge symmetry and quantum numbers. 

One possible interpretation is to view the second connection as a nondynamical reference field. This goes against the well established principle that every quantity that occurs in the action that is not a coupling constant,  mass parameter, or numerical coefficient like the dimension or a combinatorial  factor, should correspond to a dynamical quantum variable \cite{Yang}. Even if one accepts the existence of this uninvited guest, an explanation would be needed to justify its not being seen in nature. 

\subsubsection{Cobordism}

An alternative interpretation could be to assume that the spacetime is duplicated and we happen to live on one of the two parallel words where $A$ is present, while $\bar{A}$ is in the other. These two worlds need to share the same boundary, for it would be otherwise very hard to justify the condition (\ref{matching}). This picture gains support from the fact that the two connections do not interact on $M$, but only at the boundary, through the boundary term $B(A,\bar{A})$. An obvious drawback of this interpretation is that the action for $\bar{A}$ has the wrong sign, and therefore woul lead to ghosts or rather unphysical negative energy states.

Although this sounds like poor science fiction, it suggests a more reasonable option proposed in \cite{MOTZ2}: Since the two connections do not interact, they could have support on two completely different but cobordant manifolds. Then, the action really reads,
\begin{equation}
 I[A,\bar{A}]= \int_{M} {\cal C}(A) +\int_{\bar{M}} {\cal C}(\bar{A}) +\int_{\partial M} {\cal B}(A, \bar{A}),
\end{equation}
where the orientations of $M$ and $\bar{M}$ are appropriately chosen so that they join at their common boundary with $\partial M=-\partial \bar{M}$.

The picture that emeges in this interpretation is one where we live in a region of spacetime ($M$) characterized by the dynamical field $A$. At he boundary of our region, $\partial M$, there exists another field with identical properties as $A$ and matching gauge symmetry. This second field $\bar{A}$ extends on to a cobordant manifold $M$, to which we have no direct access except through the interaction of $\bar{A}$ with our $A$. If the spacetime we live in is asymptotically AdS, this could be a reasonable scenario since the boundary then is causally connected to the bulk and can be easily viewed as the common boundary of two --or more-- asymptotically AdS spacetimes \cite{Zanelli2006}.

\subsubsection{Alternative regularizations}

The boundary term (\ref{B2n}) explicitly depends on the extrinsic curvature, which would be inadequate in a formulation that uses the intrinsic properties of the geometry as boundary data. The Dirichlet problem, in which the boundary metric is specified is an example of intrinsic framework. The standard regularization procedure in the Dirichlet approach uses counterterms that are covariant functions of the intrinsic boundary geometry. In \cite{MO}, the transgression and counterterms procedures have been explicitly compared in asymptotically AdS spacetimes using an adapted coordinate frame, known as Feferman-Graham  coordinates \cite{FG}.  Both regularization techniques are shown to be equivalent in that frame, and they differ at most by a finite counterterm that does not change the Weyl anomaly. The regularization techniques were also extended for nonvanishing torsion, yielding a finite action principle and holographic anomalies in five dimensions in \cite{BMOT}.

\subsection{Further Extensions}

By embedding the Lorentz group into one of its parents, the de Sitter or anti-de Sitter group, or its uncle, the Poincar\'{e} group, one can generate
gauge theories for the spacetime geometry in any odd dimension. These theories are based on the affine ($\omega$) and metric ($e$) features of the spacetime manifold as the only dynamical fields of the system. The theory has no dimensionful couplings and is the natural continuation of gravity in 2+1
dimensions. From here one can go on to study those theories, analyzing their classical solutions, their cosmologies and black holes that live on them,
etc. Although for the author of these notes, black hole solutions in these and related theories have been a constant source of surprises \cite{BTZ94,BTZ93,BHscan,TopBH}, we will not take this path here since that would sidetrack us into a completely different industry.

One possible extension would be to investigate embeddings in other, larger groups. The results in this direction are rather disappointing. One could embed the Lorentz group $SO(D-1,1)$ in any $SO(n,m)$, if $n \geq D-1$ and $m \geq 1$, and contractions of them analogous to the limit of vanishing cosmological constant that yields the Poincar\'{e} group. There are also some accidents like the (local) identity between $SO(3)$ and $SU(2)$, which occur occasionally and which make some people smile and others explode with hysterical joy, but that is rare.

The only other natural generalization of the Lorentz group into a larger group, seems to be direct products with other groups. That yields theories constructed as mere sums of Chern-Simons actions but not very interesting otherwise. The reason for this boring scenario, as we shall see in the next chapters, is connected to the so-called \emph{No-Go Theorems} \cite{Coleman-Mandula}. Luckily, there is a very remarkable (and at the time revolutionary) way out of this murky situation that is provided by supersymmetry. In fact, despite all the propaganda and false expectations that this unobserved symmetry has generated, this is its most remarkable feature, and possibly its only lasting effect in our culture: it provides a natural way to unify the symmetries of
spacetime, with internal symmetries like the gauge invariance of electrodynamics, the weak and the strong interactions.

\newpage
\section{Chern-Simons Supergravity}

We have dealt so far with the possible ways in which pure gravity can be extended by relaxing three standard assumptions of General Relativity: \\
{\bf i}) Notion of parallelism derived from metricity, \\
{\bf ii}) Four-dimensional spacetime, and \\
{\bf iii)} Gravitational lagrangian given by the Einstein Hilbert term $\sqrt{-g}R$ only.

Instead, we demanded: \\
{\bf i')} Second order field equations for the metric components, \\
{\bf ii')}  $D$-form lagrangian constructed out of the vielbein, $e^{a}$, the spin connection, $\omega _{\;b}^{a},$ and their exterior derivatives, and \\
{\bf iii')} Invariance of the action under local Lorentz rotations in the tangent space.

In this way, a family of lagrangians containing higher powers of the curvature and torsion multiplied by arbitrary and dimensionful coefficients is obtained. In odd dimensions, the embarrassing presence of these arbitrary constants was cured by enlarging the symmetry group, thereby making the theory gauge invariant under the larger symmetry group and simultaneously fixing all parameters in the lagrangian. The cure works only in odd dimensions. The result is a highly nonlinear Chern-Simons theory of gravity, invariant under local (A)dS transformations in the tangent space. We now turn to the problem of enlarging the contents of the theory to allow for supersymmetry. This will have two effects: it will incorporate fermions and fermionic generators into the picture, and it will enlarge the symmetry by including additional bosonic generators. These additional bosonic symmetries are required by consistency (the algebra must close) and are the most important consequence of supersymmetry.

\subsection{Supersymmetry}

Supersymmetry (\textbf{SUSY}) is a curious symmetry: most theoreticians are willing to accept it as a legitimate feature of nature, although it has never
been experimentally observed. The reason for its popularity rests on its uniqueness and beauty that makes one feel it would be a pity if such an elegant feature were not realized in nature somehow. This symmetry mixes \textbf{bosons} (integer spin particles) and \textbf{fermions} (half integer spin particles). These two types of particles obey very different statistics and play very different roles in nature, so it is somewhat surprising that there should exist a symmetry connecting them\footnote{Bosons obey Bose-Einstein statistics and, like ordinary classical particles, there is no limit to the number of them that can occupy the same state. Fermions, instead, cannot occupy the same quantum state more than one at a time, something known as Fermi-Dirac statistics. All elementary particles are either bosons or fermions and they play different roles: fermions like electrons, protons, neutrons and quarks are the
constituents of matter, while the four known fundamental interactions of nature are described by gauge fields resulting from the exchange of bosons like the photon, the gluon or the $W^{\pm}$ and $Z^0$, or the graviton.}.

The simplest supersymmetric theories combine bosons and fermions on equal footing, rotating them into each other under SUSY transformations. This is possibly the most intriguing --and uncomfortable-- aspect of supersymmetry: the blatant fact that bosons and fermions play such radically different roles in nature means that SUSY is not manifest around us, and therefore, it must be strongly broken at the scale of our observations. Unbroken SUSY would predict the existence of fermionic carriers of interactions and bosonic constituents of matter as partners of the known particles. None of these two types of particles
have been observed. On the other hand, there is no clue at present as to how to break supersymmetry.

In spite of its ``lack of realism", SUSY gained the attention of the high energy community mainly because it offered the possibility of taming the ultraviolet divergences of many field theories. It was observed early on that the UV divergences of the bosons were often cancelled out by divergences coming
from the fermionic sector. This possibility seemed particularly attractive in the case of a quantum theory of gravity, and in fact, it was shown that in a
supersymmetric extension of general relativity, dubbed supergravity (\textbf{SUGRA}) the ultraviolet divergences at the one-loop level exactly
cancelled (see \cite{PvN} and references therein). This is another remarkable feature of SUSY: local (gauge) SUSY is not only compatible with gravity. In
fact, by consistency, local SUSY \emph{requires} gravity.

A most interesting aspect of SUSY is its ability to combine \emph{bosonic spacetime symmetries}, like Poincar{\'e} invariance, with other \emph{internal
bosonic} symmetries like the $SU(3)\times SU(2)\times U(1)$ invariance of the standard model. Thus SUSY supports the hope that it could be possible to
understand the logical connection between spacetime and internal invariances. SUSY makes the idea that these different bosonic symmetries might be related, and in some way necessitate each other, more natural. In this way it might be possible to understand why it is that some internal symmetries are observed and others are not. The most important lesson from supersymmetry is not the unification of bosons and fermions, but the extension of the bosonic symmetry.

From an algebraic point of view, SUSY is the simplest nontrivial way to enlarge the Poincar{\'e} group, unifying spacetime and internal symmetries, thus
circumventing an important obstruction found by S. Coleman and J. Mandula \cite{Coleman-Mandula}. The obstruction, also called \emph{no-go theorem},
roughly states that if a physical system has a symmetry described by a Lie group $\textbf{G}$, that contains the Poincar{\'e} group and some other
internal symmetry, then the corresponding Lie algebra must be a direct sum of the form $\mathcal{G} = \mathcal{P}\bigoplus \mathcal{S}$, where $\mathcal{P}$ and $\mathcal{S}$ are the Poincar{\'e} and the internal symmetry algebras, respectively \cite{WinbergIII,Freund}. Supersymmetry is nontrivial because the algebra {\em is not} a direct sum of the spacetime and internal symmetries. The way the no-go theorem is circumvented is by having both commutators (antisymmetric product, $\left[\cdot , \cdot\right ]$) and anticommutators (symmetric product, $\left\{ \cdot ,\cdot \right\} $), forming what is known as a {\bf graded Lie algebra}, also called super Lie algebra, or simply superalgebra. See, e. g., \cite{Sohnius, Freund}.

\subsection{Superalgebras}

A superalgebra has two types of generators: bosonic, ${\bf B}_{i}$, and fermionic, ${\bf \Phi}_{\alpha }$. They are closed under the (anti-)commutator
operation, which follows the general pattern 
\begin{eqnarray}
\left[ {\bf B}_{i}{\bf ,B}_{j}\right] &=&C_{ij}^{k}{\bf B}_{k}  \label{bb} \\
\left[ {\bf B}_{i}{\bf ,\Phi}_{\alpha }\right] &=&C_{i\alpha }^{\beta
}{\bf \Phi}_{\beta }  \label{bf} \\
\left\{ {\bf \Phi}_{\alpha },{\bf \Phi}_{\beta }\right\} &=&C_{\alpha \beta
}^{i}{\bf B}_i  \label{ff}
\end{eqnarray}
The generators of the Poincar\'{e} group are included in the bosonic sector, and the ${\bf \Phi}_{\alpha }$'s are the supersymmetry generators. This
algebra, however, does not close for an arbitrary bosonic group. Given a Lie group with a set of bosonic generators, it is not always possible to find a set of fermionic generators to enlarge the algebra into a closed superalgebra. The operators satisfying relations of the form (\ref{bb}-\ref{ff}), are still required to satisfy a consistency condition, the super-Jacobi identity, which is required by associativity,
\begin{equation}
\lbrack {\bf G}_I,[{\bf G}_J , {\bf G}_K]_{\pm }]_{\pm }+(-)^{\sigma (J K I)}[{\bf G}_J,[{\bf G}_K,{\bf G}_I]_{\pm }]_{\pm }+(-)^{\sigma (K I J)}[{\bf
G}_K,[{\bf G}_I,{\bf G}_J]_{\pm }]_{\pm }=0. \label{superjacobi}
\end{equation}
Here ${\bf G}_I$ represents any generator in the algebra, $[{\bf A}, {\bf B}]_{\pm}={\bf A}{\bf B}\pm {\bf B}{\bf A}$, where the sign is chosen according
the bosonic or fermionic nature of the operators in the bracket, and $\sigma (J K I)$ is the number of permutations of fermionic generators necessary for $(I J
K)\rightarrow (J K I)$ .

It is often the case that apart from extra bosonic generators, a collection of $\mathcal{N}$  fermionic operators are needed to close the algebra. This usually works for some values of $\mathcal{N}$ only, but in some cases there is simply no supersymmetric extension at all \cite{vH-VP}. This happens, for example, if one starts with the de Sitter group, which has no supersymmetric extension in general \cite{Freund}. For this reason, in what follows we will restrict to AdS theories. The general problem of classifying all possible superalgebras that extend the classical Lie algebras has been discussed in \cite{Kac}.

\subsection{Supergravity}

The name supergravity ({\bf SUGRA}) applies to any of a number of supersymmetric theories that include gravity in their bosonic sectors\footnote{Some authors would reserve the word \emph{supergravity} for supersymmetric theories whose gravitational sector is the Einstein-Hilbert (\textbf{EH}) lagrangian. This narrow definition seems untenable for dimensions $D>4$ in view of the variety of possible gravity theories beyond EH. Our point of view here is that there can be more than one system that can be called supergravity, although its connection with the standard theory remains unsettled.}. The invention/discovery of supergravity in the mid 70's came about with the spectacular announcement that some ultraviolet divergent graphs in pure gravity were canceled by the inclusion of their supersymmetric counterparts. For some time it was hoped that the supersymmetric extension of GR could be renormalizable. However, it was eventually realized that SUGRAs too might turn out to be nonrenormalizable \cite{Townsend}.

Standard SUGRAs are not gauge theories for a group or a supergroup, and the local (super-)symmetry algebra closes naturally on shell only. The algebra
could be made to close off shell by force, at the cost of introducing auxiliary fields --which are not guaranteed to exist for all $D$ and $\mathcal{N}$
\cite{RT}--, and still the theory would not have a fiber bundle structure since the base manifold is identified with part of the fiber. Whether it is the lack
of fiber bundle structure the ultimate reason for the nonrenormalizability of gravity remains to be proven. It is certainly true, however, that if GR could
be formulated as a gauge theory, the chances for its renormalizability would clearly improve. At any rate, now most high energy physicists view
supergravity as an effective theory obtained from string theory in some limit. In string theory, eleven dimensional supergravity is seen as an effective
theory obtained from ten dimensional string theory at strong coupling \cite{Theisen}. In this sense supergravity would not be a fundamental theory
and therefore there is no reason to expect that it should be renormalizable.

\subsection{From Rigid Supersymmetry to Supergravity}

Rigid or global SUSY is a supersymmetry in which the group parameters are constants throughout spacetime. In particle physics the spacetime is usually
assumed to have fixed Minkowski geometry. Then the relevant SUSY is the supersymmetric extension of the Poincar\'{e} algebra in which the supercharges
are ``square roots'' of the generators of spacetime translations, $\{{\bf \bar{Q}},{\bf Q}\}\sim \Gamma \cdot {\bf P}$. The extension of this to a local
symmetry can be done by substituting the momentum ${\bf P}_{\mu }=i\partial_{\mu }$ by the generators of spacetime diffeomorphisms, $\mathcal{H}_{\mu }$, and relating them to the supercharges by $\{{\bf \bar{Q}},{\bf Q}\}\sim \Gamma \cdot \mathcal{H}$. The resulting theory has a local supersymmetry algebra which only closes on-shell \cite{PvN}. As we discussed above, the problem with on-shell symmetries is that they are not likely to survive quantization.

An alternative approach for constructing the SUSY extension of the AdS symmetry is to work on the tangent space rather than on the spacetime manifold. This point of view is natural if one recalls that spinors are defined relative to a local frame on the tangent space and not as tensors on the coordinate basis. In
fact, spinors provide an irreducible representation for $SO(N)$, but not for $GL(N)$, which describe infinitesimal general coordinate transformations. The
basic strategy is to reproduce the 2+1 ``miracle'' in higher dimensions. This idea was applied in five dimensions \cite{Chamseddine}, as well as in higher odd
dimensions \cite {BTrZ,TrZ1,TrZ2}.

\subsection{Standard Supergravity}

In its simplest version, supergravity was conceived inthe early 70s, as a quantum field theory whose action included the Einstein-Hilbert term, representing a massless spin-2 particle (graviton), plus a Rarita-Scwinger kinetic term describing a massless spin-3/2 particle (gravitino) \cite{SUGRA}. These fields would transform into each other under local supersymmetry. Later on, the model was refined to include more ``realistic" features, like matter couplings, enlarged symmetries, higher dimensions with their corresponding reductions to 4D, cosmological constant, etc., \cite{PvN}. In spite of the  number of variations on the theme, a few features remained as the hallmark of SUGRA, which were a reflection of this history. In time, these properties have become a sort of identikit of SUGRA, although they should not be taken as a set of necessary postulates. Among these, three that will be relaxed in our construction:

\noindent
{\bf (i)} Gravity is described by the EH action (with/without cosmological constant),

\noindent
{\bf (ii)} The spin connection and the vielbein are related through the torsion equation, and,

\noindent
{\bf (iii)} The fermionic and bosonic fields in the lagrangian come in combinations such that they have equal number of propagating degrees of freedom.

The last feature is inherited from rigid supersymmetry in Minkowski space, where particles form vector representations of the Poincar\'{e} group
labelled by their spin and mass, and the matter fields form vector representations of the internal groups (multiplets). This is justified in a Minkowski background where particle states are represented by in- and out- plane waves in a weakly interacting theory. This argument, however, breaks down if the Poincar\'{e} group is not a spacetime symmetry, as it happens in asymptotically AdS spacetimes and in other cases, such as 1+1 dimensions with broken translational invariance \cite{LSV}.

The argument for the matching between fermionic and bosonic degrees of freedom goes as follows: The generator of translations in Minkowski space, $P_{\mu}=(E,\mathbf{p})$, commutes with all symmetry generators, therefore an internal symmetry should only mix particles of equal mass. Since supersymmetry changes the spin by 1/2, a supersymmetric multiplet must contain, for each bosonic eigenstate of the hamiltonian $|E>_{B}$, a fermionic one with the same energy, $|E>_{F}={\bf Q}$ $|E>_{B}$, and vice versa. Thus, it seems natural that in supergravity this would still be the case. In AdS space, however, the momentum operator is not an abelian generator, but acts like the rest of Lorentz generators and therefore the supersymmetry generator ${\bf Q}$ doesn't commute with it. Another limitation of this assumption is that it does not consider the possibility that the fields belong to a different representation of the Poincar\'{e} or AdS group, such as the adjoint representation.

Also implicit in the argument for counting the degrees of freedom is the usual assumption that the kinetic terms are those of a minimally coupled gauge
theory, a condition that is not met by any CS theory. Apart from the difference in background, which requires a careful treatment of the unitary irreducible
representations of the asymptotic symmetries \cite{Fronsdal}, the counting of degrees of freedom in CS theories follows a completely different pattern
\cite{BGH} from the counting for the same connection 1-forms in a YM theory \cite{HTZ}.

The other two issues concern the purely gravitational sector and are dictated by economy: after Lovelock's theorem, there is no reason to adopt {\bf (i)},
and the fact that the vielbein and the spin connection are dynamically independent fields on equal footing makes assumption {\bf (ii)} unnatural.
Furthermore, the elimination of the spin connection from the action introduces the inverse vielbein in the action and thereby entangling the action of the
spacial symmetries defined on the tangent space. The fact that the supergravity generators do not form a closed off-shell algebra may be traced back to these
assumptions.
\newpage

\section{AdS Superalgebras}

In order to construct a supergravity theory that contains gravity with cosmological constant, a mathematically oriented physicist would look for the
smallest superalgebra that contains the generators of the AdS algebra. This question was asked --and answered-- many years ago, at least for some
dimensions $D=2,3,4$ $mod$ $8$, in \cite{vH-VP}. But this is not all, we would also like to have an action that realizes the symmetry.

Several supergravities are known for all dimensions $D\leq 11$ \cite{Salam-Sezgin}. For $D=4$, a supergravity action that includes a cosmological constant was first discussed in \cite{Townsend77}, however, finding a supergravity with cosmological constant in an arbitrary dimension is a nontrivial task. For example, the standard supergravity in eleven dimensions has been know for a long time \cite{CJS}, however, it has been shown that it is impossible to accommodate a cosmological constant  \cite{BDHS,Deser}. Moreover, although it was known to the authors of Ref.\cite{CJS} that the supergroup containing the eleven dimensional AdS group is $SO(32|1)$, a gravity action which exhibits this symmetry was found almost twenty years later \cite{TrZ1}.

In what follows, we present an explicit construction of the superalgebras that contain AdS algebra, $so(D-1,2)$, along the lines of \cite{vH-VP}, where we have extended the method to apply it to the cases $D=5$, $7$ and $9$ as well \cite{TrZ1}. The crucial observation is that the Dirac matrices provide a
natural representation of the AdS algebra in any dimension. Thus, the AdS connection ${\bf W}$ can be written in this representation as ${\bf W}= e^{a}{\bf J}_{a}+\frac{1}{2}\omega ^{ab}{\bf J}_{ab}$, where
\begin{eqnarray}
{\bf J}_{a} &=& \frac{1}{2}(\Gamma _{a})_{\beta }^{\alpha} , \label{ja} \\
{\bf J}_{ab}&=& \frac{1}{2}(\Gamma_{ab})_{\beta }^{\alpha }. \label{jab}
\end{eqnarray}
Here ${\bf \Gamma }_{a}$, $a=1,...,D$ are $m\times m$ Dirac matrices, where $m=2^{[D/2]}$ ($[x]$ is the integer part of $x$), and ${\bf \Gamma}_{ab}=\frac{1}{2}[{\bf \Gamma }_{a},{\bf \Gamma }_{b}]$. These two classes of matrices form a closed commutator subalgebra (the AdS algebra) of the {\bf Dirac algebra}. The Dirac algebra is obtained by taking all the antisymmetrized products of ${\bf \Gamma}$ matrices
\begin{equation}
{\bf I},{\bf \Gamma }_{a},{\bf \Gamma }_{a_{1}a_{2}},...,{\bf \Gamma}_{a_{1}a_{2} \cdot \cdot \cdot a_{D}}, \label{gamma-algebra}
\end{equation}
where
\[
{\bf \Gamma }_{a_1 a_2\cdots a_k}=\frac{1}{k!}({\bf \Gamma}_{a_1}{\bf \Gamma }_{a_2} \cdots {\bf \Gamma}_{a_k}\pm [permutations]).
\]
For even $D$, the matrices in the set (\ref{gamma-algebra}) are all linearly independent, but for odd $D$ they are not, because ${\bf \Gamma }_{12\cdots
D} = \sigma {\bf I}$ and therefore half of them are proportional to the other half. Thus, the dimension of this algebra, that is, the number of independent matrices of the form (\ref{gamma-algebra}) is $m^{2}=2^{2[D/2]}$. This representation provides an elegant way to generate all $m\times m$ matrices (note that $m=2^{[D/2]}$ is not \textit{any} number).

\subsection{The Fermionic Generators} 

The supersymmetric extension of a given Lie algebra is a mathematical problem whose solution lies in the general classification of superalgebras \cite{Kac}.
Although their representations were studied more than 20 years ago in \cite{vH-VP}, the application to construct SUSY field theory actions has not been pursued much. Instead of approaching this problem as a general question of classification of irreducible representations, we will take a more practical
course, by identifying the representation we are interested in from the start. This representation is the one in which the bosonic generators take the form 
(\ref{ja}), (\ref{jab}). The simplest extension of the algebra generated by those matrices is obtained by the addition of one row and one column, as
\begin{equation}
{\bf J}_{a}=\left[ \begin{array}{cc} \frac{1}{2}(\Gamma _{a})_{\beta }^{\alpha } & 0 \\
0 & 0
\end{array}
\right] ,  \label{ja+}
\end{equation}
\begin{equation}
{\bf J}_{ab}=\left[ \begin{array}{cc} \frac{1}{2}(\Gamma _{ab})_{\beta }^{\alpha } & 0 \\
0 & 0
\end{array}
\right] .  \label{jab+}
\end{equation}
The generators associated to the new entries would have only one spinor index. Let us call ${\bf Q}_{\gamma}$ ($\gamma= 1,...,m$) the generator that has only one nonvanishing entry in the $\gamma $-th row of the last column,
\begin{equation}
{\bf Q}_{\gamma }=\left[ \begin{array}{cc} 0 & \delta _{\gamma }^{\alpha } \\
-C_{\gamma \beta } & 0
\end{array}
\right]\;\;\;\;\;\; \alpha, \beta = 1,...,m. \label{q-generator}
\end{equation}
Since this generator carries a spinorial index ($\gamma$), it is in a spin-1/2 representation of the Lorentz group.

The entries of the bottom row ($C_{\gamma \beta }$) will be so chosen as to produce the smallest supersymmetric extensions of AdS. There are essentially
two ways restrict the dimension of the representation compatible with Lorentz invariance: {\em chirality} and {\em reality}. In odd dimensions there is no
chirality because the corresponding ``$\gamma_5$'' is proportional to the identity matrix. Reality instead can be defined in any dimension and refers to whether a spinor and its conjugate are proportional up to a constant matrix, $\bar{\psi} = {\bf C} \psi$, or more explicitly,
\begin{equation}
\bar{\psi}^{\alpha }=C^{\alpha \beta }\psi _{\beta }. \label{conjugation}
\end{equation}
A spinor that satisfies this condition is said to be Majorana, and  ${\bf C}=(C^{\alpha \beta })$ is called the charge conjugation matrix. This matrix is
assumed to be invertible, $C_{\alpha \beta}C^{\beta \gamma }=\delta _{\alpha}^{\gamma }$, and plays the role of a metric in the space of Majorana spinors.

Using the form (\ref{q-generator}) for the supersymmetry generator, its Majorana conjugate ${\bf \bar{Q}}$ is found to be
\begin{eqnarray}
{\bf \bar{Q}}^{\gamma } &=& C^{\alpha \beta }{\bf Q}_{\beta } \nonumber \\
&=&\left[
\begin{array}{cc}
0 & C^{\alpha \gamma } \\
-\delta _{\beta }^{\gamma } & 0
\end{array}
\right] .  \label{qbar-generator}
\end{eqnarray}

The matrix $\mathbf{C}$ can be viewed as performing a change of basis $\psi \rightarrow \psi^{\rm{T}} = \mathbf{C}\psi$, which is turn corresponds to the change $\Gamma \rightarrow \Gamma^{\rm{T}}$. Now, since the Clifford algebra for the Dirac matrices,
\begin{equation}
\{{\bf \Gamma }^{a},{\bf \Gamma }^{b}\}=2\eta ^{ab}, \label{Dirac}
\end{equation}
is also obeyed by their transpose, $({\bf \Gamma}^{a})^{\rm{T}}$, these two algebras must be related by a change of basis, up to a sign,
\begin{equation}
({\bf \Gamma }^{a})^{\rm{T}}=\eta {\bf C\Gamma}^{a}{\bf C}^{-1} \;\;\mbox{with}\ \eta ^{2}=1. \label{c-gamma}
\end{equation}
The basis of this Clifford algebra (\ref{Dirac}) for which an operator ${\bf C}$ satisfying (\ref{c-gamma}) exists, is called the Majorana representation.
This last equation is the defining relation for the charge conjugation matrix, and whenever it exists, it can be chosen to have definite parity\footnote{The
Majorana reality condition can be satisfied in any $D$ provided the spacetime signature is such that, if there are $s$ spacelike and $t$ timelike dimensions,
then $s-t=0,1,2,6,7$ mod $8$ \cite{Sohnius,Freund}. Thus, for lorentzian signature, Majorana spinors can be defined unambiguously only for $D=2,3,4,8,9,$ mod $8$.},
\begin{equation}
{\bf C}^{\rm{T}}=\lambda {\bf C,}\rm{ with }\lambda =\pm 1. \label{c}
\end{equation}

\subsection{Closing the Algebra} 

We already encountered the bosonic generators responsible for the AdS transformations (\ref{ja}, \ref{jab}), which have the general form required by
(\ref{bb}). It is straightforward to check that commutators of the form $[{\bf J},{\bf Q}]$ turn out to be proportional to ${\bf Q}$, in agreement with the
general form (\ref{bf}). What is by no means trivial is the closure of the anticommutator $\{{\bf Q},{\bf Q}\}$ as in (\ref{ff}). Direct computation yields
\begin{eqnarray}
\{{\bf Q}_{\gamma },{\bf Q}_{\lambda }\}_{\beta }^{\alpha } &=&\left[ \begin{array}{cc}
0 & \delta _{\gamma }^{\alpha } \\
-C_{\gamma \rho } & 0
\end{array}
\right] \left[
\begin{array}{cc}
0 & \delta _{\lambda }^{\rho } \\
-C_{\lambda \beta } & 0
\end{array}
\right] +(\gamma \leftrightarrow \lambda ) \nonumber \\
&=&-\left[
\begin{array}{cc}
\delta _{\gamma }^{\alpha }C_{\lambda \beta }+\delta_{\lambda
}^{\alpha}C_{\gamma \beta } & 0 \\
0 & C_{\gamma \lambda}+C_{\lambda \gamma }
\end{array}
\right] .  \label{qq+qq}
\end{eqnarray}

The form of the lower diagonal piece immediately tells us that unless $C_{\gamma \lambda}$ is antisymmetric, the right hand side of (\ref{qq+qq}) cannot be a linear combination of ${\bf J_a}$, ${\bf J_{ab}}$ and ${\bf Q}$. In that case, new bosonic generators with nonzero entries in this diagonal block will be required to close the algebra (and possibly more than one). This relation also shows that the upper diagonal block is a collection of matrices ${\bf M}_{\gamma \lambda }$ whose components take the form
\[
\left( M_{\gamma \lambda}\right)_{\beta}^{\alpha}=-(\delta_{\gamma}^{\alpha}C_{\lambda \beta }+ \delta _{\lambda}^{\alpha}C_{\gamma \beta}).
\]
Multiplying both sides of this relation by $C$, one finds
\begin{equation}
\left( CM_{\gamma \lambda}\right) _{\alpha \beta}=-(C_{\alpha \gamma}C_{\lambda \beta }+C_{\alpha \lambda}C_{\gamma \beta}),  \label{cc+cc}
\end{equation}
which is symmetric in $(\alpha \beta )$. This means that the bosonic generators can only include those matrices in the Dirac algebra such that, when multiplied by ${\bf C}$ on the left (${\bf CI},$ ${\bf C\Gamma}_a$, ${\bf C\Gamma}_{a_1 a_2},...,$ ${\bf C\Gamma}_{a_1 a_2 \cdots a_D}$) turn out to be symmetric. The other consequence of this is that, if one wants to have the AdS algebra as part of the superalgebra, both ${\bf C\Gamma}_{a}$ and ${\bf C\Gamma}_{ab}$ should be symmetric matrices. Now, multiplying (\ref{c-gamma}) by ${\bf C}$ from the right, we have
\begin{equation}
({\bf C\Gamma }_{a})^{\rm{T}}=\lambda \eta {\bf C\Gamma }_{a}, \label{c-gamma2}
\end{equation}
which means that we need
\begin{equation}
\lambda \eta =1.  \label{eta}
\end{equation}
From (\ref{c-gamma}) and (\ref{c}), it can be seen that
\[
({\bf C\Gamma }_{ab})^{\rm{T}}=-\lambda {\bf C\Gamma }_{ab},
\]
which in turn requires
\[
\lambda =-1=\eta .
\]
This means that ${\bf C}$ is antisymmetric ($\lambda =-1$) and then the lower diagonal block in (\ref{qq+qq}) vanishes identically. However, the values of
$\lambda $ and $\eta $ cannot be freely chosen but are fixed by the spacetime dimension as is shown in the following table,\\
\begin{center}
\textbf{Table 3}\\
\[
\begin{array}{|c|c|c|}
\hline {\bf D} & \lambda & \eta \\ \hline
 3 & -1 & -1 \\
 5 & -1 & +1 \\
 7 & +1 & -1 \\
 9 & +1 & +1 \\
 11 &-1 & -1 \\ \hline
\end{array}
\] 
\end{center}
and the pattern repeats mod $8$ (see Ref.\cite{TrZ2} for details). This table shows that the simple cases occur for dimensions $3$ mod $8$, while for the
remaining cases life is a little harder. For $D=7$ mod $8$ the need to match the lower diagonal block with some generators can be satisfied quite naturally
by including several spinors labelled with a new index, $\psi _{i}^{\alpha}$, $i = 1,..., \mathcal{N}$, and the generator of supersymmetry should also carry
the same index. This means that there are actually $\mathcal{N}$ supercharges or, as it is usually said, the theory has an extended supersymmetry
($\mathcal{N} \geq 2$). For $D=5$ mod $4$ instead, the superalgebra can be made to close in spite of the fact that $\eta =+1$ if one allows complex spinor
representations, which is a particular form of extended supersymmetry since now ${\bf Q}_{\gamma}$ and ${\bf \bar{Q}}^{\gamma}$ are independent.

So far we have only given some restrictions necessary to close the algebra so that the AdS generators appear in the anticommutator of two supercharges. As we have observed, in general, apart from ${\bf J}_{a}$ and ${\bf J}_{ab}$, other matrices will occur in the r.h.s. of the anticommutator of ${\bf Q}$ and ${\bf \bar{Q}}$ which extends the AdS algebra into a larger bosonic algebra. This happens even in the cases in which the supersymmetry is not extended 
($\mathcal{N}$=1). 

The bottom line of this construction is that the supersymmetric extension of the AdS algebra for each odd dimension falls into one of these families:
\begin{itemize}
 \item $D=3$ mod $8$ (Majorana representation, $\mathcal{N}\geq $1),

 \item $D=7$ mod $8$ (Majorana representation, even $\mathcal{N}$), and

 \item $D=5$ mod $4$ (complex representations, $\mathcal{N}\geq$ 1 [or $2\mathcal{N}$ real spinors]).
\end{itemize}

The corresponding superalgebras\footnote{The algebra $osp(p|q)$ (resp. $usp(p|q)$) is that which generates the orthosymplectic (resp. unitary-symplectic) Lie group. This group is defined as the one that leaves invariant the quadratic form $G_{AB}z^A z^B = g_{ab}x^a x^b + \gamma_{\alpha \beta} \theta^{\alpha} \theta^{\beta}$, where $g_{ab}$ is a $p$-dimensional symmetric (resp. hermitean) matrix and $\gamma_{\alpha \beta}$ is a $q$-dimensional antisymmetric (resp. anti-hermitean) matrix.}were computed by van Holten and Van Proeyen for $D=2, 3, 4$ mod $8$ in Ref. \cite{vH-VP}, and in the other cases, in Refs.\cite{TrZ1,TrZ2}:

\begin{center}
\textbf{Table 4}\\
\vspace{0.3cm}
\begin{tabular}{|l|c|c|}
\hline D & S-Algebra & Conjugation Matrix \\ \hline 
3 mod $8$ & $osp(m|N)$ & $C^{T}=-C$ \\ \hline 
7 mod $8$ & $osp(N|m)$ & $C^{T}=C$ \\ \hline 
5 mod $4$ & $usp(m|N)$ & $C^{\dag }=C$ \\ \hline
\end{tabular}
\end{center}

\newpage

\section{CS Supergravity Actions}

In the previous sections we saw how to construct CS actions for the AdS connection ib $D=2n-1$ dimensions. Now, we will repeat this construction for the connection of a larger superalgebra in which AdS is embedded. Consider an arbitrary connection one-form ${\bf A}$, with values in some Lie algebra
$\cal{G}$, whose curvature is ${\bf F}=d{\bf A}+{\bf A}\wedge {\bf A}$. Then, the $2n$-form
\begin{equation}
{\cal P}_{2n}\equiv <{\bf F}\wedge \cdots \wedge {\bf F}>, \label{topclass}
\end{equation}
is invariant under the Lie group whose algebra is $\cal{G}$, provided the bracket $<\cdots >$ is an invariant tensor of the group. Furthermore, ${\cal P}_{2n} $ is closed: $d{\cal P}_{2n}=0$, and therefore can be locally written as an exact form,
\[
{\cal P}_{2n}=d{\bf L}_{2n-1}.
\]
The $(2n-1)$-form ${\bf L}_{2n-1}$ is a CS lagrangian, and therefore the problem reduces to finding the invariant bracket. The canonical --and in many
cases unique-- choice of invariant tensor with the features required here is the {\bf supertrace},
\begin{equation}
\langle {\bf F}\wedge \cdots \wedge {\bf F}\rangle =:STr\left[\Theta{\bf F}\wedge \cdots \wedge {\bf F}\right],
\label{bracket}
\end{equation}
where $\Theta$ is an invariant matrix in the supergroup, and the supertrace is defined as follows: if a matrix has the form
\[
{\bf M}=\left[ \begin{array}{cc} N_{b}^{a} & F_{\beta }^{a} \\
H_{b}^{\alpha } & S_{\beta }^{\alpha }
\end{array} \right] ,
\]
where $a,b$ are (bosonic) tensor indices and $\alpha, \beta$ are (fermionic) spinor indices, then $STr[{\bf M}]=Tr[{\bf N}]-Tr[{\bf S} ]= N_{a}^{a}-S_{\alpha}^{\alpha}$.

If we call ${\bf G}_M$ the generators of the Lie algebra, so that ${\bf A}={\bf G}_M A^M$, ${\bf F}={\bf G}_M F^M$, then
\begin{eqnarray}
{\cal P}_{2n} &=&STr\left[\Theta{\bf G}_{M_{1}}\cdots {\bf G}_{M_{n}}\right] F^{M_{1}}\cdots F^{M_{n}}  \nonumber
\\ &=&g_{_{\rm{M}_{1}\cdots \rm{M}_{n}}}F^{M_{1}}\cdots F^{M_n}=d{\bf L}_{2n-1},  \label{defCS}
\end{eqnarray}
where $g_{_{\rm{M}_{1}\cdots \rm{M}_{n}}}$ is an invariant tensor of rank $n$ in the Lie algebra. Thus, the steps to construct the CS lagrangian are
straightforward: Take the supertrace of all products of generators in the superalgebra and solve equation (\ref{defCS}) for ${\bf L}_{2n-1}$. Since the
superalgebras are different in each dimension, the CS lagrangians differ in field content and dynamical structure from one dimension to the next, although
the invariance properties are similar in all cases. The action
\begin{equation}
I_{2n-1}^{CS}[{\bf A}]=\int {\bf L}_{2n-1}  \label{generalCSaction}
\end{equation}
is invariant, up to surface terms, under the local gauge transformation
\begin{equation}
\delta {\bf A}=\nabla {\bf \Lambda},  \label{gaugetransf}
\end{equation}
where ${\bf \Lambda}$ is a zero-form with values in the Lie algebra $\textsf{g}$, and $\nabla $ is the exterior covariant derivative in the
representation of ${\bf A}$. In particular, under a supersymmetry transformation, ${\bf \Lambda}= \bar{\epsilon}^i Q_i-\bar{Q}^i \epsilon_i$, and
\begin{equation}
\delta _{\epsilon }{\bf A}=\left[ \begin{array}{cc}
\epsilon ^{k}\bar{\psi}_{k}-\psi ^{k}\bar{\epsilon}_{k} & D\epsilon_{j} \\
-D\bar{\epsilon}^{i} & \bar{\epsilon}^{i}\psi _{j}-\bar{\psi}^{i}\epsilon _{j}
\end{array} \right] ,  \label{delA}
\end{equation}
where $D$ is the covariant derivative on the bosonic connection,
\[
D\epsilon _j =\left(d+\frac{1}{2l}e^a {\bf \Gamma}_a + \frac{1}{4}\omega^{ab}{\bf \Gamma}_{ab}+\frac{1}{2r!}b^{[r]}{\bf \Gamma}_{[r]}\right) \epsilon_j -a_j^i \epsilon_i.
\]

\subsection{Examples of AdS-CS SUGRAs}

Now we examine some simple examples of anti-de Sitter Chern-Simons supergravities. For more detailed discussions and other examples, see \cite{Tr-Z,TrZ1,TrZ2,ChTrZ}.
\vskip0.5cm

{\bf A. D=3}

The simplest locally supersymmetric extension of AdS occurs in three dimensions for $\mathcal{N}$=1. In this case, there exist Majorana spinors and the
lagrangian is \cite{Achucarro-Townsend}
\begin{equation}
L=\epsilon_{abc}\left[\frac{R^{ab}e^{c}}{l}+\frac{e^{a}e^{b}e^{c}}{3l^3}\right]- \bar{\psi}\overline{\nabla}\psi, \label{L3}
\end{equation}
where $\overline{\nabla}$ stands for the AdS covariant derivative $\overline{\nabla}=d+\frac{1}{2l} e^a \Gamma_a + \frac{1}{4}\omega^{ab}\Gamma_{ab}$.  It is straightforward to show that this action is invariant --up to surface terms-- under supersymmetry,
\[
\delta_{\epsilon}\psi = \overline{\nabla} \epsilon, \;\;\; \delta_{\epsilon} e^a =\frac{1}{2}\bar{\epsilon}\Gamma^a\psi, \;\;\; \delta_{\epsilon}
\omega^{ab} =-\frac{1}{2}\bar{\epsilon}\Gamma^{ab}\psi.
\]
The proof of invariance is direct and no field equations need to be invoked (off-shell local SUSY). Action (\ref{L3}) can also be written as
\[
L= \langle \textbf{A}d\textbf{A}+\frac{2}{3} \textbf{A}^3 \rangle,
\]
where ${\bf A}$ is the connection for the superalgebra $osp(2|1)$, 
\begin{equation}
{\bf A} =  \frac{1}{l}e^a {\bf J}_a + \frac{1}{2}\omega^{ab}{\bf J}_{ab} + {\bf \bar{Q}}\psi, \label{baticonnection}
\end{equation}
and $\langle \cdots \rangle$ stands for the supertrace of the matrix representation for ${\bf J}_{a}$, ${\bf J}_{ab}$ and ${\bf Q}$ defined in (\ref{bracket}) (with an appropriate $\Theta$ insertion). The only nonvanishing brackets are:
\begin{eqnarray}
\langle{\bf J}_{ab}{\bf J}_{c}\rangle &=& 2\langle {\bf J}_a {\bf J}_b {\bf J}_c \rangle = \epsilon_{abc}, \\
\langle {\bf Q}_{\alpha}\; {\bf Q}_{\beta}\rangle &=& -C_{\alpha\beta}, \\
\langle {\bf Q}_{\alpha}\; {\bf Q}_{\beta}\; {\bf J}_{a}\rangle &=& \frac{1}{2} (C \Gamma_{a})_{\alpha\beta}, \\
\langle {\bf Q}_{\alpha}\; {\bf Q}_{\beta}\; {\bf J}_{ab}\rangle &=& \frac{1}{2} (C \Gamma_{ab})_{\alpha\beta}.
\end{eqnarray}
\vskip0.5cm

{\bf B. D=5}

In this case the supergroup is $SU(2,2|\cal{N})$. The associated connection can be written as
\begin{equation}
{\bf A} = \frac{1}{l}e^a {\bf J}_a + \frac{1}{2}\omega^{ab} {\bf J}_{ab}+ A^{K} {\bf T}_K + (\bar{\psi}^r {\bf Q}_r - {\bf \bar{Q}}^r \psi _r) + b{\bf Z},
\label{D=5connection}
\end{equation}
where the generators ${\bf J}_{a}$, ${\bf J}_{ab}$, form an AdS algebra ($so(4,2)$), ${\bf T}_{K}$ ($K=1,\cdots, {\cal N}^2 -1$) are the
generators of $su(\cal{N})$, ${\bf Z}$ generates a $U(1)$ subgroup and the spinorial supersymmetry generators ${\bf Q}_r, {\bf \bar{Q}}^r$ are in a
vector representation of $SU(\cal{N})$. The Chern-Simons Lagrangian for this gauge algebra is defined by the relation $dL=iSTr[{\bf F}^3]$, where ${\bf F}=d{\bf A} + {\bf A}^2$ is the (antihermitean) curvature. Using this definition, one obtains the lagrangian originally discussed by Chamseddine in
\cite{Chamseddine},
\begin{equation}
L=L_{G}(\omega ^{ab},e^a)+ L_{su(\cal{N})}(A_s ^r)+ L_{u(1)}(\omega^{ab}, e^a, b) + L_{F}(\omega^{ab}, e^a, A_s ^r, b,\psi_r),  \label{L}
\end{equation}
with
\begin{equation}
\begin{array}{lll} 
L_G &=& \frac{1}{8}\epsilon_{abcde} \left[R^{ab}R^{cd}e^e/l + \frac{2}{3}R^{ab}e^c e^d e^e/l^3 + \frac{1}{5}e^a e^b e^c e^d e^e/l^5 \right]\\
L_{su(\cal{N})} &=&-Tr\left[ {\bf A}(d{\bf A})^2+ \frac{3}{2}{\bf A}^3 d{\bf A }+ \frac{3}{5}{\bf A}^5 \right] \\
L_{u(1)} &=& \left( \frac{1}{16}-\frac{1}{\cal{N}^2} \right) b(db)^2 + \frac{3}{4l^2} \left[T^a T_a -R^{ab}e_a e_b -l^2 R^{ab}R_{ab}/2\right]b \\
&  & +\frac{3}\cal{N}F_s ^r F_r ^s b \\
L_f & = & \frac{3}{2i}\left[ \bar{\psi}^r {\cal R}\nabla \psi_r + \bar{\psi}^s {\cal F}_s ^r \nabla \psi_r \right] +c.c.
\end{array} ,  \label{Li}
\end{equation}
where $A_s ^r \equiv A^K ({\bf T}_K)_s ^r$ is the $su(\cal{N})$ connection, $F_{s}^{r}$ is its curvature, and the bosonic blocks of the supercurvature: ${\cal R}=\frac{1}{2}T^a {\bf \Gamma}_a + \frac{1}{4} (R^{ab}+\frac{e^a e^b}{l^2}){\bf \Gamma}_{ab}+\frac{i}{4}db{\bf Z}-\frac{1}{2}\psi_s \bar{\psi}^s$, ${\cal F}_s^r = F_s ^r + \frac{i}{\cal{N}}db\delta_s ^r - \frac{1}{2}\bar{\psi}^r \psi_s$. The cosmological constant is $ -l^{-2},$ and the $SU(2,2|\cal{N})$ covariant derivative $\nabla $ acting on $\psi _{r}$ is
\begin{equation}
\nabla \psi _{r}=D\psi_r +\frac{1}{2l}e^a{\bf \Gamma}_a \psi_r -A_{\,r}^s \psi_s +i\left(\frac{1}{4}-\frac{1}{\cal{N}}\right) b\psi_r. \label{delta}
\end{equation}
where $D$ is the covariant derivative in the Lorentz connection.

The above relation implies that the fermions carry a $u(1)$ ``electric'' charge given by $e=\left( \frac{1}{4}- \frac{1}{\cal{N}}\right)$. The purely
gravitational part, $L_{G}$ is equal to the standard Einstein-Hilbert action with cosmological constant, plus the dimensionally continued Euler
density\footnote{The first term in $L_{G}$ is the dimensional continuation of the Euler (or Gauss-Bonnet) density from two and four dimensions, exactly as
the three-dimensional Einstein-Hilbert lagrangian is the continuation of the the two dimensional Euler density. This is the leading term in the limit of
vanishing cosmological constant ($l\rightarrow \infty )$, whose local supersymmetric extension yields a nontrivial extension of the Poincar\'{e}
group \cite{BTrZ}.}.

The action is by construction invariant --up to a surface term-- under the local (gauge generated) supersymmetry transformations $\delta_{\Lambda}{\bf
A}=-(d{\bf \Lambda}+ [{\bf A},{\bf \Lambda}])$ with ${\bf \Lambda}=\bar{\epsilon}^r {\bf Q}_r-{\bf \bar{Q}}^r \epsilon_r$, or
\[
\begin{array}{lll}
\delta e^{a} & = & \frac{1}{2}\left( \overline{\epsilon}^r {\bf \Gamma}^a \psi_r- \bar{\psi}^r {\bf \Gamma}^a \epsilon_r \right) \\
\delta \omega^{ab} & = & -\frac{1}{4}\left(\bar{\epsilon}^r{\bf \Gamma}^{ab}\psi_r -\bar{\psi}^r{\bf \Gamma}^{ab}\epsilon_r \right) \\
\delta A_{\,s}^r & = & -i\left( \bar{\epsilon}^r \psi_s -\bar{\psi}^r \epsilon_s \right) \\
\delta \psi_r & = & -\nabla \epsilon_r \\
\delta \bar{\psi}^r & = & -\nabla \bar{\epsilon}^r \\
\delta b & = & -i\left( \bar{\epsilon}^r\psi_r-\bar{\psi}^r \epsilon_r \right) .
\end{array}
\]
As can be seen from (\ref{Li}) and (\ref{delta}), for $\cal{N}$=$4$ the $U(1)$ field $b$ looses its kinetic term and decouples from the fermions (the
gravitino becomes uncharged with respect to $U(1)$). The only remnant of the interaction with the field $b$ is a dilaton-like coupling with the Pontryagin
four forms for the AdS and $SU(\cal{N})$ groups (in the bosonic sector). As it is shown in Ref.\cite{ChTrZ}, the case $\cal{N}$=$4$ is also special at the
level of the algebra, which becomes the superalgebra $su(2,2|4)$ with a $u(1)$ central extension.

In the bosonic sector, for $\cal{N}$=$4$, the field equation obtained from the variation with respect to $b$ states that the Pontryagin four form of AdS and
$SU(\cal{N})$ groups are proportional. Consequently, if the spatial section has no boundary, the corresponding Chern numbers must be related. Since $\Pi
_{4}(SU(4))=0$, the above implies that the Pontryagin plus the Nieh-Yan number must add up to zero.

\vskip0.5cm
{\bf C. D=11}

In this case, the smallest AdS superalgebra is $osp(32|1)$ and the connection is
\begin{equation}
{\bf A}=e^a{\bf J}_a +\frac{1}{2}\omega^{ab}{\bf J}_{ab}+ \frac{1}{5!} b^{abcde} {\bf J}_{abcde}+\bar{Q}\psi , \label{D=11 connection}
\end{equation}
where $b^{abcde}$ is a totally antisymmetric fifth-rank Lorentz tensor one-form. Now, in terms of the elementary bosonic and fermionic fields, the CS
form in ${\bf L}_{2n-1}$reads
\begin{equation}
{\bf L}_{11}^{osp(32|1)}({\bf A})=L_{11}^{sp(32)}({\bf \Omega})+L_{f}({\bf \Omega },\psi ), \label{L11}
\end{equation}
where ${\bf \Omega }\equiv \frac{1}{2}(e^a {\bf \Gamma}_a+ \frac{1}{2} \omega^{ab}{\bf \Gamma}_{ab}+ \frac{1}{5!}b^{abcde}{\bf \Gamma}_{abcde})$ is an $sp(32)$ connection \cite{TrZ1,TrZ2,Troncoso}. The bosonic part of (\ref{L11}) can be written as
\[
L_{11}^{sp(32)}({\bf \Omega })=2^{-6}L_{G\;11}^{AdS}(\omega,e)-\frac{1}{2} L_{T\;11}^{AdS}(\omega ,e)+L_{11}^{b}(b,\omega ,e),
\]
where $L_{G\;11}^{AdS}$ is the CS form associated to the $12$-dimensional Euler density, and $L_{T\;11}^{AdS}$ is the CS form whose exterior derivative is the Pontryagin form for $SO(10,2)$ in $12$ dimensions. The fermionic lagrangian is
\begin{eqnarray*}
L_{f} &=&6(\bar{\psi}{\cal R}^4 D\psi )-3\left[ (D\bar{\psi}D\psi) + (\bar{\psi}{\cal R}\psi)\right] (\bar{\psi}{\cal R}^{2}D\psi ) \\
&&-3\left[ (\bar{\psi}{\cal R}^3\psi)+ (D\bar{\psi}{\cal R}^2 D\psi )\right] (\bar{\psi}D\psi )+ \\
&&2\left[ (D\bar{\psi}D\psi)^2+ (\bar{\psi}{\cal R}\psi)^2+ (\bar{\psi} {\cal R}\psi)(D\bar{\psi}D\psi )\right] (\bar{\psi}D\psi ),
\end{eqnarray*}
where ${\cal R}=d{\bf \Omega }+{\bf \Omega }^2$ is the $sp(32)$ curvature. The supersymmetry transformations (\ref{delA}) read
\[
\begin{array}{lll}
\delta e^a=\frac{1}{8}\bar{\epsilon}{\bf \Gamma}^a \psi & \hspace{1cm} & \delta \omega^{ab}=-\frac{1}{8}\bar{\epsilon}{\bf \Gamma}^{ab}\psi \\ &  & \\ \delta \psi =D\epsilon & \hspace{1cm} & \delta b^{abcde}=\frac{1}{8}\bar{\epsilon}{\bf \Gamma}^{abcde}\psi .
\end{array}
\]
Standard (CJS) eleven-dimensional supergravity \cite{CJS} is an $\cal{N}$=$1$ supersymmetric extension of Einstein-Hilbert gravity that does not admit a
cosmological constant \cite{BDHS,Deser}. An ${\cal N} >1$ extension of the CJS theory is not known. In our case, the cosmological constant is necessarily
nonzero by construction and the extension simply requires including an internal $so(\cal{N})$ gauge field coupled to the fermions. The resulting lagrangian is an $osp(32|\cal{N})$ CS form \cite{Troncoso}.

\subsection{In\"{o}n\"{u}-Wigner Contractions}

The Poincar\'{e} group is the symmetry of the spacetime that best approximates the world around us at low energy, Minkowski space. The Poincar\'{e} group can be viewed as the limit of vanishing cosmological constant or infinite radius ($\Lambda \sim \pm l^{-2}\rightarrow 0$) of the de Sitter or anti-de Sitter groups. This deformation is called a In\"{o}n\"{u}-Wigner(\textbf{IW}) contraction of the group that can be implemented in the algebra through a
rescaling the generators: ${\bf J}_a \rightarrow {\bf P}_a=l^{-1}{\bf J}_a$, ${\bf J}_{ab} \rightarrow{\bf J}_{ab}$. Thus, starting from the AdS symmetry in 3+1 dimensions ($SO(3,2)$), the rescaled algebra is
\begin{eqnarray}
 [{\bf P}_a,{\bf P}_b]&=&l^{-2}{\bf J}_{ab}\\ \nonumber
 [{\bf J}_{ab},{\bf P}_c]&=&{\bf P}_a \eta_{bc}-{\bf P}_b \eta_{ac}\\ \nonumber
 [{\bf J}_{ab},{\bf J}_{cd}]&\sim&{\bf J}_{ad} \eta_{bc}-\cdots,
\end{eqnarray}
and therefore, in the limit $l\rightarrow\infty$, ${\bf P}_a$ becomes a generator of translations. The resulting contraction is the Poincar\'{e} group,
$SO(3,2)\rightarrow ISO(3,1)$. A similar contraction takes the de Sitter group into Poincar\'{e}, or in general, $SO(p,q)\rightarrow ISO(p,q-1)$.

In general, the IW contractions change the structure constants and the Killing metric of the algebra without changing the number of generators, but the
resulting algebra is still a Lie algebra. Since some structure constants may go to zero under the contraction, some generators become commuting and end up
forming an abelian subalgebra. So, the contraction of a semisimple algebra is not necessarily semisimple, like in the above example. For a detailed
discussion of contractions, see, \textit{e.g.}, \cite{Gilmore}, and for a nice historical note see \cite{Inonu}. As could be expected, the contraction of a
group induces a contraction of representations and therefore it is possible to obtain a lagrangian for the contracted group by a corresponding limiting
procedure. However, as it was immediately noticed by its inventors, the IW contractions can give rise to unfaithful representations. In other words, the
limit representation may not be an irreducible faithful representation of the contracted group. Therefore, the procedure to obtain an action for the
contracted group is not the straightforward limit of the original action. This is particularly difficult in the case of the supersymmetric actions. In fact, the actions for the supersymmetric extensions of the Poincar\'{e} group were obtained in \cite{BTrZ}, and then the Chern-Simons actions for the SUSY extensions of AdS were found in \cite{TrZ1}, and it might be puzzling that the naive limit of the latter do not in general reproduce the former. However, this should not be surprising in the light of the previous discussion.

\subsection{Minimal Super-Poincar\'{e} Theory}

In \cite{BTrZ}, a general form of the CS lagrangian for the minimal SUSY extension of the Poincar\'{e} algebra was constructed. In $2+1$ dimensions the local symmetry group is super-Poincar\'{e}, whose algebra includes the Poincar\'{e} generators and one Majorana supercharge ${\bf Q}$. For $D=5$ the supercharge is complex (Dirac) spinor and the algebra also acquires a central extension (one generator which commutes with the rest of the algebra). In general, the algebra consists of the Poincar\'{e} generators, the supercharge ${\bf Q}$ (and its adjoint $\bar{{\bf Q}}$), and a fifth rank antisymmetric Lorentz tensor ${\bf Z}_{abcdf}$. Clearly, for $D=3, 5$ the general case reduces to the cases mentioned above. The connection is
\begin{equation}
{\bf A}=e^a{\bf P}_a+ \frac{1}{2}\omega^{ab}{\bf J}_{ab}+ \frac{1}{5!}b^{abcde} {\bf Z}_{abcde}+ \bar{{\bf Q}}\psi-\bar{\psi}{\bf Q}  ,
\label{Poinc-connection}
\end{equation}
where ${\bf P}_{a}$ and ${\bf J}_{ab}$ are the generators of the Poincar\'{e} group, $Z_{abcdf}$ is a fifth rank Lorentz tensor which commutes with ${\bf
P}_a$ and ${\bf Q}$, and
\begin{equation}
\{{\bf Q}^{\alpha},\bar{{\bf Q}}_{\beta}\}=-i(\Gamma^a)^{\alpha}_{\beta}{\bf P}_a- i(\Gamma^{abcde})^{\alpha}_{\beta}{\bf Z}_{abcde} \label{QQ}
\end{equation}
In the dimensions in which there exists a Majorana representation, $\bar{{\bf Q}}=C{\bf Q}$, the number of fermionic generators can be reduced in half.

The action has three terms,
\begin{equation}
I_{2n+1} =I_{G}+I_{b}+I_{\psi}.
\end{equation}
The first term describes locally Poincar\'{e}-invariant gravity,
\begin{equation}
I_{G}[e,\omega]=\int\epsilon_{a_{1} \cdots a_{2n+1}} R^{a_{1} a_{2}} \cdots R^{a_{2n-1} a_{2n}}e^{a_{2n+1}}. \label{PoincG2n+1}
\end{equation}
The second represents the coupling between the fifth-rank Lorentz tensor 1-form ${\bf b}^{abcde}$ and gravity,
\begin{equation}
I_{b} = -\frac{1}{6} \int R_{abc}R_{de}b^{abcde},
\end{equation}
and the third includes the fermionic 1-form (gravitino)
\begin{equation}
I_{\psi} = \frac{i}{6} \int R_{abc} [\bar{\psi}\Gamma^{abc}D\psi + D\bar{\psi}
\Gamma^{abc}\psi]. \label{PoincPsi2n+1}
\end{equation}
Here we have defined the symbol $R_{abc}$ as
\begin{equation}
R_{abc}:=\epsilon_{abca_1 \cdots a_{D-3}} R^{a_1 a_2} \cdots R^{a_{D-4} a_{D-3}}
\end{equation}
This action is invariant under Lorentz rotations,
\begin{eqnarray}
\delta \omega^{ab} &=& -D\lambda ^{ab}, \;\; \delta e^a = \lambda^a_{\;b}e^b,\;\; \delta \psi = \frac{1}{4} \lambda^{ab} \Gamma_{ab}\psi, \\ \nonumber 
\delta b^{abcde} &=& \lambda^a_{\;a'}b^{a'bcde}+\lambda^b_{\;b'}b^{ab'cde}+... +\lambda^e_{\;e'}b^{abcde'}
\end{eqnarray}
Poincar\'{e} translations,
\begin{equation}
\delta \omega^{ab} = 0, \;\; \delta e^a = -D\lambda^a, \;\; \delta \psi = 0,\;\; \delta b^{abcde}=0,
\end{equation}
and local SUSY transformations
\begin{equation}
\delta \omega^{ab} = 0, \;\;\delta e^a = \frac{i}{2} (\bar{\psi}\Gamma^a\epsilon -\bar{\epsilon}\Gamma^a\psi), \;\; \delta \psi = D\epsilon, \;\; \delta b^{abcde} = \frac{i}{2} (\bar{\psi}\Gamma^{abcde}\epsilon -\bar{\epsilon}\Gamma^{abcde}\psi).
\end{equation}
The bracket $\langle \cdots\rangle$ that gives rise to the action (\ref{PoincG2n+1}-\ref{PoincPsi2n+1}) in the general form is given by
\begin{equation}
\langle {\bf J}_{a_1a_2}\cdots{\bf J}_{a_{D-2}a_{D-1}}{\bf P}_{a_D}\rangle= \frac{2^n}{n+1}\epsilon_{a_1 a_2 \cdots a_D}
\end{equation}
\begin{equation}
\langle{\bf J}_{a_1a_2}\cdots{\bf J}_{a_{D-4}a_{D-3}}{\bf J}_{fg} {\bf Z}_{abcde}\rangle =-\epsilon_{a_1 \cdots a_{D-3} abc} \eta_{fd}\eta_{ge} \pm
(\mbox{permutations})
\end{equation}
\begin{equation}
\langle{\bf Q}{\bf J}_{a_1a_2}\cdots{\bf J}_{a_{D-4}a_{D-3}}{\bf \bar{Q}}\rangle =1i^n\Gamma_{a_1 \cdots a_{D-3}}.
\end{equation}
In the eleven-dimensional case, one can envision this algebra as resulting from an IW contraction of the $osp(32|1)$ superalgebra with connection (\ref{D=11
connection}). In fact, it seems natural to expect that in the limit $l\rightarrow \infty$, the generators ${\bf P}_a=l^{-1}{\bf J}_a$, ${\bf Z}_{abcde} =l^{-1}{\bf J}_{abcde}$, ${\bf Q'}=l^{-1/2}{\bf Q}$, satisfy the minimal Poincar\'{e} algebra with connection (\ref{Poinc-connection}) (with 
$\bar{{\bf Q}}=C{\bf Q}$). However, is not straightforward how to perform the limit in the lagrangian and some field redefinitions are needed in order to
make contact between the two theories. 

\subsection{M-Algebra Extension of the Poincar\'{e} Group}

The CS actions discussed previously have been constructed looking for the representations that extend the bosonic fields $e^a$ and $\omega^{ab}$
completing the SUSY multiplet. This is not as elegant as an algebraic construction from first principles in which one only has the input of an abstract algebra, but it has the advantage that the lagrangian is determined at once for the relevant fields. An additional difficulty of a formal approach is that it requires knowing the invariant tensors of the algebra which, in the explicit representation takes the form of the (super)trace $\langle\cdots \rangle$.

In eleven dimensions there is, besides the Poincar\'{e} SUGRA of the previous section, a new extension that corresponds to a supersymmetry algebra with more bosonic generators. The algebra includes, apart from the Poincar\'{e} generators ${\bf J}_{ab}$ and ${\bf P}_{a}$, a Majorana
supercharge ${\bf Q}_{\alpha}$ and two bosonic generators that close the supersymmetry algebra,
\begin{equation}
\{{\bf Q}_{\alpha},{\bf Q}_{\beta}\}=\left( C\Gamma^a \right) _{\alpha \beta}{\bf P}_a+ (C\Gamma^{ab})_{\alpha \beta }{\bf Z}_{ab}+ (C\Gamma^{abcde})_{\alpha \beta}{\bf Z}_{abcde}\;,  \label{MAlgebra}
\end{equation}
where the charge conjugation matrix $C$ is antisymmetric. The ``central charges'' ${\bf Z}_{ab}$ and ${\bf Z}_{abcde}$ are tensors under Lorentz
rotations but are otherwise Abelian generators. The algebra (\ref{MAlgebra}) is known as M-algebra because it is the expected gauge invariance of M Theory
\cite{HIPT}.

The M-algebra has more generators than the minimal super Poincar\'{e} algebra of the previous section, because the new ${\bf Z}_{ab}$ has no match there. Thus, the M-algebra has $\left(\begin{array}{cc}11 \\2 \end{array}\right)=55$ more generators than both the minimal super Poincar\'{e} and the $osp(32|1)$ algebras. This means, in particular, that the M-algebra cannot be found by a IW contraction from either one of these two algebras.

The action is a supersymmetric extension of the Poincar\'{e} invariant gravitational action (\ref{PoincG2n+1}) in eleven dimensions \cite{HTrZ},
\begin{eqnarray}
I_{\alpha } &=&I_G +I_{\psi }-\frac{\alpha}{6} \int_{M_{11}}R_{abc} R_{de}b^{abcde} +  \nonumber \\
& &8(1-\alpha)\int_{M_{11}}[R^2 R_{ab}-6(R^3)_{ab}]R_{cd} \left( \bar{\psi} \Gamma^{abcd}D\psi -12R^{[ab}b^{cd]}\right),  
\label{M-Action}
\end{eqnarray}
where $I_G$ and $I_{\psi}$ are
\begin{eqnarray}
I_{G}[e,\omega ]&=&\int_{M_{11}}\epsilon _{a_{1}\cdots a_{11}}R^{a_{1}a_{2}}\cdots R^{a_{9}a_{10}}e^{a_{11}},  \label{I_G} \\
I_{\psi} &=&-\frac{1}{3}\int_{M_{11}}R_{abc}\bar{\psi}\Gamma ^{abc}D\psi , \label{I-Psi}
\end{eqnarray}
and $R^{2}:=R^{ab}R_{ba}$ and $(R^{3})^{ab}:=R^{ac}R_{cd}R^{db}$. Here $\alpha$ is an arbitrary dimensionless constant whose meaning will be discussed below. The action is invariant under local supersymmetry transformations obtained from a gauge transformation of the M-connection (\ref{M-connection}) with parameter $\lambda =\epsilon^{\alpha }{\bf Q}_{\alpha}$,
\begin{equation}
\begin{array}{ll}
\delta_{\varepsilon}e^a=\bar{\epsilon}\Gamma^a \psi, & \delta_{\varepsilon}\psi =D\epsilon, \; \;\; \delta_{\varepsilon}\omega^{ab}=0,\\
\delta_{\varepsilon}b^{ab}=\bar{\epsilon}\Gamma^{ab}\psi, & \delta_{\varepsilon}b^{abcde}=\bar{\epsilon}\Gamma^{abcde}\psi .
\end{array} \label{susytransf}
\end{equation}
The field content is given by the components of the M-algebra connection,
\begin{equation}
{\bf A}=\frac{1}{2}\omega^{ab}{\bf J}_{ab}+e^a {\bf P}_a+ \psi^{\alpha} {\bf Q}_{\alpha}+ b^{ab}{\bf Z}_{ab}+b^{abcde}{\bf Z}_{abcde}\;,
\label{M-connection}
\end{equation}
and the action can be written in terms of the bracket $\langle \cdots\rangle$, whose only nonzero entries are
\[
\begin{array}{l}
\left\langle J_{a_1 a_2},\cdots ,J_{a_9 a_{10}},P_{a_{11}}\right\rangle = \frac{16}{3}\epsilon_{a_1 \cdots a_{11}}\;, \\
\left\langle J_{a_1 a_2},\cdots ,J_{a_9 a_{10}},Z_{abcde}\right\rangle =-\alpha \frac{4}{9}\epsilon_{a_1\cdots a_8 abc}\eta_{[a_9 a_{10}][de]}\;, \\
\left\langle J_{a_1 a_2}, J_{a_3 a_4}, J_{a_5 a_6}, J^{a_7 a_8}, J^{a_9 a_{10}},Z^{ab} \right\rangle =(1-\alpha )\frac{16}{3}\left[ \delta _{a_1\cdots a_6} ^{a_7\cdots a_{10}ab} - \delta _{a_1 \cdots a_4}^{a_9 a_{10}ab}\delta_{a_5 a_6}^{a_7 a_8}\right]  \\
\left\langle Q,J_{a_1 a_2}, J^{a_3 a_4}, J^{a_5 a_6}, J^{a_7 a_8},Q\right\rangle =\frac{32}{15}\left[ C\Gamma_{a_1 a_2}^{\;\;\quad
\;\,a_3 \cdots a_8}+ \right.  \\
\;\;\;\;\;\;\;\;\;\;\;\;\;\;\;\;\;\;\;\;\;\;\;\;\;\;\;\;\;\;\;\;\;\;\;\;\;\;\;\;\;
\qquad \;\;\left. (1-\alpha )\left(3 \delta_{a_1 a_2 ab}^{a_3 \cdots a_6}C\Gamma^{a_7 a_8 ab}+2C\Gamma^{a_3 \cdots a_6} \delta_{a_1 a_2}^{a_7 a_8} \right) \right] , \end{array}
\]
where (anti-)symmetrization under permutations of each pair of generators is understood when all the indices are lowered.

It is natural to ask whether there is a link between this theory and the vanishing cosmological constant limit of the CS theory constructed for the $osp(32|1)$ algebra. As already mentioned, these theories cannot be related through a  IW contraction, because contractions do not increase the number of generators in the algebra. In fact, the M-algebra can be obtained from the AdS algebra by a more general singular transformation called a \textit{expansion} \cite{EHTZ1}. These deformations are analytic mappings in the algebra with the only restriction that they respect the Maurer-Cartan structure equations. They have been studied in Refs. \cite{Hatsuda-Sakaguchi}, and more recently in \cite{AIPV}.

The fact that this theory contains the free parameter $\alpha$ means that there is more than one way of deforming $osp(32|1)$ which produces an action
supersymmetric under the transformations (\ref{susytransf}). For $\alpha=1$ (\ref{M-Action}) reduces to the minimal Poincar\'{e} action (\ref{PoincG2n+1}-\ref{PoincPsi2n+1}). This means that the combination $I_{\alpha=0}-I_{\alpha=1}$ is supersymmetric by itself (although it does not
describe a gravitational theory as it does not involve the vielbein).

\subsection{Field Equations}

The existence of the bracket $\langle \cdots\rangle$ allows writing the field equations for an CS theory in a manifestly covariant form as
\begin{equation}
\left\langle {\bf F}^{n}{\bf G}_{A}\right\rangle =0,  \label{FieldEqs}
\end{equation}
where ${\bf G}_{A}$ are the gauge generators. In addition, if the ($2n+1$)-dimensional spacetime is conceived as the boundary of a ($2n+2$)-dimensional manifold, $\partial \Omega_{2n+2}=M_{2n+1}$, the CS action can also be\ written as $I=\int_{\Omega_{2n+2}}\left\langle F^{n+1}\right\rangle$. This now describes a topological theory in $2n+2$ dimensions. In spite of its topological origin, the action does possess propagating degrees of freedom in the ($2n+1$)-dimensional spacetime and hence it is not a topological field theory in the lower dimension.

The field equations are nonlinear for $D\geq5$ and posses a rich dynamical structure due to the possibility of having zeros of different orders in
(\ref{FieldEqs}). The perturbative field theories that can be constructed around each of these classical configurations have different particle content
and correspond to different disconnected phases of the theory \cite{STZ}. In particular, these phases can be seen as different vacua corresponding to
different dimensional reductions of the starting theory \cite{HTrZ}.

\subsection{Overview}

Let us recap what we have found so far. The supergravities presented here have two distinctive features: The fundamental field is always the connection ${\bf A}$ and, in their simplest form, they are pure CS systems (matter couplings are briefly discussed below). As a result, these theories possess a larger gravitational sector, including propagating spin connection. Contrary to what one could expect, the geometrical interpretation is quite clear, the field structure is simple and, in contrast with the standard cases, the supersymmetry transformations close off shell without auxiliary fields. The price to pay is to have a complex classical dynamics and a richer perturbative spectrum that changes from one background to another.
\vskip0.5cm

{\bf Field content and extensions with} ${\cal N}>1$

The field contents of AdS-CS supergravities and the standard supergravities in $D=5,7,11$ are compared in the following table:
\begin{center}
\textbf{Table 5}\\
\vspace{0.2cm}
$\begin{array}{|c|c|c|c|} \hline D & {\rm Standard ~supergravity} & {\rm CS ~supergravity} & {\rm Algebra}\\ \hline 
5 & e_{\mu}^a \;\psi_{\mu}^{\alpha}\;\bar{\psi}_{\alpha \mu} & e_{\mu}^a \;\omega_{\mu}^{ab}\;A_{\mu}\;a_{\mu j}^i\;
\psi_{i\mu}^{\alpha} \;\bar{\psi}_{\alpha \mu}^i, \;i,j=1,..., \cal{N} & usp(2,2|\cal{N})\\ \hline 
7 & e_{\mu}^a\;A_{[3]}\;a_{\mu j}^i\;\lambda^{\alpha}\;\phi \;\psi_{\mu}^{\alpha i} & e_{\mu}^a\;\omega_{\mu}^{ab}\;a_{\mu j}^i \;\psi_{\mu}^{\alpha i},\; i,j=1,..., {\cal N}=2n & osp({\cal N}|8) \\ \hline 
11 & e_{\mu}^a \;A_{[3]}\;\psi_{\mu}^{\alpha} & e_{\mu}^a \;\omega_{\mu}^{ab}\; b_{\mu}^{abcde}\;\psi_{\mu}^{\alpha}\;,\;i,j =1,..., \cal{N}\; & osp(32|\cal{N}) \\ \hline
\end{array}$
\end{center}

Standard supergravity in five dimensions is dramatically different from the theory presented here: apart from the graviton ($e^a$) and the complex gravitino ($\psi_{\mu}$), in the AdS theory there is a propagating spin connection and at least a $U(1)$ gauge field ($A_{\mu}$), which have no match in
standard $D=5$ SUGRA. 

The standard $D=7$ supergravity is an ${\cal N}=2$ theory (its maximal extension is ${\cal N}=4$), whose gravitational sector is given by E-H gravity with cosmological constant and with a background invariant under $OSp(2|8)$ \cite{D=7,Salam-Sezgin83}. Standard eleven-dimensional supergravity is an ${\cal N}=1$ supersymmetric extension of Einstein-Hilbert gravity with vanishing cosmological constant \cite{CJS}. An ${\cal N}>1$ extension of this theory is not known.

In our construction, the extensions to larger ${\cal N}$ are straightforward in any dimension. In $D=7$, the index $i$ is allowed to run from $2$ to $2s$, and the lagrangian is a CS form for $osp(2s|8)$. In $D=11$, one must include an internal $so({\cal N})$ field and the lagrangian is an $osp(32|{\cal N})$ CS form \cite{TrZ1,TrZ2}. The cosmological constant is necessarily nonzero in all cases.
\vskip0.5cm

{\bf Gravity sector.}

A most remarkable result from imposing the supersymmetric extension, is the fact that if one sets all fields, except those that describe the geometry
--$e^{a}$ and $\omega ^{ab}$-- to zero, the remaining action has no free parameters. This means that the gravity sector is uniquely fixed, which is
interesting because, as we saw at least for $D=3$ and $D=7$, there are several CS actions that one can construct for the AdS gauge group. The Euler-CS form, and the so-called exotic ones that include torsion explicitly, can occur in the action with arbitrary coefficients. These coefficients are fixed in the supersymmetric theory. So, even for the purely gravitational action, the theory that does admit a supersymmetric extension is more predictive than those which do not.
\vskip0.5cm

{\bf Relation with standard (super) gravity}

In all these CS theories $\omega^{ab}$ is an independent dynamical field, something that is conspicuously absent in standard SUGRA. The spin connection
can be frozen out by imposing the torsion condition (the field equation obtained varying the action with respect to $\omega^{ab}$, which determines
$T^a$). In a generic --and sufficiently simple background--, this is an algebraic equation for $\omega^{ab}$. As we mentioned in Section 5.2,
substituting the solution $\omega^{ab}=\bar{\omega}^{ab}(e,\cdots)$ gives a classically equivalent action principle in the reduced phase space.

Standard SUGRAs have a gravity sector described by the Einstein-Hilbert action. However, the simple limit $l\rightarrow \infty$ does not turn the Lovelock action to the EH action. On the contrary, this limit yields a lagrangian which has the highest possible power of curvature ($(D-1)/2$), analogous to (\ref{odd-D-Poinc}). Perhaps General Relativity could be recovered if one were to look at the action around a special configuration where the fields behave in a way that resembles the linearized excitations in EH gravity. 

It is sometimes argued that in the limit when the curvature is not very large (small radius of curvature), the terms in the lagrangian with the lowest powers of the curvature should be more dominant. Then, pressumably, the action would be approximated by the EH term and the dynamics would be approximately described by standard GR. However this is only superficially true, because flat space is not an approximate solution of a generic Lovelock action. In the case of an AdS-CS theory for example, flat space is quite far from a classical solution, so the claim that ``around flat space'' the theory is aproximately described by GR is completely irrelevant.

Certainly, additional field identifications should be made, some of them quite natural, like the identification of the totally antisymmetric part of $\omega_{\mu}^{ab}$, $k_{\mu \nu \lambda}$ (known as the contorsion tensor), with an abelian 3-form, $A_{[3]}$ in a certain coordinate basis. (In 11 dimensions, one could also identify the totally antisymmetrized part of $A_{\mu }^{abcde}$ with an abelian 6-form $A_{[6]}$, whose exterior derivative, $dA_{[6]}$, is the dual of $F_{[4]}=dA_{[3]}$. Hence, in $D=11$ the CS theory may contain the standard supergravity as well as some kind of dual version of it.)

In trying to make contact with standard SUGRA, the gauge invariance of the CS theory in its original form would be completely entangled by the elimination of the spin connection. Moreover, the identification between the tangent space and the base manifold prduced by the elimination of the vielbein in favor of the metric tensor destroys any hope to interpret the resulting action as a gauge theory with fiber bundle structure, but this is more or less the situation in
standard SUGRAs where a part of the gauge invariance is replaced by invariance under diffeomorphisms.

Thus, the relation between the AdS-CS and standard SUGRAs is at best indirect and possibly only in one sector of the theory. One may ask, therefore, why is it necessary to show the existence of a connection between these two theories? The reason is historical (standard SUGRAs where here first) and of consistency (standard SUGRAs are known to be rather unique). So, even if it is a difficult and possibly unnecessary exercise, it would be interesting to show that the connection exists \cite{EHTZ1}.
\newpage

{\bf Classical solutions}

The field equations for these theories, in terms of the Lorentz components ($\omega $, $e$, $A$, ${\bf A}$, $\psi $), are the different tensor components of $<{\bf F}^n{\bf G}_M >=0$. These equations have a very complex space of solutions, with different sectors of radically different dynamical behaviors.

It can be easily verified that in all these theories, anti-de Sitter space with $\psi =b={\bf A}=0$, is a classical solution. This is the most symmetric vacuum: all curvature components vanish and has the maximal set of Killing vectors; moreover, since it is invariant under supersymmetry, it also has a maximal set of Killing spinors. This BPS state cannot decay into anything and what is most intriguing, it has no perturbations around it. In this sense, this does not correspond to the quantum vacuum of a dynamical theory since it cannot be populated with excitations like the vacuum in a quantum field theory. This ``extreme vacuum" is a single topological state and the action around it is effectively a surface term, so it describes a field theory at the boundary: the action in this sector describes a conformal field theory at the boundary with the same gauge SUSY as the theory in the bulk.

There exist other less symmetric classical solutions which do allow perturbations around them, and these are more interesting states to look at. For instance, in the pure gravity (matter-free) sector, there exist spherically symmetric, asymptotically AdS standard black holes\cite{JJG}, as well as topological black holes \cite{ABHPB}. The extremal form of these black holes can be shown to be BPS states \cite{AMTrZ}.
\vskip0.5cm

{\bf Spectrum}

It may be impossible to establish a complete classification of all classical solutions of the CS equations. The quantization of these systems is also an open problem at the moment, the main obstacle being the complex vacuum structure and the lack of a perturbative expansion around many of them.

Some simple classical solutions are product spaces, where one of the factors is a maximally symmetric or constant curvature space. A recurrent feature is
that when one of the factors has vanishing AdS curvature, the other factor has indeterminate local geometry. This is because the field equations are typically
a product of curvature two-forms equal to zero; therefore, if one factor vanishes, the others are not determined by the field equations. The indeterminate components correspond to the existence of gauge degrees of freedom, and it can be seen that the theory has actually fewer propagating degrees of freedom around these configurations. A dramatic example of this is found in the CS supergravity for the M-Theory algebra, where a configuration was found in the 11-dimensional theory that corresponds to the spectrum of 4-dimensional de Sitter space, plus matter couplings \cite{HTrZ}; the geometric features that describe the remaining 7 dimensional quotient being largely arbitrary.

The stability and positivity of the energy for the solutions of these theories is another highly nontrivial problem. As shown in Ref. \cite{BGH}, the number of
degrees of freedom of bosonic CS systems for $D\geq 5$ is not constant throughout phase space and different regions can have radically different
dynamical content. However, in a region where the rank of the symplectic form is maximal the theory may behave as a normal gauge system, and this condition would be stable under perturbations. As shown in \cite{ChTrZ} for $D=5$, there exists a central extension of the AdS superalgebra in anti-de Sitter space with a nontrivial $U(1)$ connection, but no other matter fields. In this background the symplectic form has maximal rank and the gauge superalgebra is realized in the Dirac brackets. This fact ensures a lower bound for the mass as a function of the other bosonic charges \cite{GH}. Moreover, it was shown in \cite{MTrZ} that if a nonabelian $SU(5)$ flux is switched on, this configuration can be a BPS state.
\vskip0.5cm

{\bf Representations}

The mismatch between fermionic and bosonic states is most puzzling for those accustomed to standard supersymmetry, where the Fermi-Bose matching is such a central feature that has been used as synonym of SUSY. In fact, it is common to read that the signal expected to emerge from accelerator experiments at very high energy (above the supposed SUSY-breaking scale) is a pairing of states with equal mass, electric charge, parity, lepton or baryon number, etc., but with different fermion number. In the theories described here, no such signal should be expected, which may be a relief after so many years of fruitless
search in this direction.

Another technical aspect related to the representation used in CS SUGRAs, is the avoidance of Fierz rearrangements (\textbf{FR}) which are a source of much suffering in standard SUSY and SUGRA. The FR are needed to express bilinear products of spinors in terms of all possible products of Dirac matrices. Since in CS theories one only deals with exterior (wedge) products, only antisymmetrized products of Dirac matrices appear in the algebra. Also, exterior products of spinors always give irreducible representations, so it is not necessary to decompose these products in smaller irreps, as it happens in standard SUSY.
\vskip0.5cm

{\bf  Couplings to matter sources}

It is possible to introduce minimal couplings to matter of the form ${\bf A}\cdot {\bf J}^{ext}$. For instance, the CS supergravity theory in five dimensions,  (\ref{L}), can couple to an electrically  charged  0--brane ($U(1)$ point charge) through the standard term $j^{\mu}A_{\mu}$; or to $SU(4)$--colored 0--branes (quarks), through $ {\bf J}^{\mu}_{rs}{\bf A}_{\mu}^{rs}$, as proposed in \cite{Wong,Balachandran, Leite:1996ff}. 

On the other hand, CS theories themselves can be viewed as a form of coupling a brane to a connection. For instance, the electromagnetic coupling between a point charge (0-brane) and the $U(1)$ connection is an integral over a 1-dimensional manifold of the CS 1-form $A$, that is a CSaction in 0+1 dmensions. Similarly, the coupling between an electrically charged 2-brane and a 2+1 CS form is just the corresponding 2+1 CS action \cite{Z-AdS-CFT10}. Thus, in general, a $2n+1$ CS theory can also be seen as the coupling between a connection and a $2n$-brane. This idea is going to be presented elsewhere \cite{Miskovic-Zanelli,Edelestein-Zanelli}. In all these cases, the gauge symmetry may be reduced to that of the subalgebra whose invariant tensor defines the CS form in the $2n+1$-worldvolume of the brane.

An alternative form of coupling for a CS theory is afforded by the so-called topologically massive gauge theories, where the CS action is complemented by a standard Yang-Mills type action \cite{Deser-Jackiw-Templeton}. These theories acquire massive excitations because the coupling involves necessarily a dimensionful constant. In spite of that, the gauge invariance of the theory is not destroyed, as would be the case if a standard mass term $m^2 |A|^2$ had been included. This idea has been also extended to supergravity in \cite{Deser:1982sw}.

The lesson one can draw from these experiments is that the possibility of including massive sources in CS theories does not require very exotic structures. These form of coupling respect gauge invariance (under a reduced group perhaps).
\vskip0.5cm

{\bf Dynamical Content}

The physical meaning of a theory is defined by the dynamics it displays both at the classical and quantum levels. In order to understand the dynamical contents of the classical theory, the physical degrees of freedom and their evolution equations must be identified. In particular, it should be possible --at least
in principle-- to separate the propagating modes from the gauge degrees of freedom, and from those which do not evolve independently at all (second class
constraints). The standard way to do this is Dirac's constrained Hamiltonian analysis and has been applied to CS systems in \cite{BGH}. In the Appendix,
that analysis is summarized. It is however, fair to say that a number of open problems remain and it is an area of research which is at a very different stage
of development compared with the previous discussion. 

\newpage
\section{Final Comments}
\vskip 0.01cm
\
{\bf 1.} Everything we know about the gravitational interaction at the classical level is described by Einstein's theory in four dimensions, which in turn is supported by a handful of experimental observations. There are many indications, however, that make it plausible to accept that our spacetime has more dimensions than those that meet the eye. In a spacetime of more than four dimensions, it is not logically necessary to consider the Einstein-Hilbert action as the best description for gravity. In fact, string theory suggests a Lovelock type action as more natural option \cite{Zwiebach}. The large number of free parameters in the Lovelock action, however, cannot be fixed by arguments from string theory. As we have shown, only in odd dimensions there is a simple symmetry principle to fix these coefficients is, and that leads to the Chern-Simons theories.

\vskip 0.5cm {\bf 2.} The CS theories have profound geometrical roots connecting them to topological invariants --the Euler and the Chern or Pontryagin classes. In the context of gravity, they appear naturally in a framework where the affine and metric structures of the geometry are taken to be independent dynamical objects. If one demands furthermore the theory to admit supersymmetry, there is, in each dimensions essentially a unique extension which completely fixes the gravitational sector, including the precise role of torsion in the action.

\vskip 0.5cm {\bf 3.} The CS theories of gravity discussed in the first part of these notes possess nontrivial black hole solutions \cite{BTZ94} which asymptotically approach spacetimes of constant negative curvature (AdS spacetimes). These solutions have a thermodynamical behavior which is unique among all possible black holes in competing Lovelock theories with the same asymptotics \cite{BHscan}. The specific heat of these black holes is positive and therefore they can always reach thermal equilibrium with their surroundings and hence, are stable against thermal fluctuations. These theories also admit solutions which represent black objects of other topologies, whose singularity is shrouded by horizons of non spherical topology \cite{TopBH}. Furthermore, these solutions seem to have well defined, quantum mechanically stable, BPS ground states \cite{AMTrZ}.

\vskip 0.5cm {\bf 4.} The higher-dimensional Chern-Simons systems remain somewhat mysterious, especially in view of the difficulties to treat them as quantum theories. However, they have many ingredients that make them likely quantum systems: They have no dimensionful couplings, the only free parameter in the classical action turns out to be quantized.  Efforts to quantize CS systems seem promising at least in the cases in which the space admits a complex structure so that the symplectic form is a K\"{a}hler form \cite{NS}.

\vskip 0.5cm {\bf 5.} It may be too soon to tell whether string theory is the correct description of all interactions and constituents of nature. If it is so, and gravity is just a low energy effective theory, there would be a compelling reason to study gravity in higher dimensions, not just as an academic exercise, but as a tool to study big bang cosmology or black hole physics. The truth is that a field theory can tell us a lot a bout the low energy phenomenology, in the same way that ordinary quantum mechanics tells us a lot about atomic physics even if we know that is all somehow contained in QED.

\vskip 0.5cm {\bf 6.} If the string scenario fails to deliver its promise, more work will be needed to understand the field theories it is supposed to represent, in order to decipher their deeper interrelations. In any case, geometry is likely to be an important clue, very much in the same way that it
is an essential element in Yang Mills and Einstein's theory. One can see the construction discussed in these lectures as a walking tour in this direction.

\vskip 0.5cm {\bf 7.} We can summarize the general features of CS theories in the following (incomplete) list:

\vskip 0.2cm $\bullet$ Truly gauge invariant theories. Their dynamical fields are connection 1-forms, their symmetry algebra closes off-shell, they have no
dimensionful coupling constants.

\vskip 0.2cm $\bullet$ Gravity is naturally included . They are fully covariant under general coordinate transformations. The vielbein and the spin connection
can be combined into a connection for the (A)dS group or the Poincar\'e group.

\vskip 0.2cm $\bullet$ No derivatives beyond second order. As the formulation is in terms of exterior forms (without the Hodge *-dual) the lagrangian has at most first derivatives of the fields, and the field equations are also first order. If the torsion condition is used to eliminate the spin connection, the field equations become at most second order for the metric.

\vskip 0.2cm $\bullet$ No spins higher than 2. All component fields in the connection carry only one spacetime index (they are 1-forms), and they are antisymmetric tensors of arbitrary rank under the Lorentz group ($\sim b_{\mu}^{ab\cdots}$). Thus, they belong to representations of the rotation
group whose Young tableaux are of the form: \\
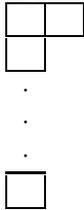
\begin{tabular}{|c|c|}
\hline  $\;$ &$\;$ \\
\hline \end{tabular}\\
 \begin{tabular}{|c|}
  $\;$  \\
  \hline
\end{tabular} \\
\begin{tabular}{c}
  $\cdot$  \\
\end{tabular} \\
\begin{tabular}{c}
  $\cdot$  \\
\end{tabular} \\
\begin{tabular}{c}
  $\cdot$  \\
  \hline
\end{tabular} \\
\begin{tabular}{|c|}
  $\;$  \\
  \hline
\end{tabular} \\
Thus, at most one second rank symmetric tensor (spin 2) can be constructed with these fields contracting the spacetime index with one of the Lorentz
indices, using the vielbein: $e_{a\mu}b^{ab\cdots}_{\nu}+e_{a\nu}b^{ab\cdots}_{\mu}=:2b^{b\cdots}_{\mu \nu}$. Therefore no fundamental fields of spin higher than two can be represented in these theories.

\vskip 0.2cm $\bullet$ No matching between fermions and bosons. Since the connection is not in the fundamental (vector) representation, but in the adjoint, and the spacetime transformations do not commute with the spinor generators, no matching should be expected between fermionic and bosonic states. This shows that it is perfectly possible to have supersymmetry and yet have no supersymmetric partners for each known particle (sleptons, squarks, gluinos,
etc.).

\vskip 0.2cm $\bullet$ Degenerate classical dynamics for $D\geq 5$ and trivial dynamics for $D\leq3$. In general for $D=2n+1$, CS theories have field equations which are polynomials of degree $n$ in curvatures. For $D\geq 5$, this gives rise to degeneracies in the symplectic structure with a corresponding breakdown in the dynamical evolution of the initial data. This produces a splitting of the phase space into several phases with different degrees of freedom within the same theory. For $D=3$, instead, the local dynamics is trivial in the sense that there are no propagating degrees of freedom. (This doesn't necessarily mean that the $D=3$ is completely uninteresting.)

\vskip 0.2cm $\bullet$ Dimensional reduction singles out $D=4$ uncompactified dimensions. Starting in any dimension $D\leq5$, CS gravity has locally propagating perturbations (gravitons) in a product spacetime in which one of the factors is four-dimensional $D=4$. This may have something to do with the observation that we live in a three-dimensional space.

\vskip 0.1cm
\noindent
------------------------------\\
Although many open questions remain to be addressed before CS theories can pretend to describe the microscopic world, hopefully the reader has been convinced that these are beautiful mathematical structures describing rich physical systems, and as such, worthy of further study. It would be really a shame if God didn't take advantage of so many interesting features somewhere in its creation.

\vskip 1.0cm

{\bf Acknowledgments}%

It is a pleasure for me to thank my collaborators, M. Hassa\"{\i}ne and R. Troncoso, who have taught me a lot about geometry, Chern-Simons theories and supergravity in all these years. Many enlightening discussions over several years with L. Alvarez-Gaum\'{e}, C. Bunster, L. Castellani, A. Dabholkar, S. Deser,  J. Edelstein, A. Gomberoff, M. Henneaux, J. Maldacena, C. Mart\'{\i}nez, O. Mi\v{s}kovi\'{c}, P. Mora, S. Mukhi, C. N\'{u}\~{n}ez, R. Olea, S. Paycha, V. Rivelles, A. Sen, C. Teitelboim, S. Theisen, F. Toppan, P. Townsend, and B. Zwiebach, have made me understand several subtle issues. Thanks are also extended to A. Anabal\'{o}n and M. Hassaine, who worked through the manuscript and corrected many misprints and suggested changes to improve the notes. This work was supported in part by FONDECYT grants 1020629, and 7020629. CECS is funded by the Chilean Government through the Millennium Science Initiative and the Centers of Excellence Base Funding Program of Conicyt. CECS is also supported by a group of private companies which includes Antofagasta Minerals, Arauco, Empresas CMPC, Indura, Naviera Ultragas and Telef\'onica del Sur.

\appendix
\section{Appendix: General dynamics of CS theories}

Chern-Simons theories are exceptions to almost any feature of a standard field theory. This is because they are extremely singular dynamical systems: \\

$\bullet$ They are \textbf{constrained systems} like any gauge theory and therefore they have a non invertible Legendre transform $p_i=p_i(q,\dot{q})$. This reflects the usual feature that some of the coordinates are non propagating gauge degrees of freedom and the corresponding constraints are the generators of gauge transformations.\\

$\bullet$ They have a \textbf{degenerate} symplectic matrix whose rank is not constant throughout phase space. This means that the separation between coordinates and momenta cannot be made uniformly throughout phase space, and there are regions where the number of degrees of freedom of the theory changes abruptly: \textbf{degenerate surfaces}. \\

$\bullet$ Their constraints are \textbf{irregular} in the sense that the number of functionally independent constraints is not constant throughout phase space. This issue is independent from degeneracy, although not totally independent from it. The \textbf{irregularity surfaces} in phase space is the locus where the constraints fail to be independent define systems with dynamically different behavior.

The standard counting of degrees of freedom for a general theory with first- and second- class constraints is a well known problem. See, e.g., \cite{HTZ} and references therein. For the reasons outlined above, this problem becomes considerably harder for CS theories. In the remaining of this appendix, we examine the Hamiltonian structure of CS systems in order to address this issue.

\subsection{Hamiltonian Analysis}

From a dynamical point of view, CS systems are described by lagrangians of the form\footnote{ Note that in this section, for notational simplicity, we assume the spacetime to be ($2n+1$)-dimensional.}
\begin{equation}
L_{2n+1}=l_{a}^i(A_j^b)\dot{A}_i^a -A_0^a K_a,
\end{equation}
where a dot ( $\dot{}$ ) represents a time derivative, and
\[
K_a=-\frac{1}{2^n n}\gamma_{a a_1....a_n}\epsilon^{i_1...i_{2n}}F_{i_1 i_2}^{a_1}\cdots F_{i_{2n-1}i_{2n}}^{a_n}.
\]
Splitting spacetime into a $2n$-dimensional space plus time, the field equations read
\begin{eqnarray}
\Omega _{ab}^{ij}(\dot{A}_j^b -D_j A_0^b) &=&0,  \label{csEq} \\
K_a &=&0,  \label{K=0}
\end{eqnarray}
where
\begin{eqnarray}
\Omega_{ab}^{ij} &=&\frac{\delta l_b^j}{\delta A_i^a}-\frac{\delta l_a^i}{\delta A_j^b}  \label{symplectic} \\
&=&-\frac{1}{2^{n-1}}\gamma _{aba_2....a_n}\epsilon^{iji_3...i_{2n}}F_{i_3 i_4}^{a_2}\cdots F_{i_{2n-1}i_{2n}}^{a_n}  \nonumber
\end{eqnarray}
is the {\bf symplectic form}. The passage to the Hamiltonian has the problem that the velocities appear linearly in the lagrangian and therefore there are a number of primary constraints
\begin{equation}
\phi _{a}^{i}\equiv p_{a}^{i}-l_{a}^{i}\approx 0.  \label{Phi's}
\end{equation}
Besides, there are secondary constraints $K_{a}\approx 0$, which can be combined with the $\phi$s into the expressions
\begin{equation}
G_{a}\equiv -K_{a}+D_{i}\phi_{a}^{i}.  \label{Ga}
\end{equation}
The complete set of constraints forms a closed Poison bracket algebra,
\[
\begin{array}{ll}
\{\phi _{a}^{i},\phi _{b}^{j}\} & =\Omega _{ab}^{ij} \\
\{\phi _{a}^{i},G_{b}\} & =f_{ab}^{c}\phi _{c}^{i} \\
\{G_{a},G_{b}\} & =f_{ab}^{c}G_{c}
\end{array},
\]
where $f_{ab}^c$ are the structure constants of the gauge algebra of the theory. Clearly the $G$s form a first class algebra which reflects the gauge invariance of the theory, while some of the $\phi $s are second class and some are first class, depending on the rank of the symplectic form $\Omega$.

\subsection{Degeneracy}

An intriguing aspect of Chern-Simons theories, not present in most classical systems is the fact that the symplectic form is not a constant matrix throughout phase space, but a function of the fields. Consequently, the rank of the symplectic form need not be constant either. In fact, it can be smaller in regions of phase space where the determinant of the matrix $\Omega_{ab}^{ij}$ develops zeros of different orders. These regions in phase space with different levels of degeneracy have different degrees of freedom. If the system reaches a degenerate configuration, some degrees of freedom become frozen in an irreversible process erasing the information about the initial conditions of the degrees of freedom that are lost.

This can also be seen from the field equations, which for $D=2n+1$, are polynomials of degree $n$. As mentioned aerlier, if some components of the curvature vanish at some point, the remaining curvature factors can take arbitrary values. This corresponds to having some degrees of freedom reduced to mere gauge directions in phase space, and therefore these configurations possess fewer dynamical degrees of freedom.

It can be shown in the context of some simplified mechanical models that the degeneracy of a system generically occur on lower-dimensional submanifolds of phase space. These regions define sets of unstable initial states or sets of stable end-points for the evolution \cite{STZ}. As it was shown in this reference, if the system evolves along an orbit that reaches a surface of degeneracy, $\Sigma$, it becomes trapped by the surface and loses the degrees o freedom that correspond to displacements away from $\Sigma$. This is an irreversible process which has been observed in the dynamics of vortices described by Burgers' equation. This equation has a symplectic form which degenerates when two vortices coalesce, an experimentally observed irreversible process.

Thus, the space of classical solutions has a rich structure, describing very distinct dynamical systems in different regions of phase space, all governed by the same action principle. As shown in \cite{HTrZ}, this phenomenon may be a way to produce dynamical dimensional reduction.

\subsection{Counting degrees of freedom}

There is a second problem and that is how to separate the first and second class constraints among the $\phi $s. In Ref.\cite{BGH} the following results are shown:
\begin{itemize}
\item The maximal rank of $\Omega_{ab}^{ij}$ is $2n(N-1)$ , where $N$ is the number of generators in the gauge Lie algebra.

\item There are $2n$ first class constraints among the $\phi$s which correspond to the generators of spatial diffeomorphisms (${\cal H}_{i}$).

\item  The generator of timelike reparametrizations ${\cal H}_{\perp }$ is not an independent first class constraint.
\end{itemize}

Putting all these facts together one concludes that, in a generic configuration (non degenerate symplectic form), the number of degrees of freedom of the theory is
\begin{eqnarray}
\Delta^{CS} &=&({\rm number} \:{\rm of} \: {\rm coordinates}) - ({\rm number}\: {\rm of}\: {\rm 1st}\: {\rm class}\: {\rm constraints}) \nonumber \\  & &-\frac{1}{2}({\rm number}\: {\rm  of}\: {\rm  2nd}\: {\rm class}\: {\rm constraints}) \nonumber \\
 &=&2nN-(N+2n)-\frac{1}{2}(2nN-2n) \label{nN-N-n} \\
 &=&nN-N-n.
\end{eqnarray}

This result is somewhat perplexing. A standard metric (Lovelock) theory of gravity in $D=2n+1$ dimensions, has \begin{eqnarray*}
\Delta^{Lovelock} &=&D(D-3)/2 \\
&=&(2n+1)(n-1)
\end{eqnarray*}
propagating degrees of freedom \cite{Te-Z}. For $D=2n+1$, a CS gravity system for the AdS group has $N=D(D+1)/2=(2n+1)(n+1)$, and therefore,
\begin{equation}
\Delta^{CS}=2n^{3}+n^{2}-3n-1.  \label{n3}
\end{equation}

In particular, for $D=5$, $\Delta^{CS}=13$, while $\Delta^{Lovelock}=5$. The extra degrees of freedom correspond to propagating modes in $\omega^{ab}_{\mu}$, which in the CS theory are independent from the metric ones, contained in $e^{a}_{\mu}$.

In \cite{BGH} it was also shown that an important simplification occurs when the group has an invariant abelian factor. In that case, the symplectic matrix $\Omega_{ab}^{ij}$ takes a partially block-diagonal form where the kernel has the maximal size allowed by a generic configuration. It is a nice surprise in the cases of CS supergravities discussed above that for certain unique choices of $\cal{N}$, the algebras develop an abelian subalgebra and make the separation of first and second class constraints possible (e.g., $\cal{N}=4$ for $D=5$, and $\cal{N}=32$ for $d=11$). In some cases the algebra is not a direct sum but an algebra with an abelian central extension ($D=5$). In other cases, the algebra is a direct sum, but the abelian subgroup is not put in by hand but it is a subset of the generators that decouple from the rest of the algebra ($D=11$).

\subsection{Irregularity}

Applying the counting of \cite{BGH} to five-dimensional CS supergravity gave the paradoxical result that the linearized theory around a generic (nondegenerate) background seems to have more degrees of freedom than those in the fully nonlinear regime \cite{ChTrZ}. This is due to the fact that the first class constraints fail to be functionally independent on some regions of the configuration space. This second type of degeneracy, that makes the linearized approximation inadequate is called \textbf{irregularity}, and has been discussed in \cite{MiZ}, where two types of irregularitiesit are distinguished :

\textbf{Type I}, those occurring when a constraint is the product of two or more regular, independent constraints (typically of the form $\Phi=\varphi_1 \varphi_2\approx 0$, with $\varphi_1$, $\varphi_1$ regular constraints).

\textbf{Type II}, those in which the gradient of a constraint vanishes on the constraint surface (typically, $\Phi=\varphi^k\approx 0$, where $\varphi$ is a regular constraint).

Type I are regularizable in the sense that they can be replaced by dynamically equivalent regular systems on different patches of phase space. In these systems the evolution is regular and smooth across boundaries from one patch to the other.  Irregularities of type II, instead, are simply ill-defined systems which cannot be consistently regularized.

Chern-Simons systems are of type I. The paradox encountered in \cite{ChTrZ} originates in an unfortunate choice of background corresponding to a configuration where two or more constraints overlap, destroying their independence. The linearized analysis gives the wrong description because some constraints disappear from the system, making the system look as if there were fewer constraints than those actually present.

\newpage

\end{document}